\documentclass[12pt]{article}
\usepackage{amsfonts,amssymb,graphicx}
\usepackage{fullpage}
\usepackage{subfig}
\usepackage{diagbox}
\usepackage{rotating}
\usepackage{sidecap}
\pdfoutput=1
\usepackage{ifpdf}
\ifpdf
 \usepackage{color}
 \definecolor{darkblue}{rgb}{0.3,0.3,0.6}
 \definecolor{darkgreen}{rgb}{0,0.6,0}
\else
 
\fi
\usepackage[numbers,compress]{natbib}
\usepackage{a4wide}
\usepackage{longtable}
\usepackage[table]{xcolor}
\usepackage{graphicx}
\usepackage[centertags]{amsmath}
\usepackage{multirow}
\usepackage{slashed}
\usepackage{pdflscape}
\usepackage{caption}
\usepackage{MnSymbol}
\usepackage{marginnote}

\usepackage{mathtools}
\usepackage{bbm}
\usepackage{dsfont}
\usepackage{hhline}

\newcommand{\sgn}{{\rm sgn}}

\def\muc{\multicolumn}

\def\Z{\mathbb{Z}}

\def\ov{\overline}

\def\ov{\overline}
\def\1{{\bf 1}}
\def\2{{\bf 2}}
\def\3{{\bf 3}}
\def\4{{\bf 4}}
\def\6{{\bf 6}}
\def\8{{\bf 8}}
\def\OR{\Omega\mathcal{R}}

\def\targ#1#2{\genfrac{[}{]}{0pt}{}{#1}{#2}}

\def\targ2#1#2{\genfrac{}{}{0pt}{}{#1}{#2}}

\definecolor{blus}{rgb}{0.1,0.1,0.8}
\definecolor{GreenYellow}{cmyk}{0.15,0,0.69,0}
\definecolor{Yellow}{cmyk}{0,0,1,0}
\definecolor{Goldenrod}{cmyk}{0,0.10,0.84,0}
\definecolor{Dandelion}{cmyk}{0,0.29,0.84,0}
\definecolor{Apricot}{cmyk}{0,0.32,0.52,0}
\definecolor{Peach}{cmyk}{0,0.50,0.70,0}
\definecolor{Melon}{cmyk}{0,0.46,0.50,0}
\definecolor{YellowOrange}{cmyk}{0,0.42,1,0}
\definecolor{Orange}{cmyk}{0,0.61,0.87,0}
\definecolor{BurntOrange}{cmyk}{0,0.51,1,0}
\definecolor{Bittersweet}{cmyk}{0,0.75,1,0.24}
\definecolor{RedOrange}{cmyk}{0,0.77,0.87,0}
\definecolor{Mahogany}{cmyk}{0,0.85,0.87,0.35}
\definecolor{Maroon}{cmyk}{0,0.87,0.68,0.32}
\definecolor{BrickRed}{cmyk}{0,0.89,0.94,0.28}
\definecolor{Red}{cmyk}{0,1,1,0}
\definecolor{OrangeRed}{cmyk}{0,1,0.50,0}
\definecolor{RubineRed}{cmyk}{0,1,0.13,0}
\definecolor{WildStrawberry}{cmyk}{0,0.96,0.39,0}
\definecolor{Salmon}{cmyk}{0,0.53,0.38,0}
\definecolor{CarnationPink}{cmyk}{0,0.63,0,0}
\definecolor{Magenta}{cmyk}{0,1,0,0}
\definecolor{VioletRed}{cmyk}{0,0.81,0,0}
\definecolor{Rhodamine}{cmyk}{0,0.82,0,0}
\definecolor{Mulberry}{cmyk}{0.34,0.90,0,0.02}
\definecolor{RedViolet}{cmyk}{0.07,0.90,0,0.34}
\definecolor{Fuchsia}{cmyk}{0.47,0.91,0,0.08}
\definecolor{Lavender}{cmyk}{0,0.48,0,0}
\definecolor{Thistle}{cmyk}{0.12,0.59,0,0}
\definecolor{Orchid}{cmyk}{0.32,0.64,0,0}
\definecolor{DarkOrchid}{cmyk}{0.40,0.80,0.20,0}
\definecolor{Purple}{cmyk}{0.45,0.86,0,0}
\definecolor{Plum}{cmyk}{0.50,1,0,0}
\definecolor{Violet}{cmyk}{0.79,0.88,0,0}
\definecolor{RoyalPurple}{cmyk}{0.75,0.90,0,0}
\definecolor{BlueViolet}{cmyk}{0.86,0.91,0,0.04}
\definecolor{Periwinkle}{cmyk}{0.57,0.55,0,0}
\definecolor{CadetBlue}{cmyk}{0.62,0.57,0.23,0}
\definecolor{CornflowerBlue}{cmyk}{0.65,0.13,0,0}
\definecolor{MidnightBlue}{cmyk}{0.98,0.13,0,0.43}
\definecolor{NavyBlue}{cmyk}{0.94,0.54,0,0}
\definecolor{RoyalBlue}{cmyk}{1,0.50,0,0}
\definecolor{Blue}{cmyk}{1,1,0,0}
\definecolor{Cerulean}{cmyk}{0.94,0.11,0,0}
\definecolor{Cyan}{cmyk}{1,0,0,0}
\definecolor{ProcessBlue}{cmyk}{0.96,0,0,0}
\definecolor{SkyBlue}{cmyk}{0.62,0,0.12,0}
\definecolor{Turquoise}{cmyk}{0.85,0,0.20,0}
\definecolor{TealBlue}{cmyk}{0.86,0,0.34,0.02}
\definecolor{Aquamarine}{cmyk}{0.82,0,0.30,0}
\definecolor{BlueGreen}{cmyk}{0.85,0,0.33,0}
\definecolor{Emerald}{cmyk}{1,0,0.50,0}
\definecolor{JungleGreen}{cmyk}{0.99,0,0.52,0}
\definecolor{SeaGreen}{cmyk}{0.69,0,0.50,0}
\definecolor{Green}{cmyk}{1,0,1,0}
\definecolor{ForestGreen}{cmyk}{0.91,0,0.88,0.12}
\definecolor{PineGreen}{cmyk}{0.92,0,0.59,0.25}
\definecolor{LimeGreen}{cmyk}{0.50,0,1,0}
\definecolor{YellowGreen}{cmyk}{0.44,0,0.74,0}
\definecolor{SpringGreen}{cmyk}{0.26,0,0.76,0}
\definecolor{OliveGreen}{cmyk}{0.64,0,0.95,0.40}
\definecolor{RawSienna}{cmyk}{0,0.72,1,0.45}
\definecolor{Sepia}{cmyk}{0,0.83,1,0.70}
\definecolor{Brown}{cmyk}{0,0.81,1,0.60}
\definecolor{Tan}{cmyk}{0.14,0.42,0.56,0}
\definecolor{Gray}{cmyk}{0,0,0,0.50}
\definecolor{Black}{cmyk}{0,0,0,1}
\definecolor{White}{cmyk}{0,0,0,0}
\definecolor{LightGray}{gray}{0.8}

\definecolor{mygr}{rgb}{0,0.6,0}
\definecolor{mygrey}{rgb}{0,0.1,0.2}
\definecolor{myblue}{rgb}{0,0.5,0.9}
\definecolor{myblue2}{rgb}{0,0.5,0.5}
\definecolor{myorange}{rgb}{1,0.5,0}
\definecolor{mypurple}{rgb}{0.6,0,1}
\definecolor{mygolden}{rgb}{1,0.8,0.2}

\newcommand{\bCaptionfonts}{\small}
\makeatletter
\long\def\@makecaption#1#2{%
 \vskip\abovecaptionskip
 \sbox\@tempboxa{{\bCaptionfonts #1: #2}}%
 \ifdim \wd\@tempboxa >\hsize
 {\bCaptionfonts #1: #2\par}
 \else
 \hbox to\hsize{\hfil\box\@tempboxa\hfil}%
 \fi
 \vskip\belowcaptionskip}
\makeatother

\makeatletter
\let\ORIGINALlatex@openbib@code=\@openbib@code
\renewcommand{\@openbib@code}{\ORIGINALlatex@openbib@code\setlength{\itemsep}{1ex plus.5ex minus.5ex}\setlength{\parsep}{0pt}}
\makeatother

\renewcommand{\arraystretch}{1.2}

\allowdisplaybreaks[4]

\begin{document}
\begin{center}
\begin{flushright}
{\small MITP/17-055\\ 
\today}

\end{flushright}

\vspace{25mm}
{\Large\bf Massless Spectra and Gauge Couplings at One-Loop

\vspace{3mm}

 on Non-Factorisable Toroidal Orientifolds}
\vspace{12mm}

{\large Mikel Berasaluce-Gonz\'alez${}^{\clubsuit}$, Gabriele Honecker${}^{\heartsuit}$ and Alexander Seifert${}^{\spadesuit}$
}

\vspace{8mm}
{
\it PRISMA Cluster of Excellence \& Institut f\"ur Physik (WA THEP), \\Johannes-Gutenberg-Universit\"at, D-55099 Mainz, Germany
\;$^{\clubsuit}${\tt mberasal@uni-mainz.de},~$^{\heartsuit}${\tt Gabriele.Honecker@uni-mainz.de},~$^{\spadesuit}${\tt alseifer@uni-mainz.de}}

\vspace{15mm}{\bf Abstract}\\[2ex]\parbox{140mm}{
So-called `non-factorisable' toroidal orbifolds can be rewritten in a factorised form as a product of three two-tori
by imposing an additional shift symmetry.
This finding of Blaszczyk {\it et al.}~\cite{Blaszczyk:2011hs}
provides a new avenue to Conformal Field Theory methods, by which the vector-like massless matter spectrum
- and thereby the type of gauge group enhancement on orientifold invariant fractional D6-branes -  and 
the one-loop corrections to the gauge couplings in Type IIA orientifold theories can be computed in addition to 
the well-established chiral matter spectrum derived from topological intersection numbers among three-cycles.
We demonstrate  this framework for the $\mathbb{Z}_4 \times \Omega\mathcal{R}$ orientifolds on the $A_3 \times A_1 \times B_2$-type torus. 
As observed before for factorisable backgrounds, also here the one-loop correction can drive the gauge groups to stronger coupling
as demonstrated by means of a four-generation Pati-Salam example.
}
\end{center}
\thispagestyle{empty}
\clearpage 

\tableofcontents
\setlength{\parskip}{1em plus1ex minus.5ex}
%

%%%%%%%%%%%%%%%%%%%%%%%%%%%%%%%%%%%%%%%%%%%%%%%%%%%%%%%%%%%%%%%%%%%%%%%%%%%%%%%%%%%
%%%%%%%%%%%%%%%%%%%%%%%%%%%%%%%%%%%%%%%%%%%%%%%%%%%%%%%%%%%%%%%%%%%%%%%%%%%%%%%%%%%

%\include{introduction}
%%%%%%%%%%%%%%%%%%%%%%%%%%%%%%%%%%%%%%%%%%%%%%%%%%%%%%%%%%%%%%%%%
\section{Introduction}

D6-brane model building in Type IIA string theory and the development of the associated low-energy effective field theory using Conformal Field Theory (CFT) on the worldsheet
have to date mainly focused on Abelian toroidal orbifold/orientifold backgrounds, where the underlying six-torus can be factorised into a product of three two-tori~\cite{Blumenhagen:1999ev,Forste:2000hx,Ibanez:2001nd,Blumenhagen:2001te,Cvetic:2001tj,Cvetic:2001nr,Blumenhagen:2002gw,Blumenhagen:2002wn,Abel:2003vv,Honecker:2003vq,Cremades:2003qj,Cvetic:2003ch,Abel:2003yx,Lust:2004cx,Honecker:2004kb,Gmeiner:2005vz,Abel:2005qn,Bailin:2006zf,Bailin:2007va,Gmeiner:2007zz,Bailin:2008xx,Gmeiner:2008xq,Honecker:2012qr,Ecker:2014hma,Ecker:2015vea},\footnote{For further models in the dual Type IIB orientifold set-up with magnetised D-branes see e.g.~\cite{Pradisi:1999ii,Larosa:2003mz,Angelantonj:2005hs,Abe:2008fi,Angelantonj:2009yj,Camara:2009uv,Angelantonj:2011hs,Abe:2017gye}.}
see e.g. the review articles~\cite{Angelantonj:2002ct,Blumenhagen:2005mu,Blumenhagen:2006ci,Schellekens:2013bpa,Honecker:2016gyz} and textbooks~\cite{Ibanez:2012zz,Blumenhagen:2013fgp}
for more comprehensive lists of references. To our best knowledge, the earliest and for many years only works using other background lattices, so-called `non-factorisable' tori, 
were~\cite{Blumenhagen:2004di,Forste:2007zb,Forste:2008ex,Bailin:2013sya}, with a first investigation of chiral spectra on the non-factorisable $\Z_4$ orientifolds in~\cite{Seifert:2015fwr,Berasaluce-Gonzalez:2016kqb}.\footnote{In the context of heterotic string theory, a systematic treatment of orbifolds with non-factorisable background lattices was advertised in~\cite{Fischer:2012qj} and added to the `orbifolder' software~\cite{Nilles:2011aj} which computes massless spectra and superpotential couplings from input data such as shift vectors and Wilson lines.} Based on the counting of {\it special Lagrangian (sLag)} three-cycles and inspections of their non-trivial intersection numbers, the $A_3 \times A_3$ background lattice could be excluded, while the $A_3 \times A_1 \times B_2$ background lattice showed first promising results in view of phenomenologically appealing spectra.

The void in the quest for phenomenologically appealing four-dimensional Type II string vacua with control over the associated low-energy effective action beyond the mere dimensional reduction of the (closed string)  supergravity and (purely gauge sector) Dirac-Born-Infeld actions, which all belong to the tree-level, is particularly astonishing in view of the postulated richness of flux vacua with moduli stabilization, see e.g. the reviews~\cite{Grana:2005jc,Koerber:2010bx,Tsimpis:2016bbq}. Vacuum expectation values  ({\it vev}s) of higher $p$-form fluxes are naturally associated to (twisted) $p$-dimensional tori $T^p$, see e.g.~\cite{Marchesano:2006ns}, which includes in particular the NS-NS three-form $H_3$ on a three-torus $T^3$, motivating again the $\Z_4$ orientifolds involving some $A_3$-type torus background.

Even before contemplating closed string background fluxes, non-factorisable lattices generically have by construction a reduced number of geometric moduli compared to the factorisable toroidal backgrounds. For the $\Z_4$ invariant toroidal backgrounds (before orientifolding), these are enumerated by $(h_{1,1}^{twisted} \, , \, h_{2,1}^{twisted}) = (26,6)$, $(22,2)$ and $(20,0)$ for the $B_2 \times B_2 \times A_1^2$, the $A_3 \times A_1 \times B_2$ and the $A_3 \times A_3$ lattices, respectively.

Unexpected symmetries between different lattice orientations obtained by some rotation over a non-supersymmetric angle have in recent years
provided means to reduce the computational effort of systematic computer scans for phenomenologically interesting models on factorisable toroidal orbifold backgrounds on the one hand 
and served as non-trivial cross-checks for expressions to compute the massless matter spectrum and terms in the low-energy effective action on the other hand~\cite{Gmeiner:2008xq,Forste:2010gw,Honecker:2012qr,Ecker:2014hma}. First hints that this concept can be transferred to non-factorisable toroidal orientifolds were found in~\cite{Berasaluce-Gonzalez:2016kqb}  for the $\Z_4 \times \OR$ case on the $A_3 \times A_1 \times B_2$ lattice further discussed here.

Despite the folklore that computations of the associated low-energy effective action are straightforward, to date only vacuum amplitudes to determine the 
RR tadpole cancellation conditions and by magnetic gauging along the non-compact directions~\cite{Bachas:1996zt,Antoniadis:1999ge} the one-loop gauge thresholds 
could be generalised from pure bulk D6-branes on the factorisable six-torus~\cite{Lust:2003ky,Akerblom:2007np} to fractional D6-branes on factorisable toroidal orbifolds~\cite{Blumenhagen:2007ip,Gmeiner:2009fb,Honecker:2011sm,Honecker:2011hm}. The derivation provides an indirect and - albeit seemingly tedious - very robust framework to determine 
the massless vector-like open string spectrum via the associated one-loop beta function coefficients. A special case, the identification of gauge group enhancements  $U(1) \hookrightarrow USp(2)$, is of particular importance not only to model the Standard Model weak interactions, but also in order to determine possibly non-trivial D2-brane $O(1)$ instanton corrections to the effective action, to compute the K-theory constraints and to calculate the ultimate coupling selection rules due to remnant discrete gauge symmetries $\Z_n \subset U(1)_{\text{massive}}$~\cite{BerasaluceGonzalez:2011wy,Ibanez:2012wg,BerasaluceGonzalez:2012zn,Honecker:2013sww,Honecker:2013kda,Honecker:2015ela}. 
The finding in~\cite{Blaszczyk:2011hs} that any non-factorisable toroidal orbifold can be formulated in a factorisable toroidal background by imposing an additional shift symmetry 
will be of crucial importance here in order to generalise the known results on gauge threshold corrections.

This article is organised as follows: in section~\ref{nonfactorisable_geometry}, we briefly review the metric degrees of freedom and three-cycles of the $\Z_4$ orbifold on the non-factorisable lattice of type $A_3 \times A_1 \times B_2$. We then proceed in section~\ref{factorisable_geometry} with a detailed discussion of a factorised description of the metric and three-cycles for all eight possible orientations of the $A_3 \times A_1 \times B_2$-type lattice under the anti-holomorphic involutions ${\cal R}_{i=1,2,3,4}$ accompanying the worldsheet parity operation. After discussing the supersymmetry, RR tadpole cancellation and K-theory conditions, we proceed  with a discussion of the massless closed string spectrum, which has to our best knowledge not been computed before and supports the hint on pairwise relations among the different lattice orientations first proposed in~\cite{Berasaluce-Gonzalez:2016kqb}; then we very briefly review the counting of chiral states in terms of topological intersection numbers, leaving the correct identification of enhanced $USp(2M)$ versus $SO(2M)$ gauge groups and the counting of massless vector-like (open string) matter states for the later section~\ref{amplitudes}. Section~\ref{duality} contains a detailed discussion of the conjectured pairwise duality relations among lattice orientations in terms of the factorised three-cycle geometry. Finally, in section~\ref{amplitudes} we are ready to compute one-loop corrections to the gauge couplings, using the conjectured duality relations as a guiding principle whenever unexpected subtleties due to non-trivial discrete D6-brane data and the shift symmetry in the factorised lattice description occur. 
The separation of the one-loop amplitudes into contributions from massless open string states and threshold corrections due to massive string excitations then provides us with the correct identification of the gauge group enhancement and the vector-like matter spectrum. 
In section~\ref{example}, we apply the formalism to an explicit four-generation Pati-Salam model, before concluding in section~\ref{conclusion} with an outlook. 
%on the extension of our strategy to find a factorised CFT description to other non-factorisable backgrounds. 
Technical details on the implementation of a correction factor in the factorised intersection numbers and vacuum amplitudes due to the shift symmetry are for completeness presented in appendix~\ref{kappa}.

%%%%%%%%%%%%%%%%%%%%%%%%%%%%%%%%%%%%%%%%%%%%%%%%%%%%%%%%%%%%%%%%%%%%
\section{Non-factorisable $\Z_4$ orbifold with the lattice of type $A_3\times A_1 \times B_2$}\label{nonfactorisable_geometry}

\begin{figure}[h!]
\begin{center}
\includegraphics[width=16cm]{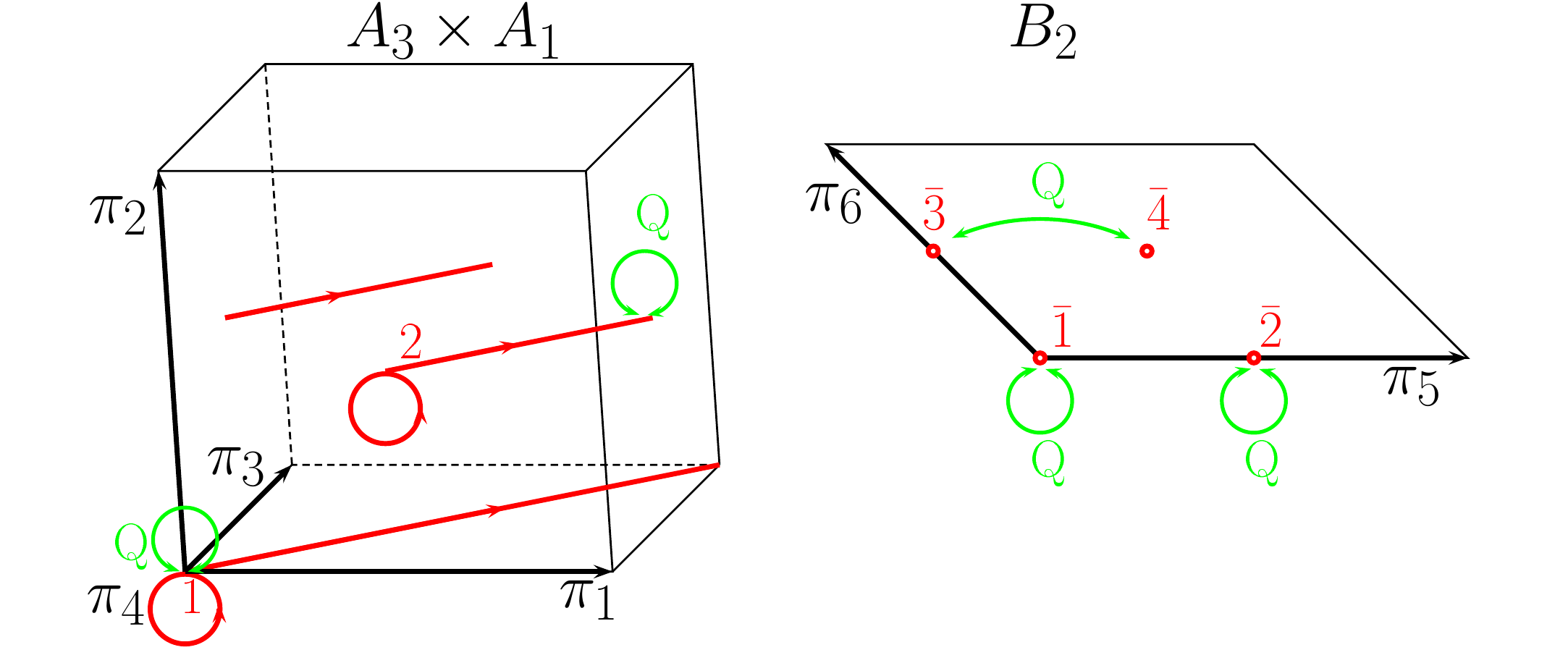}
\caption{$T^6/\Z_4$ orbifold on the $A_3\times A_1\times B_2$ lattice and its $\Z_2$ fixed lines (\textcolor{red}{in red}). The two-torus is a usual square torus spanned (up to overall scaling) by the two simple roots of the Lie algebra $B_2$ and with the complex structure $\frac{-1+i}{2}$. }
\label{fig:non-factorised-picture}
\end{center}
\end{figure}

There exist two different so-called non-factorisable $\Z_4$ orbifold backgrounds: with the lattice of type $A_3\times A_3$ and of type $A_3\times A_1\times B_2$. Due to the 
three-cycle topology and geometry, only orbifolds of the second type are interesting  for model building in Type IIA string theory~\cite{Seifert:2015fwr,Berasaluce-Gonzalez:2016kqb}, see figure  \ref{fig:non-factorised-picture}.
The $\Z_4$ action  is generated by the Coxeter element which acts on the root lattice spanned by simple roots $\{e_i\}_{i=1,\dots,6}$. Defining the fundamental one-cycles $\pi_i$ along the simple roots $e_i$, they have then the same transformation behaviour under the $\Z_4$ action:
\begin{equation}\label{Q-action}
\begin{array}{llll}
&\mathcal{Q}\pi_1=\pi_2\,,  &\mathcal{Q}\pi_2=\pi_3\,,   &\mathcal{Q}\pi_3=-\pi_1-\pi_2-\pi_3\,,\\
&\mathcal{Q}\pi_4=-\pi_4\,,  &\mathcal{Q}\pi_5=\pi_5+2\pi_6\,,   &\mathcal{Q}\pi_6=-\pi_5-\pi_6\,.
\end{array}
\end{equation}
Since this action generates a discrete subgroup of $SU(3)$, it preserves $\mathcal{N}=2$ supersymmetry in four dimension in Type II string theory compactifications. The Hodge numbers per untwisted and twisted sectors are given by (cf. e.g.~\cite{Lust:2006zh}):
\begin{equation}\label{hodge-numbers}
\begin{split}
h_{2,1}&=h_{2,1}^{untw} +h^{\Z_2}_{2,1} \hspace{10.5mm} =1+2\,,\\
h_{1,1}&=h_{1,1}^{untw}+h^{\Z_2}_{1,1} +h^{\Z_4}_{1,1}=5+6+16\,.
\end{split}
\end{equation}  
From solving the equation $\mathcal{Q}^tg\,\mathcal{Q}=g$ we obtain the following shape of the metric of the underlying six-torus:
\begin{equation}\label{non-factorised-metric}
g:=e_i\cdot e_j=\small\begin{pmatrix}
R_3^2&\hat{a}R_3^2&-(1+2\hat{a})R_3^2&\hat{d}R_3R_1&\hat{b}R_3R_2&\hat{c}R_3R_2\\
\hat{a}R_3^2&R_3^2&\hat{a}R_3^2&-\hat{d}R_3R_1&-(\hat{b}+2\hat{c})R_3R_2&(\hat{b}+\hat{c})R_3R_2\\
-(1+2\hat{a})R_3^2&\hat{a}R_3^2&R_3^2&\hat{d}R_3R_1&-\hat{b}R_3R_2&-\hat{c}R_3R_2\\
\hat{d}R_3R_1&-\hat{d}R_3R_1&\hat{d}R_3R_1&R_1^2&0&0\\
\hat{b}R_3R_2&-(\hat{b}+2\hat{c})R_3R_2&-\hat{b}R_3R_2&0&2R_2^2&-R_2^2\\
\hat{c}R_3R_2&(\hat{b}+\hat{c})R_3R_2&-\hat{c}R_3R_2&0&-R_2^2&R_2^2\end{pmatrix}.
\end{equation}
The real positive moduli $R_3$, $R_1$ and $R_2$ describe the radii of $A_3\times A_1 \times B_2$, respectively, and $\hat{a}$, $\hat{b}$, $\hat{c}$ and $\hat{d}$ specify the cosines of angles between the vectors of the lattice. More precisely, $\hat{a}$ is the cosine of the angle between the root vectors $e_1$ and $e_2$, $\hat{d}$ the cosine of the angle between $e_1$ and $e_4$, $\hat{b}$ ($\hat{c}$) is the cosine of the angle between $e_1$ and $e_5$ ($e_6$).\footnote{Mathematically rigorously, the direct product of Lie algebra lattices $A_3\times A_1 \times B_2$ requires $\hat{a}=-\frac{1}{2}$ and $\hat{b}=\hat{c}=\hat{d}=0$, whereas we loosely use the notation to denote a $\Z_4$ invariant six-torus.}

By taking orbits of the $\mathcal{Q}$-action, we can define a basis of the $\Z_4$-invariant \textit{bulk} three-cycles:
\begin{equation}\label{2.bulk_basis} 
\begin{split}
\gamma_1&:=-\sum_{i=0}^3 \mathcal{Q}^i\pi_{136}=
2(\pi_{125}+\pi_{126}-\pi_{136}-\pi_{235}-\pi_{236})\,,\\
\gamma_2&:=-\sum_{i=0}^3 \mathcal{Q}^i\pi_{125}=
2(\pi_{126}+\pi_{135}+\pi_{136}-\pi_{236})\,,\\
\bar{\gamma}_1&:=\sum_{i=0}^3 \mathcal{Q}^i\pi_{146}=
\pi_{145}+2\pi_{146}+2\pi_{245}+2\pi_{246}+\pi_{345}\,,\\
\bar{\gamma}_2&:=\sum_{i=0}^3 \mathcal{Q}^i\pi_{246}=
-\pi_{145}+2\pi_{246}+\pi_{345}+2\pi_{346}\,.
\end{split}
\end{equation} 
Here we used the shorthand notation $\pi_{ijk} \equiv \pi_i\wedge\pi_j\wedge\pi_k$. 
Note that for this particular choice of  basis, the linear combinations $\frac{1}{2}(\gamma_1\pm\gamma_2) \in H_3(T^6/\Z_4,\Z)$,
along which some O6-plane orbits will be extended according to table~\ref{tab:oplanes},
are also bulk three-cycles.\\
We can define the intersection number between two toroidal three-cycles $\pi_a^\text{torus}$ and $\pi_b^\text{torus}$ on the six-torus $T^6$ 
as $\pi^{\text{torus}}_a\circ\pi^{\text{torus}}_b=\pi^{\text{torus}}_a \wedge\pi^{\text{torus}}_b/\text{Vol}(T^6)$. 
Using the formula for the bulk intersection numbers~\cite{Blumenhagen:2002gw},
\begin{equation}\label{Eq:IntersectionNumber-with-Q}
\pi^\text{bulk}_a\circ\pi^\text{bulk}_b=\frac{1}{4}\bigl(\sum_{i=0}^{3}\mathcal{Q}^i\pi^{\text{torus}}_a\bigr)
\circ\bigl(\sum_{i=0}^{3}\mathcal{Q}^i\pi^{\text{torus}}_b\bigr),
\end{equation}
we compute the intersection numbers in the bulk basis $\{\gamma_{1/2}\,,\;\bar{\gamma}_{1/2}\}$:
\begin{equation}
\gamma_i\circ \bar{\gamma}_j=-2\delta_{ij}\,,\quad \gamma_i\circ\gamma_j=\bar{\gamma}_i\circ\bar{\gamma}_j=0 .
\end{equation}
Besides the four bulk three-cycles $\gamma_i,\bar{\gamma}_i$ (with $i=1,2$), there exist also 
four \textit{exceptional} three-cycles appearing in the $\mathbbm{Z}_2$ twisted sector of the orbifold. It is easy to see that $\mathcal{Q}^2$ acts trivially on the sub-manifold spanned by two one-cycles $\pi_1+\pi_3$ and $\pi_4$. One can show that there exist eight such $\mathcal{Q}^2$-invariant sub-manifolds, which are indicated in red in figure $\ref{fig:non-factorised-picture}$. We numerate them by $\alpha \bar{\beta}$ where the first index denotes the $\Z_2$ invariant two-tori ({\color{red}1,\;2}) on the $A_3\times A_1$-torus, also denoted as $T^4_{(3)}$ later on, and the second index denotes the $\Z_2$ invariant points ({$\color{red}\bar{1},\,\bar{2},\,\bar{3},\,\bar{4}$}) on the $B_2$-torus, also denoted as $T^2_{(3)}$ in the following.
Since the  $\mathcal{Q}^2$-invariant sub-manifolds transform under the $\mathcal{Q}$-action as
\begin{equation}\label{fix-point-trafo}
e_{i\bar{1}}\,\overset{\mathcal{Q}}{\longrightarrow}\,e_{i\bar{1}}\,,\quad e_{i\bar{2}}\,\overset{\mathcal{Q}}{\longrightarrow}\,e_{i\bar{2}}\,,\quad e_{i\bar{3}}\,\overset{\mathcal{Q}}{\longrightarrow}\,e_{i\bar{4}}\,,\quad e_{i\bar{4}}\,\overset{\mathcal{Q}}{\longrightarrow}\,e_{i\bar{3}} \quad\text{for } i=1,2\,,
\end{equation}
they can be arranged in six congruence classes: $\{e_{1\bar{1}}\},\,\{e_{1\bar{2}}\},\,\{e_{1\bar{3}},\,e_{1\bar{4}}\},\,\{e_{2\bar{1}}\},\,\{e_{2\bar{2}}\},\,\{e_{2\bar{3}},\,e_{2\bar{4}}\}. $ 
The resolution of these six $\Z_2$ singular sub-manifolds gives rise to six four-dimensional sub-manifolds, where the exceptional two-cycle $\mathbf{e}_{\alpha \bar{\beta}}$ describes the $S^2$-part, and the two one-cycles $\pi_1+\pi_3$ and $\pi_4$ span a two-torus. The index of $\mathbf{e}_{\alpha \bar{\beta}}$ is inherited from the enumeration of the $\Z_2$ invariant two-tori. Finally by splitting the $\Z_2$ invariant two-torus into one-cycles $\pi_1+\pi_3$ and $\pi_4$, we construct $\mathcal{Q}$-invariant exceptional three-cycles. Due to the transformation \eqref{fix-point-trafo} and the inversion of the one-cycles under the action of $\mathcal{Q}$, only the exceptional two-cycles $\mathbf{e}_{1\bar{3}},\, \mathbf{e}_{1\bar{4}},\,\mathbf{e}_{2\bar{3}},\,\mathbf{e}_{2\bar{4}}$ provide non-trivial results in the construction. Thus, the exceptional three-cycles are given by: 
\begin{equation}\label{2.except_cycles} 
\begin{split}
\gamma_3&:=(\mathbf{e}_{1\bar{3}}-\mathbf{e}_{1\bar{4}})\wedge(\pi_1+\pi_3)\,,\qquad \bar{\gamma}_3:=(\mathbf{e}_{1\bar{3}}-\mathbf{e}_{1\bar{4}})\wedge\pi_4\,,\\
\gamma_4&:=(\mathbf{e}_{2\bar{3}}-\mathbf{e}_{2\bar{4}})\wedge(\pi_1+\pi_3)\,,\qquad \bar{\gamma}_4:=(\mathbf{e}_{2\bar{3}}-\mathbf{e}_{2\bar{4}})\wedge\pi_4 \,,
\end{split}
\end{equation} 
with the intersection numbers:
\begin{equation}\label{2.intersection_number_exc_cycles} 
\gamma_i\circ\bar{\gamma}_j=2\delta_{ij}\,,\quad\gamma_i\circ\gamma_j=\bar{\gamma}_i\circ\bar{\gamma}_j=0\qquad i=3,4\,.
\end{equation}
Due to the non uni-modular intersection form of $\{\gamma_i,\,\bar{\gamma}_i\}_{i=1,2,3,4}$ these three-cycles do not build the minimal integral basis. In \cite{Berasaluce-Gonzalez:2016kqb} using the fractional three-cycles which consist of half a bulk cycle and simultaneously of half an exceptional cycle, 
\begin{equation}\label{fractional-cycle}
\pi^\text{frac}=\frac{1}{2}\pi^\text{bulk}+\frac{1}{2}\pi^\text{exc} ,
\end{equation}
we constructed an unimodular basis which forms an $F_4\oplus F_4$-lattice.

%%%%%%%%%%%%%%%%%%%%%%%%%%%%%%%%%%%%%%%%%%%%
\section{Fundamentals of the factorisable picture}\label{factorisable_geometry}

%%%%%%%%%%%%%%%%%%%%%%
\subsection{Factorisation and geometry}\label{factor+geometry}
Compared to the factorisable orbifolds considered e.g. in~\cite{Blumenhagen:2002gw,Honecker:2004kb,Bailin:2006zf,Bailin:2007va,Gmeiner:2007zz,Bailin:2008xx,Gmeiner:2008xq},
 the non-factorisable structure of the $A_3\times A_1\times B_2$ lattice demands more effort to construct fractional three-cycles and to calculate the orientifold projections and supersymmetry conditions. However, this difficulty can be resolved by a trick.
 
\begin{figure}[h!]
\begin{center}
\includegraphics[width=16cm]{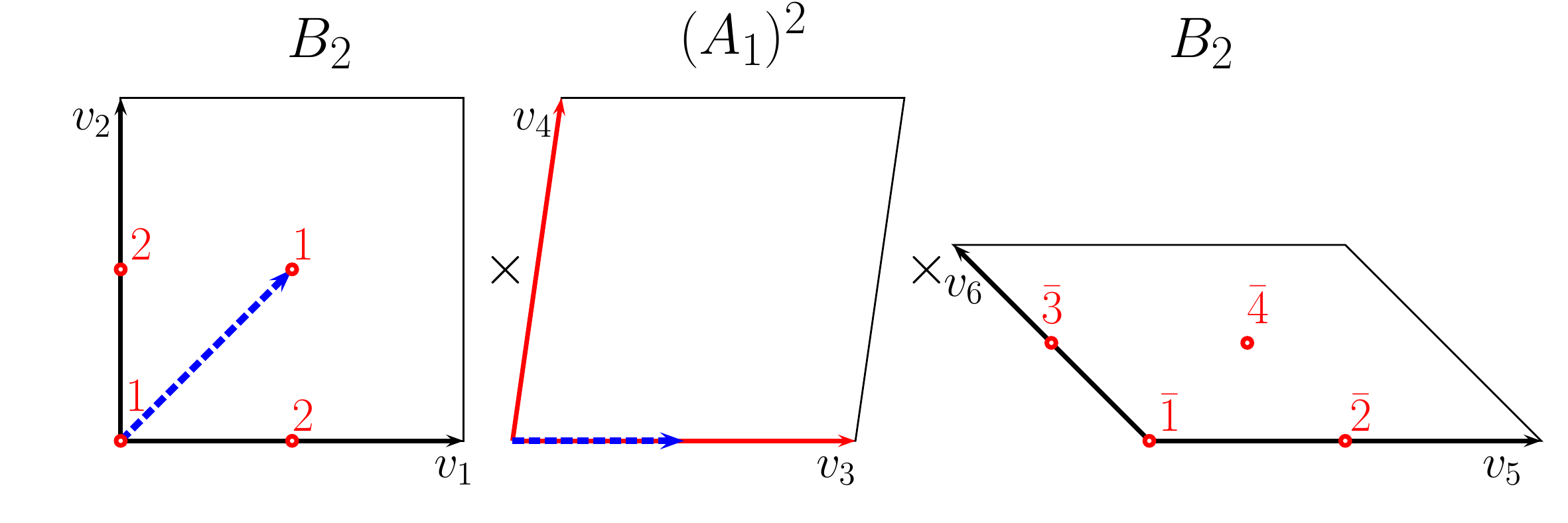}
\caption{Factorised form of the $A_3\times A_1\times B_2$ orbifold. The two $B_2$-tori differ in the choice of the complex structure. Since the last one has the complex structure $ \frac{-1+i}{2}$, the first one has $i$ as the complex structure. The shift symmetry on the four-torus $A_3\times A_1$ acts along $\frac{v_1+v_2+v_3}{2}$ (\textcolor{blue}{in blue}). The $\Z_2$ invariant lines are depicted in \textcolor{red}{red}. The shape of the $(A_1)^2$-torus remains arbitrary for the $\Z_4$ orbifold and is only fixed  by the 
anti-holomorphic involution ${\cal R}$ pertaining to the orientifold projection $\Omega\mathcal{R}$.}
\label{fig:factorised-picture}
\end{center}
\end{figure}

It was shown in \cite{Blaszczyk:2011hs} that non-factorisable toroidal orbifolds can be written in a factorisable form by imposing an extra shift symmetry. In the particular case of the $A_3\times A_1\times B_2$ lattice, in which we are interested, this is achieved by introducing a new basis of one cycles,
\begin{equation}\label{new-basis}
\begin{array}{llll}
&v_1:=\pi_1+\pi_2\,,&v_3:=\pi_1+\pi_3\,,&v_5:=\pi_5\,,\\
&v_2:=\pi_2+\pi_3\,,&v_4:=\pi_4\,,&v_6:=\pi_6\,,
\end{array}
\end{equation}
where $\{\pi_i\}$ is the original basis of one-cycles introduced in equation~\eqref{Q-action}. 
With respect to the new basis $\{v_i\}$, the $A_3\times A_1\times B_2$ torus is decomposed into three two tori $\left(T^2\right)^3$ (see figure \ref{fig:factorised-picture})
with the metric:
\begin{equation}\label{eq:g-factorised}
g=\text{diag}\left(
\begin{array}{cc}
2(1+\hat{a})R_3^2& 0 \\
 0 & 2(1+\hat{a})R_3^2 \\
\end{array}
\right)\oplus\left(
\begin{array}{cc}
-4\hat{a}R_3^2 & 2\hat{d}R_1R_3 \\
2\hat{d}R_1R_3 & R_1^2 \\
\end{array}
\right)\oplus\left(
\begin{array}{cc}
2R_2^2 & -R_2^2 \\
 -R_2^2 & R_2^2 \\
\end{array}
\right)\,.
\end{equation}
On the new basis, the $\Z_4$ action \eqref{Q-action} acts as follows:
\begin{equation}\label{Q-action-factorised}
\begin{array}{llll}
&\mathcal{Q}v_1=v_2\,,  &\mathcal{Q}v_3=-v_3\,,   &\mathcal{Q}v_5=v_5+2v_6\,,\\
&\mathcal{Q}v_2=-v_1\,,  &\mathcal{Q}v_4=-v_4\,,   &\mathcal{Q}v_6=-v_5-v_6\,.
\end{array}
\end{equation}

But, as we already mentioned, the basis change gives rise to an additional shift symmetry, which identifies points on the factorised torus: 
\begin{equation}\label{shift-symmetry}
p \simeq p+\frac{1}{2}(v_1+v_2+v_3)\quad \text{for any }p\in T^6\,.
\end{equation}
The first advantage of the factorisation is the description of  {\it Lagrangian} ({\it Lag}) three-cycles on the torus. In the non-factorised picture such three-cycles are described by ten wrapping numbers and several additional non-linear constraints ({\it Lag} condition), see \cite{Berasaluce-Gonzalez:2016kqb} for further details. Using the new basis $\{v_i\}$, any {\it Lag} three-cycle on the torus in the factorised picture can be written just as a product of three one-cycles, one on each two-torus, in terms of six toroidal wrapping numbers $(n^i,\,m^i)_{i=1,2,3}$:
\begin{equation}\label{fractorised-fact-cycle}
\pi^\text{torus}=(n^1v_1+m^1v_2)\wedge (n^2v_3+m^2v_4)\wedge (n^3v_5+m^3v_6)\,.
\end{equation}
As in the usual factorisable backgrounds, the wrapping numbers $(n^i,m^i)$ have to be coprime. Additionally, if the toroidal three-cycles traverse the $\Z_2$ invariant points on the four-torus $T^4_{(2)} \equiv T^2_{(1)}\times T^2_{(3)}$, they are $\Z_2$ invariant.

It is noteworthy that, when depicting the toroidal three-cycles in the factorisable picture, sometimes one needs to introduce their replicas under the shift symmetry defined in equation~\eqref{shift-symmetry}, which reflects the fact that a cell in the factorisable picture actually corresponds to two unit cells in the non-factorisable picture. These replicas will play a crucial role in the following  calculations.

Since any three-cycle with wrapping numbers  $(n^1,m^1;n^2,m^2)=$(odd,\,odd;\,odd,\,even) on the four-torus  $T^2_{(1)}\times T^2_{(2)}$  coincides with its replica, it passes through the same points twice.  To avoid this doubling, one has to allow the wrapping number $(n^1,m^1)$ or $n^2$ to take  half-integer values. In the following we choose the convention that $n^2\in\frac{\mathbb{Z}}{2}$ if $n^1+m^1$ is even.\footnote{Later in this article we will need all wrapping numbers to be integers. For cycles with half-integer $n^2$ we will double the wrapping numbers in the second two-torus and compensate for it by introducing the factor $\kappa$ when computing intersection numbers in section~\ref{ShiftSymm}. \label{Foot:integer}}

Any bulk three-cycle inherited from the toroidal three-cycle expressed as in equation~\eqref{fractorised-fact-cycle} can be decomposed with respect to the bulk basis \eqref{2.bulk_basis} as:
\begin{equation}\label{bulk-cycle}
\pi^\text{bulk}=P\gamma_1+Q\gamma_2 +\bar{P}\bar{\gamma}_1+\bar{Q}\bar{\gamma}_2\,,
\end{equation}
where the bulk wrapping numbers $P,\,Q,\,\bar{P},\,\bar{Q}$, expressed in terms of the toroidal wrapping numbers $(n^i,\,m^i)_{i=1,2,3}$, are given by:
\begin{equation}\label{bulk-wrapping-numbers}
\begin{array}{lll}
&P=((m^1-n^1)m^3-2m^1n^3)n^2\,, & Q=(2n^1n^3-(m^1+n^1)m^3)n^2\,,\\
&\bar{P}=((n^1-m^1)m^3+2m^1n^3)m^2\,,  &
\bar{Q}=((m^1+n^1)m^3-2n^1n^3)m^2\,.
\end{array}
\end{equation}
A general exceptional three-cycle at some $\mathbb{Z}_2$ invariant locus is given by:
\begin{subequations}\label{exc-three-cycles}
\begin{align}
\pi^\text{exc}&=(-1)^{\tau^0}2\sum_{k=0}^1\mathcal{Q}^k[(\textbf{e}_{\alpha\bar{\beta}_1}
+(-1)^{\tau^3}\textbf{e}_{\alpha\bar{\beta}_2})\wedge(n^2v_3+m^2v_4)]
\hspace{0.9cm}\text{for $n^1+m^1=$even},\\
\pi^{\text{exc}}&=(-1)^{\tau^0}\sum_{k=0}^1\mathcal{Q}^k[(\textbf{e}_{1\bar{\beta}_1}
+(-1)^{\tau^1}\textbf{e}_{2\bar{\beta}_1}
+(-1)^{\tau^3}\textbf{e}_{1\bar{\beta}_2}+(-1)^{\tau^1+\tau^3}
\textbf{e}_{2\bar{\beta}_2})\wedge(n^2v_3+m^2v_4)]\nonumber\\
&\hspace{10cm}\quad\text{for $n^1+m^1=$odd}.
\end{align}
\end{subequations}
$\textbf{e}_{\alpha\bar{\beta}}$ are the exceptional two-cycles stemming from the orbifold singularities where the index $\alpha=1,2$ now numerates the singular points on the first two-torus, $T^2_{(1)}$, and $\bar{\beta}=\bar{1},\bar{2},\bar{3},\bar{4}$ on the third  one, $T^2_{(3)}$. This notation agrees with the enumeration of $\Z_2$ invariant sub-manifolds in section~\ref{nonfactorisable_geometry}. Like on the factorisable orbifolds, the orientation of a reference exceptional two-cycle is described by the $\Z_2$ eigenvalue $(-1)^{\tau^0}$ with
$\tau^0\in\{0,1\}$. $\tau^{1,3}\in\{0,1\}$ encode the discrete Wilson lines on the corresponding two-torus $T^2_{(1),(3)}$. The shifts $\tilde{\sigma}^{k=1,2,5,6}\in\{0,1\}$ parametrise
the displacements from the origin along the basis one-cycles $v_{1,2}$ and $v_{5,6}$  on the tori $T^2_{(1),(3)}$, where $\tilde{\sigma}^i =1$  corresponds to the displacement by 
$\frac{v_i}{2}$.
Because displacing along the three-cycle itself does not change the exceptional part, one can restrict the consideration to the cases where the toroidal cycle is displaced in some non-parallel direction. Here and below, we choose the following notation for the spatial displacement $\sigma^3$ on the two torus $T^2_{(3)}$:
\begin{equation}\label{dicplacements}
\sigma^3=1\,\Leftrightarrow\;(\tilde{\sigma}^{5},\tilde{\sigma}^{6})=\left\lbrace
\begin{array}{rl}
(1,0)&\quad\text{for }(n^3,\,m^3)=\text{(odd, odd)}\,,\\
(0,1)&\quad\text{for }(n^3,\,m^3)=\text{(odd, even)}\,,\\
(1,0)&\quad\text{for }(n^3,\,m^3)=\text{(even, odd)}\,.\\
\end{array}
\right.
\end{equation}

\begin{table}[h!]
\centering
\begin{tabular}{|c||ccccc|}
\hline 
\multicolumn{6}{|c|}{\textbf{Wrapping numbers and $\Z_2$ fixed points}}\\
\hline\hline
$(n^3,m^3)$& (odd, odd) &\vline & (odd, even) &\vline & (even, odd) \\ 
\hline 
& && $ \sigma^3=0$ & & \\ 
\cline{2-6} 
& $e_{1\bar{1}}$, $e_{1\bar{4}}$, $e_{2\bar{1}}$, $e_{2\bar{4}}$ &\vline & $e_{1\bar{1}}$, $e_{1\bar{2}}$, $e_{2\bar{1}}$, $e_{2\bar{2}}$ &\vline & $e_{1\bar{1}}$, $e_{1\bar{3}}$, $e_{2\bar{1}}$, $e_{2\bar{3}}$ \\ 
\cline{2-6}
$n^1+m^1=$odd& && $\sigma^3=1$ &&  \\ 
\cline{2-6}
&$e_{1\bar{2}}$, $e_{1\bar{3}}$, $e_{2\bar{2}}$, $e_{2\bar{3}}$& \vline & $e_{1\bar{3}}$, $e_{1\bar{4}}$, $e_{2\bar{3}}$, $e_{2\bar{4}}$ &\vline & $e_{1\bar{2}}$, $e_{1\bar{4}}$, $e_{2\bar{2}}$, $e_{2\bar{4}}$  \\ 
\hline
 \hline
& && $ \sigma^3=0$ &&  \\ 
\cline{2-6} 
$n^1+m^1=$even& $e_{\alpha\bar{1}}$, $e_{\alpha\bar{4}}$ &\vline & $e_{\alpha\bar{1}}$, $e_{\alpha\bar{2}}$ &\vline & $e_{\alpha\bar{1}}$, $e_{\alpha\bar{3}}$ \\ 
\cline{2-6}
& && $\sigma^3=1$ &  &\\ 
\cline{2-6}
& $e_{\alpha\bar{2}}$, $e_{\alpha\bar{3}}$& \vline & $e_{\alpha\bar{3}}$, $e_{\alpha\bar{4}}$& \vline & $e_{\alpha\bar{2}}$, $e_{\alpha\bar{4}}$ \\
\hline
\end{tabular} 
\caption{Fixed points on $T^2_{(1)}\times T^2_{(3)}$ which are traversed by the toroidal cycles in dependence of the even-/oddness of wrapping numbers and of 
the displacement parameter $\sigma^3$. The first lower index takes the values $\alpha=1$ for $\sigma^1=0$ and $\alpha=2$ for $\sigma^1=1$ in the case of $n^1+m^1=$even.
}\label{Tab:fixed points 1b}
\end{table}
However, the non-factorisable structure (or the shift symmetry along $T_{(1)}^2$) leads to some differences compared to  factorisable orbifolds. If one takes a closer look at the structure of the exceptional three-cycles \eqref{exc-three-cycles}, one notes the absence of the discrete Wilson line $\tau^1$ in (\ref{exc-three-cycles}a). Moreover, in  table \ref{Tab:fixed points 1b} the reader can see which fixed points are passed through by the toroidal three-cycle for all possible displacements $\sigma^3$ on the third torus. Due to the shift symmetry \eqref{shift-symmetry}, on the first torus the displacements $\tilde{\sigma}^1$ and $\tilde{\sigma}^2$ coincide, so that we  can choose only one displacement $\sigma^1$  
 on the first torus along the direction $v_1=\pi_1+\pi_2$.
Moreover,  for the toroidal three-cycles with $n^1+m^1=$odd the non-parallel displacement does not change the brane configuration on the first two-torus due to the shift symmetry \eqref{shift-symmetry}(said otherwise, because of the shift symmetry we can always choose a configuration where the displacement occurs in the second two-torus).

Like the bulk three-cycles, any exceptional three-cycle which can be expand in the basis $\{\gamma_i,\,\bar{\gamma}_i\}_{i=3,4}$ of equation  \eqref{2.except_cycles} takes the form:
\begin{equation}\label{exceptional-three-cycles}
\pi^\text{exc}=p\,\gamma_3+q\,\gamma_4+\bar{p}\,\bar{\gamma}_3+\bar{q}\,\bar{\gamma}_4  \, ,
\end{equation}
with integer coefficients $p,\,q,\,\bar{p},\,\bar{q}$. These \textit{exceptional wrapping numbers} can be deduced from equation~\eqref{exc-three-cycles} and are given by:
\begin{equation}\label{exceptional-wrapping-numbers}
\begin{split}
& \left(\begin{array}{c}p\\ \bar{p}\\q\\ \bar{q}\end{array}\right)  
 =(-)^{\tau^0}2
 \left(\begin{array}{c}n^2 \delta_{\sigma^1,0}\\ m^2 \delta_{\sigma^1,0}\\n^2 \delta_{\sigma^1,1}\\ m^2 \delta_{\sigma^1,1}\end{array}\right)\times 
\left\lbrace \begin{array}{ll}
 1-(-1)^{\sigma^3\tau^3}&(n^3,\,m^3)=\text{(odd,\,even)}\\
 (-1)^{\tau^3+\sigma^3+1}&(n^3,\,m^3)=\text{(odd,\,odd)}\\
(-1)^{\tau^3+\sigma^3}&(n^3,\,m^3)=\text{(even,\,odd)}
 \end{array} \right\rbrace\quad n^1+m^1=\text{even}\,,\\
& \left(\begin{array}{c}p\\ \bar{p}\\q\\ \bar{q}\end{array}\right)  
 =(-)^{\tau^0}
 2\left(\begin{array}{c}n^2 \\(-1)^{\tau^1} m^2 \\n^2 \\ (-1)^{\tau^1}m^2  \end{array}\right)\times 
\left\lbrace \begin{array}{ll}
 1-(-1)^{\sigma^3\tau^3}&(n^3,\,m^3)=\text{(odd,\,even)}\\
 (-1)^{\tau^3+\sigma^3+1}&(n^3,\,m^3)=\text{(odd,\,odd)}\\
(-1)^{\tau^3+\sigma^3}&(n^3,\,m^3)=\text{(even,\,odd)}
 \end{array} \right\rbrace\quad n^1+m^1=\text{odd}\,.
 \end{split}
\end{equation}

%%%%%%%%%%%%%%%%%%%%%%%%%%%%%%%%%%%%%%%%%%%%%
\subsection{Orientifolding and Supersymmetry}\label{orientifolding}
In order to obtain $\mathcal{N}=1$ supersymmetry in four dimensions by compactifying Type IIA string theory,
one needs to quotient the orbifold by $\Omega\mathcal{R}$, where $\Omega$ is the worldsheet parity operator and $\mathcal{R}$ the complex conjugation,
\begin{equation}\label{involution}
\mathcal{R}:\,z^i\,\rightarrow\; e^{i\theta_i}\bar{z}^i  
\qquad \text{with} \quad i=1,2,3 \,,
\end{equation}
for some suitable real parameters $\theta_i$. The complex coordinates in the factorisable picture are given by
\begin{equation}\label{A3A1B2complex_coord} 
z^1=y^1+iy^2\,,\quad
z^2=y^3+\mathcal{U}y^4\,,\quad
z^3=y^5-\frac{y^6}{2}+i\frac{y^6}{2} \, ,
\end{equation}
where $y^i$ are the coordinates along the new basis vectors $v_i$, with the complex structure,
\begin{equation}\label{2.complex_structure} 
\mathcal{U}:=u_1+iu_2:=-\frac{R_1}{2\hat{a}R_3}(\hat{d}+i\sqrt{-\hat{a}-\hat{d}^2})\,,
\end{equation}
defined in terms of the real parameters introduced in equation~\eqref{non-factorised-metric}.

There exist four possible orientifold projections which  are given by:
\begin{equation}\label{orietifold-projections}
\begin{split}
&\textbf{A$_{\textbf{a/b}}$AB}\quad\left\lbrace\begin{array}{llll}
&\mathcal{R}_1v_1=-v_2\,, & \mathcal{R}_1v_3=v_3\,, & \mathcal{R}_1v_5=v_5\,, \\
&\mathcal{R}_1v_2=-v_1\,, & \mathcal{R}_1v_4=2u_1v_3-v_4\,, & \mathcal{R}_1v_6=-v_5-v_6\,,
\end{array}\right\rbrace\;\;\quad\vec{\theta}=(-\frac{\pi}{2},0,0) \, ,\\
&\textbf{A$_{\textbf{a/b}}$AA}\quad\left\lbrace\begin{array}{llll}
&\mathcal{R}_2v_1=-v_2\,, & \mathcal{R}_2v_3=v_3\,, & \mathcal{R}_2v_5=-v_5-2v_6\,, \\
&\mathcal{R}_2v_2=-v_1\,, & \mathcal{R}_2v_4=2u_1v_3-v_4\,, & \mathcal{R}_2v_6=v_6\,,
\end{array}\right\rbrace\quad\vec{\theta}=(-\frac{\pi}{2},0,-\frac{\pi}{2}) \, ,\\
&\textbf{A$_{\textbf{a/b}}$BA}\quad\left\lbrace\begin{array}{llll}
&\mathcal{R}_3v_1=-v_1\,, & \mathcal{R}_3v_3=v_3\,, & \mathcal{R}_3v_5=-v_5-2v_6\,, \\
&\mathcal{R}_3v_2=v_2\,, & \mathcal{R}_3v_4=2u_1v_3-v_4\,, & \mathcal{R}_3v_6=v_6\,,
\end{array}\right\rbrace\quad\vec{\theta}=(\pi,0,-\frac{\pi}{2}) \, ,\\
&\textbf{A$_{\textbf{a/b}}$BB}\quad\left\lbrace\begin{array}{llll}
&\mathcal{R}_4v_1=-v_1\,, & \mathcal{R}_4v_3=v_3\,, & \mathcal{R}_4v_5=v_5\,, \\
&\mathcal{R}_4v_2=v_2\,, & \mathcal{R}_4v_4=2u_1v_3-v_4\,, & \mathcal{R}_4v_6=-v_5-v_6\,.
\end{array}\right\rbrace\;\;\quad\vec{\theta}=(\pi,0,0)  \, .\\
\end{split}
\end{equation}
Since the orientifold projection has to act on the lattice crystallographically, the real part of the complex structure can take only two values, $u_1=0,\,\frac{1}{2}$.  Here we used the notation of the orientifold projections and lattice orientations in correspondence to our previous article~\cite{Berasaluce-Gonzalez:2016kqb}. An \textbf{a} or \textbf{b} as a subscript after the first \textbf{A} specifies  the choice $u_1=0$ or $u_1=\frac{1}{2}$, respectively. For more information about the complex coordinates and orientifold projections in terms of non-factorisable coordinates and one-cycles we refer the reader to the original article~\cite{Berasaluce-Gonzalez:2016kqb}.

The corresponding actions on the homological basis $\{\gamma_i,\,\bar{\gamma}_i\}_{i=1,2,3,4}$ are presented in table \ref{tab:involutions_Z4_A3xA1xB2}.
\begin{table}[h!]
\begin{center}
\begin{tabular}{|c|c|c|c|c|}
\hline
\multicolumn{5}{|c|}{\textbf{Orientifold actions on the basis of the three-homology $H_3(T^6/\Z_4,\,\Z)$}}\\
\hline
\hline
lattice & \textbf{A$_{\textbf{a/b}}$AB} & \textbf{A$_{\textbf{a/b}}$AA} \quad & \textbf{A$_{\textbf{a/b}}$BA} & \textbf{A$_{\textbf{a/b}}$BB}\\
\hline
$\Omega\mathcal{R}:\,\gamma_1$ & $\gamma_2$ & $\gamma_1$  & $-\gamma_2$ & $\gamma_1$\\
\hline
$\Omega\mathcal{R}:\,\gamma_2$ & $\gamma_1$ & $-\gamma_2$  & $-\gamma_1$ & $-\gamma_2$ \\
\hline
$\Omega\mathcal{R}:\,\gamma_3$ & $\gamma_3$ & $-\gamma_3$ & $-\gamma_3$ & $\gamma_3$ \\
\hline
$\Omega\mathcal{R}:\,\gamma_4$ & $\gamma_4$ & $-\gamma_4$ & $-\gamma_4$ & $\gamma_4$ \\
\hline
$\Omega\mathcal{R}:\,\bar{\gamma}_1$ &\quad $-\bar{\gamma}_2-2u_1\gamma_2$ \quad &\quad $-\bar{\gamma}_1-2u_1\gamma_1$ \quad &  $\bar{\gamma}_2+2u_1\gamma_2$ \qquad & $-\bar{\gamma}_1-2u_1\gamma_1$ \\
\hline
$\Omega\mathcal{R}:\,\bar{\gamma}_2$  & $-\bar{\gamma}_1-2u_1\gamma_1$ & $\bar{\gamma}_2+2u_1\gamma_2$  & $\bar{\gamma}_1+2u_1\gamma_1$ & $\bar{\gamma}_2+2u_1\gamma_2$\\
\hline
$\Omega\mathcal{R}:\,\bar{\gamma}_3$  & $-\bar{\gamma}_3+2u_1\gamma_3$ & $\bar{\gamma}_3-2u_1\gamma_3$ & $-\bar{\gamma}_3+2u_1\gamma_3$ & $\bar{\gamma}_3-2u_1\gamma_3$ \\
\hline
$\Omega\mathcal{R}:\,\bar{\gamma}_4$  & $-\bar{\gamma}_4+2u_1\gamma_4$ & $\bar{\gamma}_4-2u_1\gamma_4$ & $\bar{\gamma}_4-2u_1\gamma_4$  & $-\bar{\gamma}_4+2u_1\gamma_4$\\
\hline
\end{tabular}
\caption{Orientifold actions on the homological basis three-cycles of the $\Z_4$ orbifolds of lattice type $A_3\times A_1\times B_2$.}\label{tab:involutions_Z4_A3xA1xB2}
\end{center}
\end{table}
There exist two orbits of cycles invariant under the projections $\mathcal{R}\mathcal{Q}^{2k}$ and $\mathcal{R}\mathcal{Q}^{2k+1}$, respectively, and wrapped by the O6-planes, see table \ref{tab:oplanes} for details.
\begin{table}[h!]
\begin{center}
\begin{tabular}{|c|c|c|c|}
\hline
\multicolumn{4}{|c|}{\textbf{O6-planes for $\Z_4$ orbifolds}}\\
\hline \hline
lattice & $\phi_{O6}$& $N_{O6}\times(n^1,\,m^1;\,n^2,\,m^2;\,n^3,\,m^3)$ & hom. cycle \\ 
\hline 
\textbf{A$_{\textbf{a/b}}$AB}& $-\frac{\pi}{4}$ & $2\times(1,-1;\frac{1}{2},0;1,0)$  & $\gamma_1+\gamma_2$ \\
& & $(4-4u_1)\times(0,-1;2u_1,-1-2u_1;0,1)$& $-2u_1(\gamma_1-\gamma_2)
-2(\bar{\gamma}_1-\bar{\gamma}_2)$\\ 
\hline 
\textbf{A$_{\textbf{a/b}}$AA} & $\frac{\pi}{2}$ & $4\times(1,-1;\frac{1}{2},0;0,1)$ & $-2\gamma_1$  \\
& & $(2-2u_1)\times(0,1;2u_1,-1-2u_1;1,2)$& $-2u_1\gamma_2-2\bar{\gamma}_2$\\ 
\hline
\textbf{A$_{\textbf{a/b}}$BA}& $\frac{\pi}{4}$ & $(4-4u_1)\times(0,-1;1,0;0,1)$ & $(2-2u_1)(-\gamma_1+\gamma_2)$  \\
& &$2\times(1,1;u_1,-1;1,2)$ & $-2u_1(\gamma_1+\gamma_2)-2\bar{\gamma}_1-2\bar{\gamma}_2$\\ 
\hline
\textbf{A$_{\textbf{a/b}}$BB}& $-\frac{\pi}{2}$ & $(2-2u_1)\times(0,1;1,0;-1,0)$ & $(2-2u_1)\gamma_1$ \\
& &$4\times(-1,-1;u_1,-1;0,1)$& $4u_1\gamma_2+4\bar{\gamma}_2$ \\ 
\hline
\end{tabular} 
\caption{O6-planes for the $\Z_4$ orbifolds with $A_3 \times A_1 \times B_2$ lattice. The two lines for each lattice correspond to the orbits of three-cycles wrapped by the O-planes invariant under the projection $\mathcal{R}\mathcal{Q}^{2k}$ (first line) and $\mathcal{R}\mathcal{Q}^{2k+1}$ (second line). The toroidal wrapping numbers represent the toroidal cycles for $k=0$. The number of identical O-planes is given as a prefactor in the third column.}\label{tab:oplanes}
\end{center}
\end{table}

The supersymmetry conditions on bulk cycles can be deduced from 
\begin{equation}\label{Omega}
Z:=e^{-i\phi_{O6}}(n^1+im^1)(n^2+(u_1+iu_2)m^2)(n^3-\frac{1-i}{2}m^3)\,.
\end{equation}
For the \textbf{A$_{\textbf{a/b}}$AA}, \textbf{A$_{\textbf{a/b}}$AB}, \textbf{A$_{\textbf{a/b}}$BA} and \textbf{A$_{\textbf{a/b}}$BB} lattices the calibration phase $\phi_{O6}$ is  fixed by the sum of the angles between the corresponding O6-plane and the $v_{1,3,5}$-axes to  $\frac{\pi}{2}$, $-\frac{\pi}{4}$, $\frac{\pi}{4}$ and $-\frac{\pi}{2}$, respectively. A D6-brane is supersymmetric if it wraps a {\it bulk} three-cycle which satisfies the {\it special Lagrangian} conditions 
\begin{equation}\label{slags}
\text{Im}\,Z=0\;(\text{necessary})\,\quad \text{and} \quad \text{Re}\,Z>0\;(\text{sufficient})\,,
\end{equation}
and the {\it exceptional} three-cycle takes the shape detailed in equation~\eqref{exc-three-cycles} in section~\ref{factor+geometry}.

The conditions~\eqref{slags} can be written in terms of the bulk wrapping numbers. Thus, the necessary condition is given by 
\begin{equation}\label{necessary-conditions}
\begin{array}{lll}
& \textbf{A$_{\textbf{a/b}}$AB}: 
& -P+Q+(\bar{P}-\bar{Q})u_1-(\bar{P}+\bar{Q})u_2=0\,,\\
&\textbf{A$_{\textbf{a/b}}$AA}:
&-Q+\bar{Q}u_1+\bar{P}u_2=0\,,\\
&\textbf{A$_{\textbf{a/b}}$BA}:
& P+Q-(\bar{P}+\bar{Q})u_1+(-\bar{P}+\bar{Q})u_2=0\,,\\
&\textbf{A$_{\textbf{a/b}}$BB}:
&-Q+\bar{Q}u_1+\bar{P}u_2=0\,.
\end{array}
\end{equation}
The sufficient condition chooses  one of two global orientations so that anti-D-branes are excluded:
\begin{equation}\label{sufficient-conditions}
\begin{array}{lll}
& \textbf{A$_{\textbf{a/b}}$AB}: 
& P+Q-(\bar{P}+\bar{Q})u_1-(\bar{P}-\bar{Q})u_2>0\,,\\
&\textbf{A$_{\textbf{a/b}}$AA}:
&-P+\bar{P}u_1-\bar{Q}u_2>0\,,\\
&\textbf{A$_{\textbf{a/b}}$BA}:
& P-Q-(\bar{P}-\bar{Q})u_1+(\bar{P}+\bar{Q})u_2<0\,,\\
&\textbf{A$_{\textbf{a/b}}$BB}:
&-P+\bar{P}u_1-\bar{Q}u_2<0\,.
\end{array}
\end{equation}
The tangent of the angles between the one-cycles on each torus and the axes along the basis vectors $v_{1,3,5}$ are given by:
\begin{equation}\label{anlgles}
\begin{split}
\tan\phi^1&=\frac{m^1}{n^1}\,,\\
\tan\phi^2&=u_2\frac{m^2}{n^2+u_1m^2}\,,\\
\tan\phi^3&=\frac{m^3}{2n^3-m^3}\,.
\end{split} 
\end{equation}
Using these expressions, one can easily verify that the conditions \eqref{necessary-conditions} and \eqref{sufficient-conditions} can be rewritten as:
\begin{equation}\label{susy-condition}
\phi_1+\phi_2+\phi_3=\phi_{O6} \;\text {mod }2\pi\,,
\end{equation}
in compliance with the standard factorisable picture. 

%%%%%%%%%%%%%%%%%%%%%%%%%%%%%%%%%%%
\subsection{RR-tadpoles}\label{RRtcc}
As was already mentioned above, any fractional three-cycle can be written as \mbox{$\pi_a=\frac{1}{2}\pi_a^\text{bulk}+\frac{1}{2}\pi_a^\text{exc}$} with homological wrapping numbers $(P_a,\,Q_a,\,\bar{P}_a,\,\bar{Q}_a)$ for the bulk part and $(p_a,\,q_a,\,\bar{p}_a,\,\bar{q}_a)$ for the exceptional part defined in equations~\eqref{bulk-wrapping-numbers} and~\eqref{exceptional-wrapping-numbers}, respectively.
For each orientifold projection \eqref{orietifold-projections}, the tadpole cancellation condition,
\begin{equation}\label{tadpole-def}
\sum_a N_a(\pi_a+\pi^\prime_a)=4\pi_\text{O6},
\end{equation}
can be expressed in terms of these homological wrapping numbers. It is useful, however, to first define the quantities $\textbf{P}\equiv\sum_a N_a P_a$ and similar ones for the other wrapping numbers in order to shorten the notation. In terms of these new quantities the RR tadpole cancellation conditions are written as
\begin{equation}
\begin{split}
\textbf{A}_\textbf{a/b}\textbf{AB}:&
\left\lbrace\begin{array}{lr}
\textbf{P}+\textbf{Q}-2u_1\bar{\textbf{Q}}=8(1-2u_1)\,
&\hspace{0.5cm}\textbf{p}+u_1\bar{\textbf{p}}=0\\
\bar{\textbf{P}}-\bar{\textbf{Q}}=-16\,
&\hspace{0.5cm}\textbf{q}+u_1\bar{\textbf{q}}=0
\end{array}\right.\,,\\
\textbf{A}_\textbf{a/b}\textbf{AA}:&
\left\lbrace\begin{array}{lr}
\textbf{P}-u_1\bar{\textbf{P}}=-8\,
&\hspace{3.8cm}\bar{\textbf{p}}=0\\
\bar{\textbf{Q}}=-8
&\hspace{3.8cm}\bar{\textbf{q}}=0\\
\end{array}\right.\,,\\
\textbf{A}_\textbf{a/b}\textbf{BA}:&
\left\lbrace\begin{array}{lr}
\textbf{P}-\textbf{Q}+2u_1\bar{\textbf{Q}}=-16
&\hspace{2.7cm}\bar{\textbf{p}}=0\\
\bar{\textbf{P}}+\bar{\textbf{Q}}=-16
&\hspace{2.7cm}\bar{\textbf{q}}=0\\
\end{array}\right.\,,\\
\textbf{A}_\textbf{a/b}\textbf{BB}:&
\left\lbrace\begin{array}{lr}
\textbf{P}-u_1\bar{\textbf{P}}=8-8u_1\,
&\hspace{2.1cm}\textbf{p}+u_1\bar{\textbf{p}}=0\\
\bar{\textbf{Q}}=16\,
&\hspace{2.1cm}\textbf{q}+u_1\bar{\textbf{q}}=0\\
\end{array}\right.\,.
\end{split}
\end{equation}
The first column gives the RR tadpole cancellation conditions for the bulk part, the second one those for the  exceptional part. 

%%%%%%%%%%%%%%%%%%%%%%%%%%%%%%%%%%%%%%%%%%%%
\subsection{K-theory constraints}\label{Kconstraints}
D-branes are not fully characterised by their homology but rather K-theory~\cite{Witten:1998cd}. This gives rise to additional constraints on four-dimensional model building which take values in $\Z_2$ and match field theoretical anomaly considerations~\cite{Witten:1982fp}. 
The K-theory constraints can be formulated in terms of intersection numbers of three-cycles wrapped by D6-branes~\cite{Uranga:2000xp}:
\begin{equation}\label{K-theory-def}
\sum_x N_x\pi_x\circ\pi_\text{probe}=0\;\text{mod}\; 2\,,
\end{equation}
where we sum over all D6-branes in a given model and $\pi_\text{probe}$ is any three-cycle wrapped by a D6-brane carrying an $SU(2)$ or $USp(2N)$ gauge group.

As we will show in section \ref{amplitudes}, for each orientifold projection there are three types of three-cycles that give rise to $USp(2N)$ gauge groups when wrapping D6-branes on them. Table \ref{Preliminary-KTheory-constraints} shows all constraints coming from equation \eqref{K-theory-def} for each type $a,b,c$ of the orientifold invariant three-cycles.
\begin{table}[h]
\begin{center}
\begin{tabular}{|c|c|}
\hline
\multicolumn{2}{|c|}{\textbf{K-theory constraints $=0$ mod 2}}\\
\hline\hline
brane & \textbf{$\mathcal{R}_1$} \\ 
 \hline
$a$ & $\frac{1}{2}\left(\bar{\textbf{P}}+\bar{\textbf{Q}}\pm(\bar{\textbf{p}}\pm \bar{\textbf{q}})\right)$ \\ 
\hline 
$b$ & $\frac{1}{2}\left(\bar{\textbf{P}}+\bar{\textbf{Q}}\right)$,\quad $\bar{\textbf{p}}$,\quad $\bar{\textbf{q}}$,\\
\hline
$c$ & $\textbf{P}-\textbf{Q}-u_1(\bar{\textbf{P}}-\bar{\textbf{Q}})$\\
\hline\hline
\multicolumn{2}{|c|}{\textbf{$\mathcal{R}_2$}}\\
\hline
$a$& $\textbf{Q}-u_1\bar{\textbf{Q}}\pm(\textbf{p}+u_1\bar{\textbf{p}})$,\quad $\textbf{Q}-u_1\bar{\textbf{Q}}\pm(\textbf{q}+u_1\bar{\textbf{q}})$ \\
\hline
$b$ & $\textbf{Q}+2u_1(\textbf{Q}-\bar{\textbf{Q}})$,\quad $(\textbf{p}\pm \textbf{q})(1+2u_1)+2u_1(\bar{\textbf{p}}\pm\bar{\textbf{q}}))$\\
\hline 
$c$ & $\bar{\textbf{P}}$\\
\hline\hline
\multicolumn{2}{|c|}{\textbf{$\mathcal{R}_3$}}\\
\hline
$a$& $\frac{1}{2}\left( (\textbf{P}+\textbf{Q})(1+2u_1)-2u_1(\bar{\textbf{P}}+\bar{\textbf{Q}})\pm((\textbf{p}\pm \textbf{q})(1+2u_1)+2u_1(\bar{\textbf{p}}\pm\bar{\textbf{q}}))\right)$ \\
\hline
$b$ & $\textbf{P}+\textbf{Q}-u_1(\bar{\textbf{P}}+\bar{\textbf{Q}})$,\quad $2(\textbf{q}+u_1 \bar{\textbf{q}})$,\quad $2(\textbf{p}+u_1 \bar{\textbf{p}})$\\
\hline 
$c$ & $\frac{1}{2}(\bar{\textbf{P}}-\bar{\textbf{Q}})$\\
\hline\hline
\multicolumn{2}{|c|}{\textbf{$\mathcal{R}_4$}}\\
\hline
$a$& $\frac{1}{2}(\bar{\textbf{P}}\pm \bar{\textbf{p}})$,\quad  $\frac{1}{2}(\bar{\textbf{P}}\pm \bar{\textbf{q}})$\\
\hline
$b$ & $\bar{\textbf{P}}$,\quad $\bar{\textbf{p}}\pm\bar{\textbf{q}}$\\
\hline 
$c$ & $-\textbf{Q} +2u_1(\bar{\textbf{Q}}-\textbf{Q})$\\
\hline
\end{tabular}
\caption{Constraints arising from solving equation~\eqref{K-theory-def} for each kind of orientifold invariant three-cycle $a,b,c$ per background lattice orientation under the anti-holomorphic involution ${\cal R}_i$.}
\label{Preliminary-KTheory-constraints}
\end{center}
\end{table}
These formal constraints are not necessarily new ones. It might happen that some (if not all) of them are already satisfied by imposing the RR tadpole cancellation conditions. In addition, there are some restrictions to the parity of the homological wrapping numbers that may cause some of the constraints to be automatically satisfied. On the one hand, the wrapping numbers $(\bar{P}_a,\bar{Q}_a,\bar{p}_a,\bar{q}_a)$ are all odd or all even integer numbers. Therefore, $\bar{\textbf{P}}=\bar{\textbf{Q}}=\bar{\textbf{p}}=\bar{\textbf{q}} \mod 2$. On the other hand, among the wrapping numbers $(P_a,Q_a,p_a,q_a)$ there are two odd and two even numbers. This implies that $P_a+Q_a=p_a+q_a \mod 2$, and therefore, $\textbf{P}+\textbf{Q}=\textbf{p}+\textbf{q}\mod 2$.\\\\

After taking into account both the RR tadpole cancellation conditions and the remarks above about the parity of the wrapping numbers, the non-trivial new constraints arising from K-theory for each of the orientifold projections \eqref{orietifold-projections} are:
\begin{itemize}
\item $\textbf{A}_{\textbf{a}}\textbf{AB}$. The new constraints are satisfied if $\bar{\textbf{P}}$, $\bar{\textbf{Q}}$, $\bar{\textbf{p}}$ and $\bar{\textbf{q}}$ are all even, and
\begin{equation}
\frac{1}{2}(\bar{\textbf{P}}+\bar{\textbf{Q}})=0\mod 2,\quad \frac{1}{2}(\bar{\textbf{p}}+\bar{\textbf{q}})=0\mod 2.
\end{equation}
\item $\textbf{A}_\textbf{b}\textbf{AB}$. The new constraints correspond to just the last two for $\textbf{A}_\textbf{a}\textbf{AB}$. $\bar{\textbf{P}}$, $\bar{\textbf{Q}}$, $\bar{\textbf{p}}$ and $\bar{\textbf{q}}$ are allowed to be odd.
\item $\textbf{A}_\textbf{a}\textbf{AA}$. For the K-theory constraints to be satisfied, $\textbf{Q}$, $\textbf{p}$ and $\textbf{q}$ have to be even.
\item $\textbf{A}_\textbf{b}\textbf{AA}$. For the K-theory constraints to be satisfied, $\textbf{Q}$, $\textbf{p}$ and $\textbf{q}$ need to have the same parity.
\item $\textbf{A}_\textbf{a}\textbf{BA}$. The new constraints are
\begin{equation}
\textbf{P}+\textbf{Q}=0\mod 2, \quad \frac{1}{2}\left((\textbf{P}+\textbf{Q})\pm(\textbf{p}\pm \textbf{q})\right)=0\mod 2.
\end{equation}
\item $\textbf{A}_\textbf{b}\textbf{BA}$. There are no new constraints arising from K-theory.
\item $\textbf{A}_\textbf{a}\textbf{BB}$. For the K-theory constraints to be satisfied, $\textbf{Q}$ has to be even and
\begin{equation}
\frac{1}{2}\left(\bar{\textbf{P}}\pm\bar{\textbf{p}}\right)=0\mod 2,\quad \frac{1}{2}\left(\bar{\textbf{P}}\pm\bar{\textbf{q}}\right)=0\mod 2.
\end{equation}
\item $\textbf{A}_\textbf{b}\textbf{BB}$. The new constraints are the same as for $\textbf{A}_\textbf{a}\textbf{BB}$ minus $\textbf{Q}$ having to be even.
\end{itemize}

%%%%%%%%%%%%%%%%%%%%%%%%%%%%%%%%%%%%%%
\subsubsection{Relation to discrete $\Z_n$ gauge symmetries}\label{Sss:Zn}

The linear combination $U(1)_X=\sum_aq_aU(1)_a$ with $q_a\in\mathbb{Q}$ remains massless and anomaly-free provided that: 
\begin{equation}
\pi^\text{even}\circ\left(\sum_{a}q_aN_a \pi_a \right)=0\,,
\end{equation}
for any $\Omega\mathcal{R}_i$-invariant homological three-cycle $\pi^\text{even}$. 
Otherwise, a non-vanishing value on the right hand side indicates a St\"uckelberg coupling leading to a mass proportional to the string scale $M_\text{string}$.\footnote{
More precisely, as first noted in~\cite{Ghilencea:2002da}
the masses of the Abelian gauge factors are calculated from the matrix
$M^2_{ab} = M^2_{\text{string}}  g_{U(1)_a} g_{U(1)_b} \sum_{i=0}^{h_{2,1}} c^i_a c^i_b$,
where $c^i_a$ are the coefficients of the orientifold-odd contributions to the  three-cycle $\pi_a$ in a given basis, and the 
dependence on the gauge couplings $g_{U(1)_a} = g_{SU(N_a)} /\sqrt{2N_a}$ % \propto \sqrt{g_{\text{string}} / N_a\text{Vol}(\pi_a)}$ 
arises when transforming the effective action from the string frame to  canonically normalised kinetic terms of the massive gauge factors.
Notice that the tree-level value $g_{SU(N_a)}\propto \sqrt{g_{\text{string}} / \text{Vol}(\pi_a)}$ can be
 rewritten in terms of $\sqrt{M_{\text{string}}/M_{\text{Planck}}}$, cf. equation~\eqref{eq:g-tree-example} for a Pati-Salam example, but that
 one-loop corrections might significantly change the dependence on volumes and/or mass scales.
}
Any $U(1)_{\text{massive}}$ remains as a perturbative global symmetry, but is broken to some  discrete $\Z_n$ symmetry by non-perturbative effects, in particular D-brane instantons~\cite{BerasaluceGonzalez:2011wy,Ibanez:2012wg,Anastasopoulos:2012zu,Honecker:2013sww}. 
The existence of such a $\Z_n$ symmetry can also be expressed in terms of homological intersection numbers:
\begin{equation}
\pi^\text{even}\circ\left(\sum_{a}k_aN_a \pi_a \right)=0\;\text{mod }n  \quad  \text{ with }\;\Z_n\subset\sum_ak_aU(1)_a\,,
\end{equation}
where the integer coefficients satisfy the conditions $0\leq k_a<n$ and \mbox{gcd$(k_a,k_b,\dots,n)=1$}.  
Generally, cycles $\pi^\text{even}$ can either support $SO(2N)$ or $USp(2N)$ gauge groups, but anticipating the result of section~\ref{USP_or_SO}, that on the $\Z_4 \times \OR$ 
orientifolds of the $A_3 \times A_1 \times B_2$-type lattice discussed in this article any gauge group enhancement is of $USp$-type, the set $\{\pi^\text{even}\}$ equals the set $\{ \pi_\text{probe}\}$ 
used in equation~\eqref{K-theory-def}, and therefore the K-theory constraint implies a $\Z_2$ symmetry with coefficients $(k_a,k_b,\ldots)=(1,1,\ldots)$.

%%%%%%%%%%%%%%%%%%%%%%%%%%%%%%%%%%%%%%%%%%%
\subsection{Closed string spectrum}\label{ClosedSpectrum}

The massless closed string spectrum for factorisable $\Z_4$ orientifolds was computed in~\cite{Blumenhagen:1999ev,Blumenhagen:2004di}, see also~\cite{Forste:2010gw}. We give here for the first time the massless closed string spectra of the $A_3\times A_1\times B_2$ orientifold backgrounds. In addition  to the universally present ${\cal N}=1$ supergravity and axio-dilaton multiplets, which are contained in the untwisted sector, there are also non-universal chiral and vector multiplets which depend on the orientifold background. The amount of these multiplets is determined by the  Hodge numbers $h_{1,1}=h^+_{1,1}+h^-_{1,1}$ and $h_{2,1}$, cf. the Hodge numbers before orientifolding in equation~\eqref{hodge-numbers}. More precisely, one obtains  $h^-_{1,1}+h_{2,1}$ chiral and $h^+_{1,1}$  vector multiplets. The complete list of ${\cal N}=1$ supersymmetric massless multiplets arising from the closed string sector is presented in table \ref{tab:closed-string-spectrum}.
%%%%%%%%%%%%%%%%%%%%%%%%%%
\begin{table}[h!]
\begin{center}
\begin{tabular}{|c|c|c|c|c|}
\hline
\multicolumn{5}{|c|}{\textbf{Massless closed string spectrum on $T^6/(\Z_4\times \Omega\mathcal{R})$}}\\
\hline \hline
sector & \textbf{A$_{\textbf{a/b}}$AB} & \textbf{A$_{\textbf{a/b}}$AA} &
 \textbf{A$_{\textbf{a/b}}$BA} & \textbf{A$_{\textbf{a/b}}$BB} \\ 
\hline 
untwisted& \multicolumn{4}{|c|}{SUGRA\,+\,[Axio-Dilaton]\,+\,5C\,+\,1V}\\
\hline
$\mathcal{Q}^1+\mathcal{Q}^3$ & \multicolumn{2}{|c|}{12C\,+\,4V } & \multicolumn{2}{|c|}{(16-8$u_1$)C\,+\,8$u_1$V } \\ 
\hline 
$\mathcal{Q}^2$& \multicolumn{4}{|c|}{8C}\\
\hline 
\end{tabular} 
\vspace{.3cm}
\caption{Massless closed string spectrum for the eight different lattice orientations on the non-factorisable $\Z_4$ orbifolds with lattice of type $A_3 \times A_1 \times B_2$.  C and V denote  ${\cal N}=1$ supersymmetric chiral and vector  multiplets, respectively. The lower index \textbf{a} and \textbf{b} corresponds to the value of the real part   of the complex structure parameter  $u_1=0$ and  $\frac{1}{2}$, respectively.}\label{tab:closed-string-spectrum}
\end{center}
\end{table}

%%%%%%%%%%%%%%%%%%%%%%%%%%%%%%%%%%%%
\subsection{Chiral open string spectrum}

For completeness we also briefly review the well-known massless chiral open string spectrum. The states of this spectrum arise at the intersection points among stacks of D6-branes and their orientifold images. 
The amount of such states is determined by the topological intersection numbers between the corresponding fractional three-cycles. For a general gauge group $\prod_aU(N_a)\times \prod_cUSp(2M_c)\times\prod_dSO(2M_d)$, the massless chiral spectrum is given in table \ref{tab:open-chiral-string-spectrum}.  
\begin{table}[h!]
\begin{center}
\begin{tabular}{|c|c||c|c|}
\hline
\multicolumn{4}{|c|}{\textbf{Chiral spectrum of $U(N_a)\times U(N_b)\times USp(2M_c)\times SO(2M_d)$}}\\
\hline \hline
representation & net chirality& representation & net chirality\\ 
\hline 
(\textbf{Sym})$_a$ & $\frac{1}{2}(\pi_a\circ \pi_a^\prime-\pi_a\circ\pi_\text{O6})$ & \multicolumn{2}{|c|}{} \\ 
\cline{1-2} 
(\textbf{Anti})$_a$ & $\frac{1}{2}(\pi_a\circ \pi_a^\prime+\pi_a\circ\pi_\text{O6})$& \multicolumn{2}{|c|}{} \\ 
\hline  
(\textbf{N}$_a,\,\overline{\textbf{N}}_b$) & $\pi_a\circ\pi_b$ &(\textbf{N}$_a,\,2\textbf{M}_c$) & $\pi_a\circ\pi_c$\\ 
\hline 
(\textbf{N}$_a\,\textbf{N}_b$) & $\pi_a\circ\pi_b^\prime$ &(\textbf{N}$_a,\,2\textbf{M}_d$) & $\pi_a\circ\pi_d$\\ 
\hline 
\end{tabular} 
\caption{Net chirality via  intersection numbers of homological three-cycles $\pi_a$. The topological three-cycles $\pi_\text{O6}$ wrapped by the O6-planes are given in table \ref{tab:oplanes}.}\label{tab:open-chiral-string-spectrum}
\end{center}
\end{table}

Since in general the non-chiral massless matter fields cannot be determined from topology, to obtain the complete massless spectrum, we need to compute the one-loop beta function coefficients using the annulus and M\"obius strip amplitudes. This will be done in  section \ref{amplitudes}. Alternatively, one could explicitly compute the Chan-Paton labels of massless open string states. The subtleties of sign factors due to non-trivial Wilson lines and orientifold projections, however, are more easily applied and implemented in computer codes in the computation of beta function coefficients via toroidal and $\Z_2$-invariant intersections numbers among pairs of D6-branes as well as the counting of orientifold-invariant intersection numbers between D6-branes and O6-planes as detailed in sections~\ref{Annulus-1-insertion} to~\ref{MS-insertions}.

%%%%%%%%%%%%%%%%%%%%%%%%%%%%%%%%%%%%%%%%%%%%%%
\section{Duality relations among lattice orientations}\label{duality}

In \cite{Berasaluce-Gonzalez:2016kqb}, we found the first hints indicating that not all possible lattice orientations are independent. We computed the number of the corresponding fractional cycles not overshooting the bulk RR tadpole cancellation conditions and the number of possible complex structure values $u_2$. The results from \cite{Berasaluce-Gonzalez:2016kqb} are reproduced here in table \ref{tab:T6} for convenience.
\begin{table}[!h] 
\centering
\begin{tabular}{|c|c|c|c|}
\hline 
\muc{4}{|c|}{\bf $A_3\times A_1\times B_2$-orientifolds}\\
\hline\hline
Lattice & Orien. proj. & \# of frac. cycles & \# of $u_2$\\ 
\hline
\textbf{A}$_\textbf{a}$\textbf{AA} & $\mathcal{R}_2$ $(u_1=0)$ & 2126 & 96\\ 
\hline 
\textbf{A}$_\textbf{a}$\textbf{AB} & $\mathcal{R}_1$ $(u_1=0)$ & 2126 & 96 \\ 
\hline 
\textbf{A}$_\textbf{a}$\textbf{BA} & $\mathcal{R}_3$ $(u_1=0)$ & 5134 & 210\\ 
\hline 
\textbf{A}$_\textbf{a}$\textbf{BB} & $\mathcal{R}_4$ $(u_1=0)$ & 5134 & 210\\ 
\hline\hline
\textbf{A}$_\textbf{b}$\textbf{AA} & $\mathcal{R}_2$ $(u_1=\frac{1}{2})$ & 2410 & 118\\ 
\hline 
\textbf{A}$_\textbf{b}$\textbf{AB} & $\mathcal{R}_1$ $(u_1=\frac{1}{2})$ & 3646 & 140 \\ 
\hline 
\textbf{A}$_\textbf{b}$\textbf{BA} & $\mathcal{R}_3$ $(u_1=\frac{1}{2})$ & 3646 & 140\\ 
\hline 
\textbf{A}$_\textbf{b}$\textbf{BB} & $\mathcal{R}_4$ $(u_1=\frac{1}{2})$ & 2410 & 118\\ 
\hline
\end{tabular} 
\caption{The numbers of supersymmetric fractional cycles bounded by the bulk RR tadpole cancellation conditions and the number of possible complex structure values $u_2$ for different choices of orientifold axes.}
\label{tab:T6}
\end{table}

Looking at the number of fractional cycles for each given value of $u_2$, we were able to postulate the following pairwise relations between the different lattice orientations:
\begin{equation}\label{2.dualities}
\begin{split}
\text{\textbf{A}$_\textbf{a}$\textbf{AA} dual to \textbf{A}$_\textbf{a}$\textbf{AB} and }\; u_2=\frac{1}{2u_2^\prime}\,, \\
\text{\textbf{A}$_\textbf{a}$\textbf{BA} dual to \textbf{A}$_\textbf{a}$\textbf{BB} and }\; u_2=\frac{1}{2u_2^\prime}\,, \\
\text{\textbf{A}$_\textbf{b}$\textbf{AA} dual to \textbf{A}$_\textbf{b}$\textbf{BB} and }\; u_2=\frac{1}{4u_2^\prime}\,, \\
\text{\textbf{A}$_\textbf{b}$\textbf{BA} dual to \textbf{A}$_\textbf{b}$\textbf{AB} and }\; u_2=\frac{1}{4u_2^\prime}\,.
\end{split}
\end{equation}

Since in \cite{Berasaluce-Gonzalez:2016kqb} we only mentioned these pairwise relations, here we will take a closer look at them. 
We now performed a thorough analysis of the fractional cycles for each pair of (conjectured to be) dual backgrounds. We found that, for a suitable choice of the geometric parameters $\hat{a},\, R_1$ and $R_3$ the dual candidate backgrounds have the same number of cycles with identical length. Comparing then the wrapping numbers of toroidal three-cycles, we found the following duality maps.

\subsection{$u_1=0$}\label{u1=0}
For $u_1=0$, the proposed duality between lattice orientations is given by
\begin{equation}
\begin{split}
&\text{\textbf{A$_\textbf{a}$XA} dual to \textbf{A$_\textbf{a}$XB} and }u_2^\prime=\frac{1}{2u_2}\quad \text{for \textbf{X}$\in\{$\textbf{A,B}$\}$}\,.
\end{split}
\end{equation} 
Since any fractional three-cycle stems from a toroidal one, in order to find a duality map it suffices to give a map between the six wrapping numbers $(n^i,\,m^i)^{i=1,2,3}$, the $\Z_2$ eigenvalue $\tau^0$, discrete Wilson lines $(\tau^1,\, \tau^3)$ and discrete displacements $(\sigma^1,\sigma^3)$. Thus, the map $\mathcal{F}$ between \textbf{A$_\textbf{a}$XA} and \textbf{A$_\textbf{a}$XB} lattice orientations is given by:
\begin{equation}\label{dualitymap1}
\begin{split}
&\hspace{1cm}\mathcal{F}:\text{\textbf{A$_\textbf{a}$XA}}\;\rightarrow\;\text{\textbf{A$_\textbf{a}$XB}}\quad \text{for \textbf{X}$\in\{$\textbf{A,B}$\}$}\,,\\
&\left(\begin{array}{rr}
n^1, m^1\\
n^2, m^2\\
n^3, m^3
\end{array}\right)
\;\mapsto\;
\left(\begin{array}{rr}
n^{\prime 1}, m^{\prime 1}\\
n^{\prime 2}, m^{\prime 2}\\
n^{\prime 3}, m^{\prime 3}
\end{array}\right)=
\left(\begin{array}{cc}
n^1+m^1,& -n^1+m^1\\
\frac{1}{2}m^2,& -n^2\\
n^{3},& m^{3}
\end{array}\right)\,,\\
&\hspace{0.0cm}(\tau^0,\,\tau^1,\,\tau^3)\;\mapsto\;(\tau^{0\prime},\,\tau^{1\prime},\,\tau^{3\prime})=(\tau^0,\,\sigma^1,\,\tau^3)\,,\\
&\hspace{0.6cm}(\sigma^1,\,\sigma^3)\;\mapsto\;(\sigma^{1\prime},\,\sigma^{3\prime})=(\tau^1,\,\sigma^3)\,.
\end{split}
\end{equation}
This map gives rise  to the transformation of bulk wrapping numbers :
\begin{equation}\label{trafo-bulk}
(P,\,Q,\,\bar{P},\,\bar{Q})\mapsto (P^\prime,\,Q^\prime,\,\bar{P}^\prime,\,\bar{Q}^\prime)=(-\frac{1}{2}(\bar{P}+\bar{Q}),\,\frac{1}{2}(\bar{P}-\bar{Q}),\,P+Q,\,Q-P)\,.
\end{equation}
For the exceptional part we obtain:
\begin{equation}\label{trafo-exc}
(p,\,q,\,\bar{p},\,\bar{q})\mapsto (p^\prime,\,q^\prime,\,\bar{p}^\prime,\,\bar{q}^\prime)=(\frac{1}{2}(\bar{p}+\bar{q}),\,\frac{1}{2}(\bar{p}-\bar{q}),\,-p-q,\,q-p)\,.
\end{equation}
An easy check of the supersymmetry conditions \eqref{necessary-conditions} and \eqref{sufficient-conditions} verifies that the transformation $\mathcal{F}$  preserves the {\it sLag} property of a three-cycle, i.e. any supersymmetric three-cycle on the \textbf{A$_\textbf{a}$XA}-orientifold transforms under $\mathcal{F}$ to a supersymmetric three-cycle on the \textbf{A$_\textbf{a}$XB}-orientifold.

It is noteworthy that for $u_1=0$ the transformation \eqref{dualitymap1} requires an exchange of the discrete Wilson line $\tau^1$ and displacement parameter $\sigma^1$ along the first two-torus $T^2_{(1)}$.

\subsection{$u_1=1/2$}\label{u1=12}
The duality map between \textbf{A$_\textbf{b}$AA} and \textbf{A$_\textbf{b}$BB} is given by
\begin{equation}\label{dualitymap2a}
\begin{split}
&\hspace{1cm}\mathcal{F}:\text{\textbf{A$_\textbf{b}$AA}}\;\rightarrow\;\text{\textbf{A$_\textbf{b}$BB}}\,,\\
&\left(\begin{array}{rr}
n^1, m^1\\
n^2, m^2\\
n^3, m^3
\end{array}\right)
\;\mapsto\;
\left(\begin{array}{rr}
n^{\prime 1}, m^{\prime 1}\\
n^{\prime 2}, m^{\prime 2}\\
n^{\prime 3}, m^{\prime 3}
\end{array}\right)=
\left(\begin{array}{cc}
m^1,& -n^1\\
n^2+m^2,& -2n^2-m^2\\
n^{3},& m^{3}
\end{array}\right)\,,\\
&\hspace{0.0cm}(\tau^0,\,\tau^1,\,\tau^3)\;\mapsto\;(\tau^{0\prime},\,\tau^{1\prime},\,\tau^{3\prime})=(\tau^0,\,\tau^1,\,\tau^3)\,,\\
&\hspace{0.6cm}(\sigma^1,\,\sigma^3)\;\mapsto\;(\sigma^{1\prime},\,\sigma^{3\prime})=(\sigma^1,\,\sigma^3)\,.
\end{split}
\end{equation}
This gives rise to the transformation of the bulk wrapping numbers:
\begin{equation}
(P,\,Q,\,\bar{P},\,\bar{Q})\mapsto (P^\prime,\,Q^\prime,\,\bar{P}^\prime,\,\bar{Q}^\prime)=(Q-\bar{Q},\,-P+\bar{P},\,2Q-\bar{Q},\,-2P+\bar{P})\,.
\end{equation}
The exceptional part transforms as follows
\begin{equation}
(p,\,q,\,\bar{p},\,\bar{q})\mapsto (p^\prime,\,q^\prime,\,\bar{p}^\prime,\,\bar{q}^\prime)=(p+\bar{p},\,q+\bar{q},\,-2p-\bar{p},\,-2q-\bar{q})\,.
\end{equation}

And finally the map between the remaining orientifolds \textbf{A$_\textbf{b}$AB} and \textbf{A$_\textbf{b}$BA} is given by
\begin{equation}\label{dualitymap2b}
\begin{split}
&\hspace{1cm}\mathcal{F}:\text{\textbf{A$_\textbf{b}$AB}}\;\rightarrow\;\text{\textbf{A$_\textbf{b}$BA}}\,,\\
&\left(\begin{array}{rr}
n^1, m^1\\
n^2, m^2\\
n^3, m^3
\end{array}\right)
\;\mapsto\;
\left(\begin{array}{rr}
n^{\prime 1}, m^{\prime 1}\\
n^{\prime 2}, m^{\prime 2}\\
n^{\prime 3}, m^{\prime 3}
\end{array}\right)=
\left(\begin{array}{cc}
n^1,& m^1\\
-n^2-m^2,& 2n^2+m^2\\
n^{3},& m^{3}
\end{array}\right)\,,\\
&\hspace{0.0cm}(\tau^0,\,\tau^1,\,\tau^3)\;\mapsto\;(\tau^{0\prime},\,\tau^{1\prime},\,\tau^{3\prime})=(\tau^0,\,\tau^1,\,\tau^3)\,,\\
&\hspace{0.6cm}(\sigma^1,\,\sigma^3)\;\mapsto\;(\sigma^{1\prime},\,\sigma^{3\prime})=(\sigma^1,\,\sigma^3)\,.
\end{split}
\end{equation}
The corresponding bulk and exceptional wrapping numbers transform as follows:
\begin{equation}
\begin{split}
(P,\,Q,\,\bar{P}\,,\bar{Q})&\mapsto (P^\prime,\,Q^\prime,\,\bar{P}^\prime,\,\bar{Q}^\prime)=(-P+\bar{P},\,-Q+\bar{Q},\,-2P+\bar{P},\,-2Q+\bar{Q})\,,\\
(p,\,q,\,\bar{p}\,,\bar{q})&\mapsto (p^\prime,\,q^\prime,\,\bar{p}^\prime,\,\bar{q}^\prime)=(-p-\bar{p},\,-q-\bar{q},\,2p+\bar{p},\,2q+\bar{q})\,.
\end{split}
\end{equation}
Similarly to the map for $u_1=0$, these transformations also preserve the {\it sLag} property of all three-cycles under consideration.

It is worth to note that for $u_1=1/2$ our proposed transformations \eqref{dualitymap2a} and \eqref{dualitymap2b} preserve the discrete Wilson lines and displacement parameters on each two-torus.

%%%%%%%%%%%%%%%%%%%%%%%%%%%%%%%%%
\subsection{Limitations of the duality map for $u_1=0$}

In the sections~\ref{u1=0} and~\ref{u1=12} we saw that, from the geometrical point of view, there are pairwise relations between fractional three-cycles on different lattice orientations. Since the duality maps preserve the topological intersection number, the related models have the same chiral spectrum.  In this section we will, however, show that for $u_1=0$ there exist some D-brane configurations for which the conjectured duality does not hold for the non-chiral spectrum if Wilson lines and/or displacements are turned on. In the remainder of this section we will firstly ignore any replica arising due to the shift symmetry and secondly turn a blind eye on the third two-torus $T^2_{(3)}$ (of $B_2$ shape) since both are irrelevant for the discussion.

The setup we will consider is the following: on the one hand, two branes wrapping three-cycles $(0,1;0,1;n_a^3,m_a^3)$ and $(1,0;0,1;n_b^3,m_b^3)$, which intersect in the first two-torus but are parallel in the second one; and on the other hand, the branes wrapping three-cycles $(1,1;\frac{1}{2},0;n_a^3,m_a^3)$ and $(1,-1;\frac{1}{2},0;n_b^3,m_b^3)$ conjectured to be dual to the two previous ones. To illustrate better our point, we will study both the configuration without Wilson lines or displacements and a configuration where they have been turned on.

\begin{figure}[h]
\begin{minipage}{\textwidth}
\begin{minipage}{\dimexpr0.5\textwidth}
\includegraphics[width=6cm]{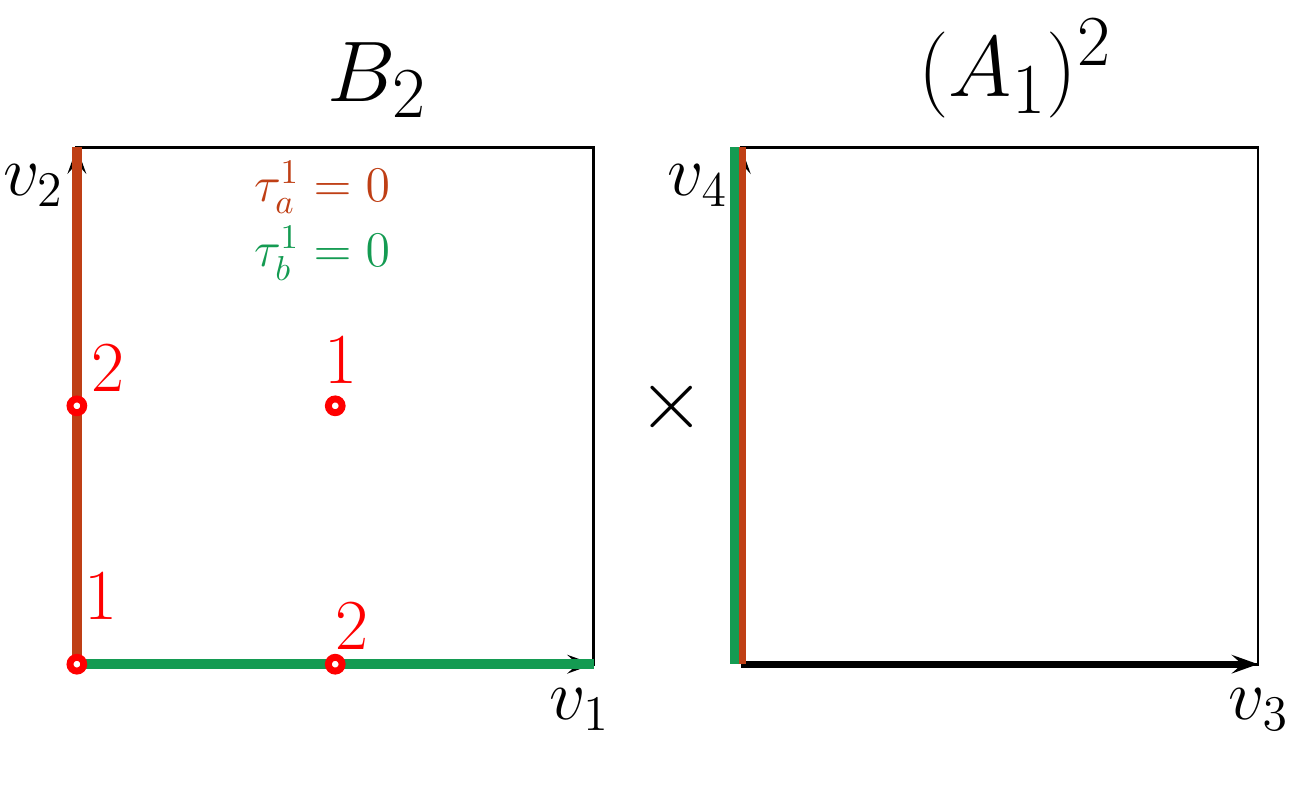}
\end{minipage}
\begin{minipage}{\dimexpr0.5\textwidth}
\includegraphics[width=6cm]{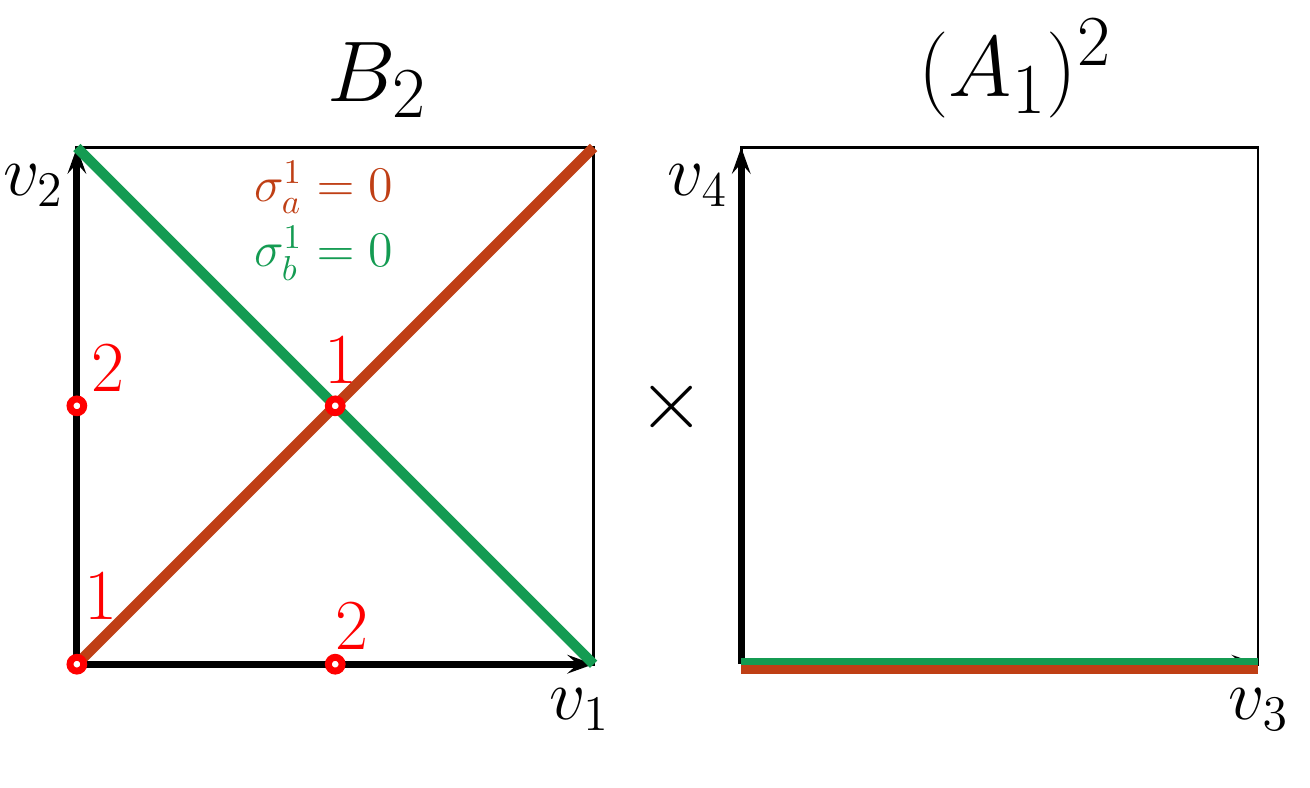}
\end{minipage} 
\end{minipage}
\caption{Left: Branes $a$ (orange) and $b$ (green) with vanishing Wilson lines and displacements, i.e. $(\tau^1_a,\sigma^1_a;\tau^1_b,\sigma^1_b)=(0,0;0,0)$. Right: Dual picture; branes $a$ (orange) and $b$ (green) with again vanishing Wilson line and displacement parameters $(\tau_a^{1'},\sigma_a^{1'};\tau_b^{1'},\sigma_b^{1'})=(0,0;0,0)$.}
\label{dualitytauzero}
\end{figure}

Let us start with the case where no brane has either Wilson line or displacement, like the one shown in figure \ref{dualitytauzero}. In other words, on the left hand side, $(\tau^1_a,\sigma^1_a;\tau^1_b,\sigma^1_b)=(0,0;0,0)$, and on the right hand side, $(\tau_a^{1'},\sigma_a^{1'};\tau_b^{1'},\sigma_b^{1'})=(0,0;0,0)$. It is clear that for this configuration, in both cases,  $I_{ab}^{\mathbb{Z}_2,(1)}=1$ (even though the branes intersect twice in the right hand side picture, the two intersection points are identified under the shift symmetry, leaving only one independent intersection point).

\begin{figure}[h]
\begin{minipage}{\textwidth}
\begin{minipage}{\dimexpr0.5\textwidth}
\includegraphics[width=6cm]{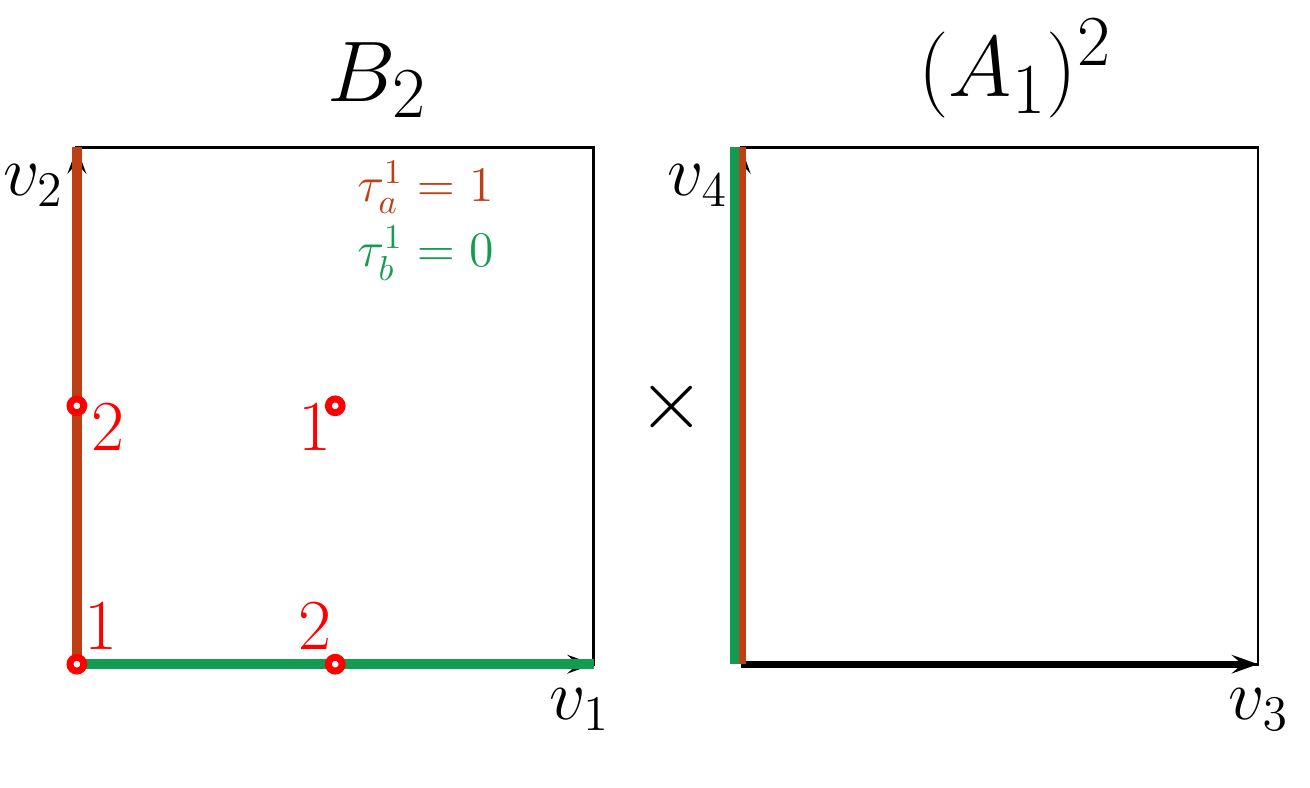}
\end{minipage}
\begin{minipage}{\dimexpr0.5\textwidth}
\includegraphics[width=6cm]{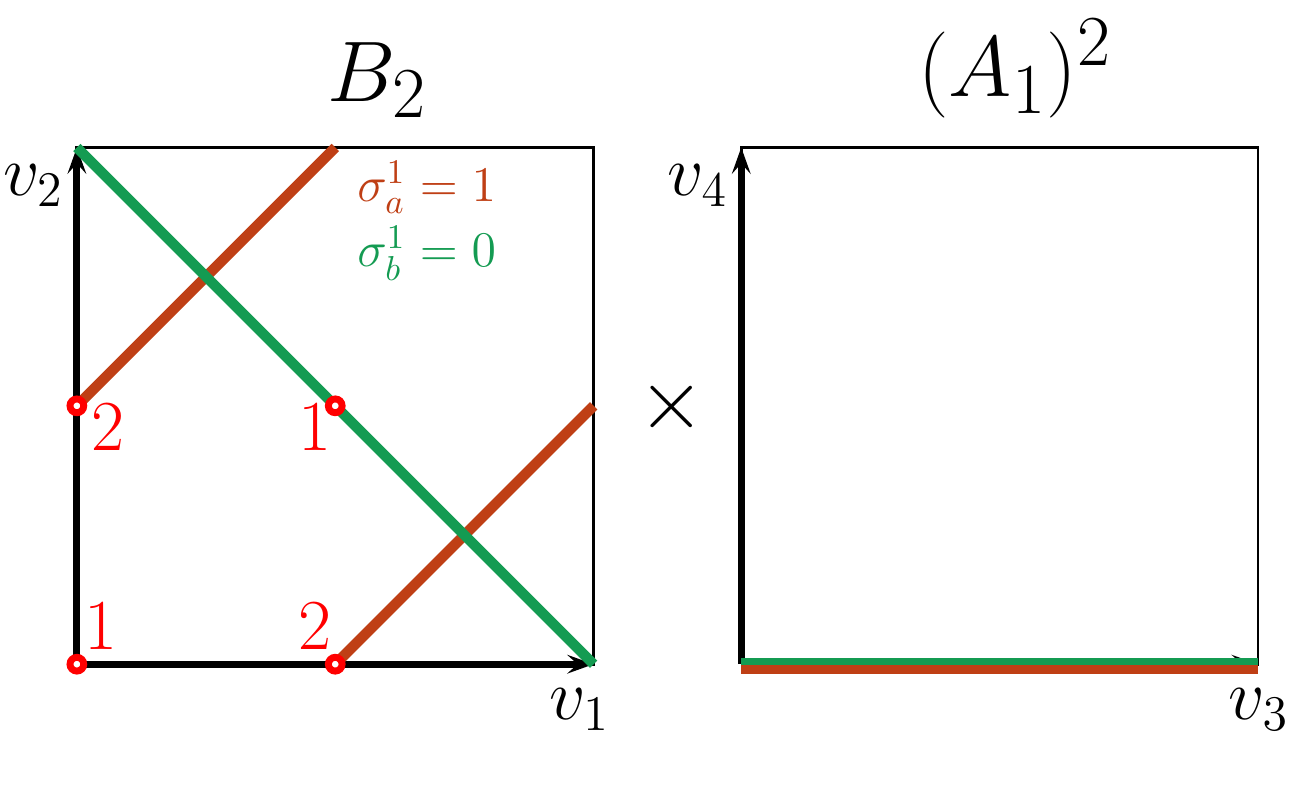}
\end{minipage} 
\end{minipage}
\caption{Left: Branes $a$ (orange) and $b$ (green) with a non-vanishing Wilson line, i.e. $(\tau^1_a,\sigma^1_a;\tau^1_b,\sigma^1_b)=(1,0;0,0)$. Right: Dual picture; branes $a$ (orange) and $b$ (green) with a non-vanishing displacement, i.e. $(\tau_a^{1'},\sigma_a^{1'};\tau_b^{1'},\sigma_b^{1'})=(0,1;0,0)$.}
\label{dualitytauone}
\end{figure}

We will now turn on a discrete Wilson line $\tau_a^1=1$ (which implies turning on a displacement $\sigma_a^{1'}$ in the conjectured dual geometry); that is, on the left hand side we have $(\tau^1_a,\sigma^1_a;\tau^1_b,\sigma^1_b)=(1,0;0,0)$, while on the right hand side we have $(\tau_a^{1'},\sigma_a^{1'};\tau_b^{1'},\sigma_b^{1'})=(0,1;0,0)$. This is shown in figure \ref{dualitytauone}. One can clearly see that now $I_{ab}^{\mathbb{Z}_2,(1)}=1$ on the left while $I_{ab}^{\mathbb{Z}_2,(1)}=0$ on the right. Therefore, it seems not possible to construct the same model on both backgrounds without modifying or refining the conjectured duality relations from section~\ref{u1=0}.

One can see that this mismatch is caused by the exchange of $\tau^1$ and $\sigma^1$ by the duality map \eqref{dualitymap1}. Since the analysis of the toroidal three-cycles of the same length required $\tau^1_a\,\leftrightarrow\,\sigma^1_b$ for the backgrounds to be dual at the topological level, i.e. in terms of the twisted RR tadpole cancellation conditions and the chiral spectrum, we conclude that the conjectured duality does not hold true for $u_1=0$ and $(\tau^1,\sigma^1)\neq(0,0)$ at the level of the non-chiral spectrum without so far unknown modifications, but works perfectly for all lattice orientations and three-cycles with $u_1=1/2$ discussed in section~\ref{u1=12}.

The discussion above concludes that the proposed map \eqref{dualitymap1} - which is based on the chiral spectrum, RR  tadpole cancellation and supersymmetry conditions as well as K-theory constraints - in general does not ensure duality between the  lattice orientations \textbf{A$_{\bf a}$XA} and \textbf{A$_{\bf a}$XB} at the level of the vector-like spectrum. A modification
 either  of the map \eqref{dualitymap1} or
 in the formulae used to compute the beta function coefficients and thus also the associated one-loop gauge threshold corrections might potentially lift the discrepancy.
 However, our preliminary considerations  show that such a modification will be very subtle, and we postpone a dedicated investigation to the future.

%%%%%%%%%%%%%%%%%%%%%%%%%%%%%%%%%%%%%%%%
\section{One-loop vacuum amplitudes and gauge couplings}\label{amplitudes}

The gauge coupling of an $SU(N_a)$ factor at energy scale $\mu$ up to one-loop is in the string frame given by:
\begin{equation}
\frac{8\pi^2}{g_a^2(\mu)}=\frac{8\pi^2}{g_{a,\textup{string}}^2}+\frac{b_a}{2}\ln\left(\frac{M_s}{\mu}\right)^2+\frac{\Delta_a}{2}
\qquad
\text{with}
\qquad 
\frac{4\pi}{g^2_{a,\textup{string}}}=\frac{M_{\textup{Planck}}}{2\sqrt{2} k_a \, M_{\textup{string}}}\frac{\prod_{i=1}^3\sqrt{V_{aa}^{(i)}}}{2}
,
\end{equation}
where the tree-level value $g_{a,\textup{string}}$ for a non-Abelian gauge group supported on a stack of D6$_a$-branes is obtained from the product of the  dimensionless (length$)^2\text{/Vol}_6$ of the wrapped three-cycles,
\begin{equation}\label{Eq:product-Vs}
\begin{split}
\prod_{i=1}^3 V_{aa}^{(i)} 
&=\left((n^1_a)^2+(m^1_a)^2\right)
\times\frac{ \frac{-2\hat{a}R_3}{R_1}(n^2_a)^2+2\hat{d}n^2_am^2_a+\frac{R_1}{2R_3}(m^2_a)^2}{\sqrt{-\hat{a}-\hat{d}^2}}
\times \left(2(n^3_a)^2-2n^3_am^3_a+(m^3_a)^2 \right)\\
&=\frac{1}{\sqrt{-\hat{a}-\hat{d}^2}}\left(-2\hat{a}\frac{R_3}{R_1}(P_a^2+Q_a^2)+\frac{R_1}{2R_3}(\bar{P}_a^2+\bar{Q}_a^2)-2\hat{d}(P_a\bar{P}_a+Q_a\bar{Q}_a)\right)\,,
\end{split}
\end{equation}
with $R_1/R_3$, $\hat{a}$ and $\hat{d}$ encoding the complex structure
in the factorisable description of the $A_3 \times A_1 \times B_2$ lattice (see equation~\eqref{2.complex_structure})
and $k_a=1$ for $SU(N_a)$ gauge factors (for $SO(2N_a)$ and $USp(2N_a)$ gauge groups - such as the left-/right-symmetric and hidden gauge groups in the example in section~\ref{example} -  one has $k_a=2$ since the Dirac-Born-Infeld action 
of a stack of D6-branes then corresponds to its own orientifold image). The beta function coefficient $b_a$ describes the usual field theoretical one-loop running due to massless open string states. For a $SU(N_a)$ gauge factor it is of the form:
\begin{equation}\label{beta-SUN}
b_{SU(N_a)}=-N_a(3-\varphi^{\textbf{Adj}_a})+\sum_{b\neq a}\frac{N_b}{2}(\varphi^{ab}+\varphi^{ab'})+\frac{N_a-2}{2}\varphi^{\textbf{Anti}_a}+\frac{N_a+2}{2}\varphi^{\textbf{Sym}_a},
\end{equation}
where e.g. $\varphi^{ab}$ counts the number of ${\cal N}=1$ chiral multiplets in the representation $(\textbf{N}_a,\ov{\textbf{N}}_b)$ or $(\ov{\textbf{N}}_a,{\textbf{N}}_b)$ (i.e. irrespective of the chirality), while for $USp(2N_x)$ and $SO(2N_y)$ the beta function coefficients are given by
\begin{equation}\label{usp_theory}
b_{USp(2N_x)}=(N_x+1)(-3+\varphi^{\textbf{Sym}_x})+(N_x-1)\varphi^{\textbf{Anti}_x}+\sum_{a\neq x}\frac{N_a}{2}\varphi^{ax}
\end{equation}
and 
\begin{equation}\label{so_theory}
b_{SO(2N_y)}=(N_y-1)(-3+\varphi^{\textbf{Anti}_y})+(N_y+1)\varphi^{\textbf{Sym}_y}+\sum_{a\neq y}\frac{N_a}{2}\varphi^{ay},
\end{equation}
respectively. \\
Massive string states contribute at one-loop to the gauge threshold correction $\Delta_a$.

Since we can describe our non-factorisable lattice background in a factorised way as detailed in section \ref{factorisable_geometry}, we should be able to apply the techniques and formulae to compute string vacuum and gauge threshold amplitudes derived in \cite{Lust:2003ky,Akerblom:2007np,Blumenhagen:2007ip,Gmeiner:2009fb,Honecker:2011sm} for D6-branes in Type IIA orientifolds on factorisable toroidal orbifold backgrounds. More precisely, we will be computing the annulus and M\"obius strip amplitudes with magnetic gauging $\nu$ along the non-compact directions as  required to obtain the one-loop gauge thresholds:
\begin{equation}
\begin{split}
\mathcal{T}_{\mathcal{A}}&\sim\frac{\partial^2}{\partial\nu^2}\tilde{\mathcal{A}}\big|_{\nu=0}=\frac{\partial^2}{\partial\nu^2} \left[
\int_0^\infty dl\sum_{a,b}\langle B_a |e^{-2\pi l H_{cl}}|B_b \rangle\right]_{\nu=0},\\
\mathcal{T}_{\mathcal{M}}&\sim\frac{\partial^2}{\partial\nu^2}\tilde{\mathcal{M}}\big|_{\nu=0}=\frac{\partial^2}{\partial\nu^2}\left[
\int_0^\infty dl\sum_{a}\left(\langle C|e^{-2\pi l H_{cl}}|B_a \rangle+h.c.\right)\right]_{\nu=0},
\end{split}
\end{equation}
where $|B_a (\nu)\rangle$ are the (magnetically gauged) boundary states corresponding to D$6_a$-branes wrapped along fractional cycles $\pi_a$  and $|C(\nu)\rangle$ is the overall crosscap state pertaining to the three-cycle $\pi_{O6}$.

These amplitudes take into account both massless and massive string excitations. In general, the result of a given amplitude will be of the form
\begin{equation}
\mathcal{T}=c\int dl + b\ln\left(\frac{M_s}{\mu}\right)^2+\Delta,
\end{equation}
where $c$, the  coefficient of $\int dl$, gives the contribution of the amplitude to the RR tadpole cancellation conditions; $b$, the coefficient of $\ln\left(\frac{M_s}{\mu}\right)^2$, corresponds to the contribution to the beta function coefficient due to massless strings in the loop; finally $\Delta$ is the contribution to the gauge threshold corrections due to massive string excitations.

For a given model, after all relevant amplitudes for each sector have been computed and the cancellation of the prefactors of $\int \!dl$  has been cross-checked, comparing the obtained contribution to the beta function coefficient for each sector with its field theoretical counterpart in equations~\eqref{beta-SUN} and~\eqref{usp_theory}, and taking into account the chiral massless spectrum computed from topological intersection numbers according to table~\ref{tab:open-chiral-string-spectrum}, 
we can derive the missing vector-like part of the massless open string spectrum.
We will briefly summarise the findings of contributions to the $SU(N)$ beta function coefficients in section~\ref{Summary-b-SUN} and
use the identification of contributions to~\eqref{usp_theory} to distinguish enhanced gauge groups of type $USp(2N)$ or $SO(2N)$ in section~\ref{USP_or_SO},
before applying the formalism to obtain the full massless spectrum of an example in section~\ref{example}.

From a computational point of view, the main difference to the examples considered in \cite{Gmeiner:2009fb,Honecker:2011sm} is the existence of replicas arising from the shift symmetry when transitioning from the original non-factorisable to a factorisable representation of the toroidal three-cycles. Naively, the original cycles and the replicas should be considered independently, and one should take into account all possible contributions from both of them when computing a given amplitude. The total contribution should then be multiplied by a factor of 1/2, since each actual contribution has been considered twice because of the shift symmetry.
In this section, however, we will take a slightly different approach. Our aim is to write the amplitudes in such a way that we will only need to consider one representative of each pair to obtain the full contribution.

As a final remark, in the remainder of this section we will only consider three-cycles with integer wrapping numbers. Therefore, for cycles with half-integer wrapping numbers appearing due the shift symmetry on the second two-torus $(n^2/2,m^2)$ with $n^2\in2\mathbb{Z}+1$ when transitioning from the non-factorisable to the factorisable formulation as detailed below equation~\eqref{fractorised-fact-cycle}, we will double them to $(\tilde{n}^2,\tilde{m}^2)\equiv(n^2,2m^2)$ and account for this doubling in the correction factor $\kappa$ defined below in equation~\eqref{defkappa}.

%%%%%%%%%%%%%%%%%%%%%%%%%%%%%%%%%%%%%
\subsection{Classification of cycles}\label{sec:cycleclassification}

As we have already mentioned, because of the shift symmetry in equation \eqref{shift-symmetry} (which reflects the fact that a cell in the factorised picture corresponds to two unit cells in the non-factorisable one), when we wrap a stack of D6-branes along a three-cycle in the factorised picture a copy of the stack may appear. If no replica appears it means that the stack is wrapped on a cycle twice as long as in the non-factorisable picture. 
The three-cycles can be grouped into four classes with respect to the shift symmetry, and some of the computations later on will change slightly depending on the type of cycles the branes are wrapped on as detailed below.

The class to which a cycle belongs is determined by its (factorised) wrapping numbers. Since the shift symmetry in eq. \eqref{shift-symmetry} only involves the first two two-tori $T^2_{(1)} \times T^2_{(2)}$, the wrapping numbers in the third two-torus $T^2_{(3)}$ do not play any role in the classification, and we will only consider the first two-tori $T^2_{(1)} \times T^2_{(2)}$ in the following. As already mentioned in footnote~\ref{Foot:integer} of section~\ref{factor+geometry}, 
the wrapping numbers are taken to be fully integer, i.e. if there are half-integer wrapping numbers in the second two-torus, use $(\tilde{n}^2,\tilde{m}^2)$ as defined above (notice that to simplify the notation we will drop the tilde from now on). After all these considerations, the different types are:
\begin{enumerate}
\item $n^1+m^1=$odd, $m^2$=odd. In this kind of configuration none of the one-cycles wrapped by the branes is invariant under the shift symmetry, giving rise to a separated copy of the branes in both two-tori.
\item $n^1+m^1=$odd, $m^2$=even. The one-cycle in the second two-torus is invariant under the shift symmetry but the one-cycle in the first one is not invariant. Therefore there is a separate copy of the branes in the first two-torus while both copies coincide in the second one.
\item $n^1+m^1=$even, $m^2=$odd. The one-cycle in the first two-torus is invariant under the shift symmetry but the one-cycle in the second one is not invariant. Therefore there is a separate copy of the branes in the second two-torus while both copies coincide in the first one.
\item $n^1+m^1=$ even, $m^2$=even. Since both one-cycles are invariant under the shift symmetry in this kind of configurations, no additional copy of the stack of branes appears. Instead, the wrapped cycle is twice as long as in the non-factorisable picture. This is the case that corresponds to the half-integer wrapping numbers described above. Since the wrapping numbers in each two-torus have to be coprime, $n^2$ needs to be odd, and when we divide by two to get the actual cycle, the half-integer wrapping number arises.
\end{enumerate}

\begin{figure}[h!]
\begin{center}
\includegraphics[width=12cm]{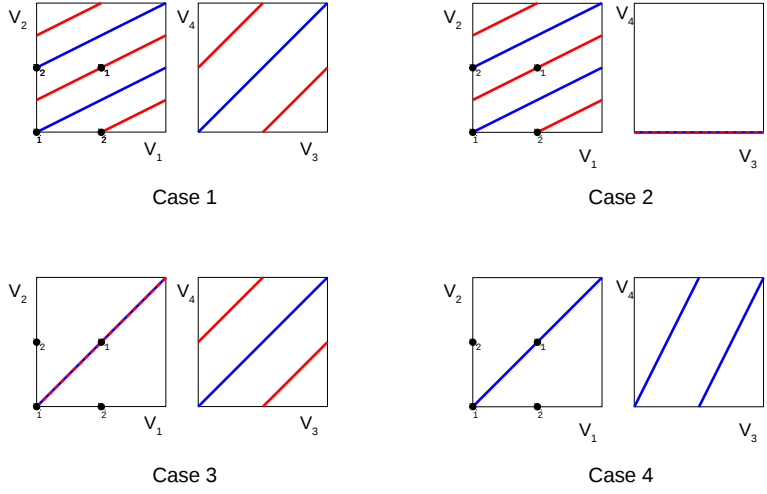}
\caption{Particular examples for each of the possible configurations with/without copies in one or two two-tori; the wrapping numbers are listed as $(n^1,m^1)\times(n^2,m^2)$. Top-left: Case 1, $(2,1)\times (1,1)$. Top-right: Case 2, $(2,1)\times (1,0)$. Bottom-left: Case 3, $(1,1)\times (1,1)$. Bottom-right: Case 4, $(1,1)\times (1,2)$.}
\label{duplicatedcycles}
\end{center}

\end{figure}

Figure \ref{duplicatedcycles} shows one example for each of all four kinds of configurations. While the pictures correspond to a background with real part of the complex structure $u_1=0$, similar ones can be drawn for $u_1=1/2$.

%%%%%%%%%%%%%
\subsection{Subtleties of the shift symmetry}\label{ShiftSymm}

Most of the time, both branes in a replica pair give the same contribution to a given annulus or M\"obius strip amplitude (while keeping the other end of the open string on another given brane). In that case, the effect of the replicas is taken into account by introducing a correction factor, which we will denote by $\kappa$, that ensures we count the correct multiplicity of contributions to that amplitude. The value of $\kappa$ depends on what types of three-cycles (according to the classification above) are being wrapped by the two branes giving rise to the open string sector we are considering. These values are:
\begin{equation}\label{defkappa}
\kappa=\begin{cases}  2\quad&\textup{no cycles of type 4,}\\ 1\quad&\textup{one cycle of type 4,}\\1/2\quad&\textup{both cycles of type 4.}\end{cases}
\end{equation}
For more details on why $\kappa$ takes those values, we refer the reader to appendix~\ref{kappa}.\\

This correction factor allows us, given two stacks of branes $a$ and $b$, to obtain the correct intersection number as $\kappa I_{ab}$ where $I_{ab}=\prod_{i=1}^3(n_a^im_b^i-n_b^im_a^i)$ is the usual intersection number between two stacks of branes on a factorisable toroidal background.\footnote{
Notice that we follow here the usual sign convention, $\pi^{\text{torus}}_a \circ \, \pi^{\text{torus}}_b = - I_{ab}$, which is required to reproduce the correct net-chiralities when matching spectra computed by means of  bulk (cf. table~\ref{tab:open-chiral-string-spectrum})
versus toroidal \& $\Z_2$-invariant  (cf. table~\ref{tab:contributions-to-beta-function})
intersection numbers.
}

Additionally, $\kappa$ can also be used for the computation of $\mathbb{Z}_2$ invariant points, i.e. the following formula holds: $\kappa I_{ab}^{\mathbb{Z}_2}=\kappa (-1)^{\tau_0^a+\tau_0^b}I_{ab}^{\mathbb{Z}_2,(1)}I_{ab}^{(2)}I_{ab}^{\mathbb{Z}_2,(3)}$, where $I_{ab}^{\mathbb{Z}_2,(i)}$ denote the number of $\mathbb{Z}_2$-invariant intersection points on the first $(i=1)$  and on the third torus $(i=3)$, as given in tables \ref{tab:number-Z2-invariant points on T2_1} \& \ref{tab:number-Z2-invariant points on T2_1-parallel} and \ref{tab:number-Z2-invariant points on T2_3}, respectively.
%%%%%%%%%%%%%%%%%%%%%%%%%%%%%
\begin{table}[h]
\begin{center}
\begin{tabular}{|c|c|c|}
\hline
\multicolumn{3}{|c|}{\textbf{$\mathbb{Z}_2$ invariant intersection numbers $I_{ab}^{\mathbb{Z}_2,(1)}$}}\\
  \multicolumn{3}{|c|}{\textbf{on the two-torus $T^2_{(1)}$}}\\
 \multicolumn{3}{|c|}{\textbf{for intersecting three-cycles on $T^2_{(2)}$ }}\\
\hline\hline
\diagbox{type of $b$}{type of $a$} & 1/2 & 3/4 \\
\hline
1/2&$\delta_{\tau^1_a,\tau^1_b}$ &$(-1)^{\tau^1_b\sigma^1_a}$\\
\hline
3/4& $(-1)^{\tau^1_a\sigma^1_b}$ &$2\delta_{\sigma^1_a\sigma^1_b}$ \\
\hline
\end{tabular}
\caption{$\mathbb{Z}_2$-invariant intersection numbers $I_{ab}^{\mathbb{Z}_2,(1)}$ on the first two-torus $T^2_{(1)}$ for non-parallel three-cycles on  the second two-torus $T^2_{(2)}$, by taking into account the relative Wilson lines and displacements. The table is symmetric along the diagonal upon the exchange of $a \leftrightarrow b$.
}\label{tab:number-Z2-invariant points on T2_1}
\end{center}
\end{table}
%%%%%%%%%%%%%%%%%%%%%%%%%%%%%%%%%%%%%%%%%%%
\begin{table}[h]
\begin{center}
\begin{tabular}{|c|c|c|c|c|}
\hline
\multicolumn{5}{|c|}{\textbf{$\mathbb{Z}_2$ invariant intersection numbers $I_{ab}^{\mathbb{Z}_2,(1)}$ on the two-torus $T^2_{(1)}$}}\\
 \multicolumn{5}{|c|}{\textbf{for parallel three-cycles on $T^2_{(2)}$ }}\\
\hline\hline
\diagbox{type of $b$}{type of $a$} & 1 & 2 & 3 & 4 \\
\hline
1&$\begin{array}{cc}
1&\text{if }I^{(1)}_{ab}\text{ odd }\\
2\delta_{\tau^1_a,\tau^1_b}&\text{if }I^{(1)}_{ab}\text{ even }
\end{array}$& $\emptyset$  & $(-1)^{\tau^1_b\sigma^1_a}$&$\emptyset$\\
\hline
2&  &$2\delta_{\tau^1_a,\tau^1_b}$ &$\emptyset$ &$(-1)^{\tau^1_b\sigma^1_a}$\\
\hline
3& & & $2\delta_{\sigma^1_a\sigma^1_b}$ &$\emptyset$ \\
\hline
4& & & &$\delta_{\sigma^1_a\sigma^1_b}$ \\
\hline
\end{tabular}
\caption{$\mathbb{Z}_2$-invariant intersection numbers $I_{ab}^{\mathbb{Z}_2,(1)}$ on the first two-torus $T^2_{(1)}$ for parallel three-cycles on the second two-torus $T^2_{(2)}$, by taking into account the relative Wilson lines and displacements. The $\emptyset$-sign denotes the case where the branes of two different types are not parallel (e.g. 1 and 2) on the second torus. The table is symmetric along the diagonal upon exchanging $a \leftrightarrow b$.
}\label{tab:number-Z2-invariant points on T2_1-parallel}
\end{center}
\end{table}
%%%%%%%%%%%%%%%%%%%%%%%%%%%
\begin{table}[h]
\begin{center}
\resizebox{16cm}{!}{
\begin{tabular}{|c||c||c|c|c||c|c|c|}
\hline
\multicolumn{8}{|c|}{\textbf{$\mathbb{Z}_2$ invariant intersection numbers $I_{(ab)}^{\mathbb{Z}_2,(3)}$ on a the $\mathbb{Z}_4$ invariant $B_2$-type torus $T^2_{(3)}$}}\\
\hline \hline
\diagbox{$(n^3_b,\,m^3_b)$}{$(n^3_a,\,m^3_a)$} & (odd, even) & \multicolumn{3}{|c||}{(odd, odd)} & \multicolumn{3}{|c|}{(even, odd)} \\ 
\hline \hline
 &  & \diagbox{$\sigma^3_b$}{$\sigma^3_a$} & 0 & 1 & \diagbox{$\sigma^3_b$}{$\sigma^3_a$} & 0 & 1 \\ 
\cline{3-8} 
(odd, even) & $(1+(-1)^{\tau_{ab}^3})\delta_{\sigma_{ab}^3,0}$ & 0 & 1 & $(-1)^{\tau^3_b}$ &0 & 1 & $(-1)^{\tau_b^3}$ \\ 
\cline{3-8}
 &  & 1 & $(-1)^{\tau_{ab}^3}$ & $(-1)^{\tau^3_a}$ & 1 & $(-1)^{\tau^3_a}$ & $(-1)^{\tau_{ab}^3}$ \\ 
\hline \hline
 &  & \multicolumn{3}{|c||}{} & \diagbox{$\sigma^3_b$}{$\sigma^3_a$} & 0 & 1 \\ 
\cline{6-8}
(odd, odd) & & \multicolumn{3}{|c||}{ $(1+(-1)^{\tau_{ab}^3})\delta_{\sigma_{ab}^3,0}$ }  & 0 & 1 & $(-1)^{\tau_{ab}^3}$ \\ 
\cline{6-8} 
 &  &  \multicolumn{3}{|c||}{} & 1 & $(-1)^{\tau_{ab}^3}$ & 1 \\
\hline \hline
& & \multicolumn{3}{|c||}{}&\multicolumn{3}{|c|}{}\\
(even, odd)& & \multicolumn{3}{|c||}{}&\multicolumn{3}{|c|}{ $(1+(-1)^{\tau_{ab}^3})\delta_{\sigma_{ab}^3,0}$ }\\
& & \multicolumn{3}{|c||}{}&\multicolumn{3}{|c|}{}\\
\hline
\end{tabular} }
\caption{$\mathbb{Z}_2$-invariant intersection numbers with relative Wilson lines and displacements for a two-torus $T^2_{(3)}$ with  $\mathbb{Z}_4$ symmetry. Here we used the shorthand notation: $\sigma_{ab}^3:=|\sigma^3_a-\sigma^3_b|$ and $\tau_{ab}^3:=|\tau^3_a-\tau^3_b|$. The table is symmetric along the diagonal upon exchanging $a \leftrightarrow b$.
}\label{tab:number-Z2-invariant points on T2_3}
\end{center}
\end{table}
%%%%%%%%%%%%%%%%%%%%%%%%%%%%%

\begin{figure}[h!]
\begin{center}
\includegraphics[width=15cm]{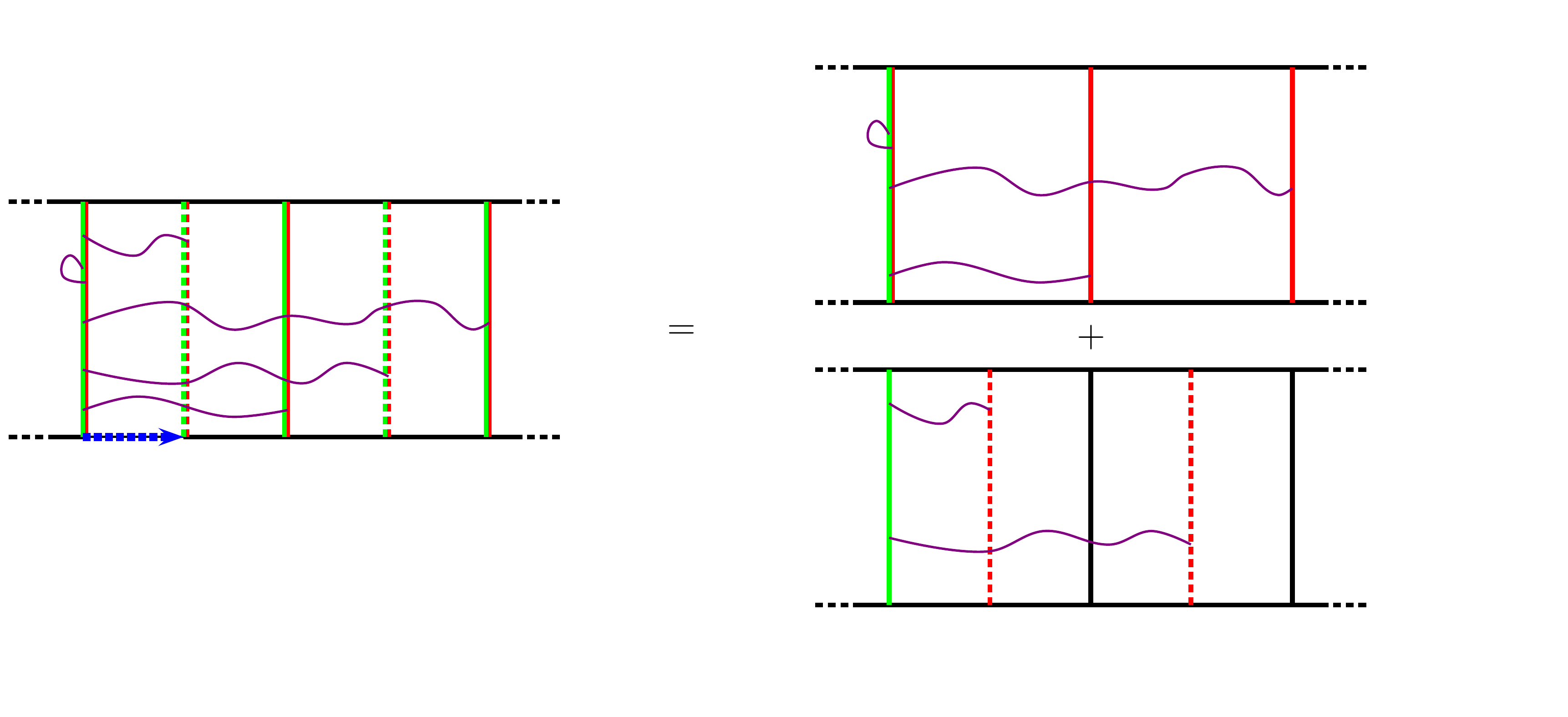}
\caption{Strings contributing to the annulus amplitude between two stacks of branes (red and green) parallel on the second two-torus (the first and third two-tori are not shown). The blue arrow represents the effect of the shift symmetry. The strings can be divided into two sets depending on which brane they end on. The formulae 
in \cite{Gmeiner:2009fb} for the so-called lattice sums of Kaluza-Klein and winding states can be applied to each set on the right hand side separately.}
\label{fig:doubling}
\end{center}
\end{figure}

There are a few cases where each component of a replica pair gives a different contribution to the amplitude (again, with the other end of the string being fixed on a given brane). This might happen when the two stacks of branes involved are parallel on a two-torus. In those cases, instead of being corrected by $\kappa$, the amplitude will just be given by the sum of the two different types of Kaluza-Klein and winding sum contributions. A particular case is shown in figure \ref{fig:doubling} where the two brane stacks are parallel on the second two-torus, and intersect on the first and third two-tori (which are not shown for simplicity). Each colour represents a different stack, and the dashed lines are the replicas that arise because of the shift symmetry. Two unit cells of the two-torus are shown for a better depiction of higher winding modes. The total annulus amplitude will take into account strings starting from a solid green brane to any red one (both solid and dashed). Strings starting from the dashed green brane can be related to strings from the solid green brane by the shift symmetry; therefore, to avoid double counting, we should not consider them. As shown in figure \ref{fig:doubling}, the strings contributing to the amplitude can be divided into two sets, depending on which brane image they end on (solid or dashed red). For each of these sets we can apply directly the formulae for the Kaluza-Klein and winding sums 
 from~\cite{Gmeiner:2009fb}, and the total amplitude is then given by the sum of these two different lattice sums.

Finally, since it will be useful later, let us formally define the weighted intersection number $\tilde{I}_{a}^{\Omega\mathcal{RQ}^{-k},(i)}=2(1-b_i)I_{a}^{\Omega\mathcal{RQ}^{-k},(i)}$ 
and  weighted (length$)^2$ of a one-cycle $\tilde{V}^i_{aa^\prime}=2(1-b_i)V_{aa^\prime}^i$ in analogy to \cite{Gmeiner:2009fb}. However, only the case $i=3$ agrees with the definition in \cite{Gmeiner:2009fb} where $b_3=1/2$ corresponds to a tilted two-torus and $b_3=0$ to an untilted two-torus, whereas for the first and second two-torus in the formulation of equation~\eqref{eq:g-factorised}, we need to  first distinguish the two choices $u_1=0$ or $u_1=1/2$:
\begin{itemize}
\item $u_1=0$. In this case, for all four lattice orientations, $b_1+b_2=1/2$, so one can have either $(b_1,b_2)=(0,1/2)$ or $(b_1,b_2)=(1/2,0)$. When the brane and the O-plane are at non-trivial angles on $T^4_{(3)}=T^2_{(1)} \times T^2_{(2)}$ both choices lead to the same result. However, when the brane and O-plane are parallel along  $T^2_{(2)}$, and the O-plane wraps a cycle of type 1 or 4, we have to choose $(b_1,b_2)=(0,1/2)$. In all other cases, $(b_1,b_2)=(1/2,0)$ is applied. This choice will be further justified later on in section \ref{USP_or_SO}.
\item $u_1=1/2$. For all four lattice orientation, the O-plane only wraps cycles of type 2 or 4. In the first case, one always has $(b_1,b_2)=(1/2,1/2)$. For the O-plane wrapping cycles of type 4, $(b_1,b_2)$  either takes the value  $(1/2,0)$ or $(0,1/2)$. As in the  case $u_1=0$, both choices lead to the same physical quantities if the brane and O-plane are at non-trivial angles on $T^4_{(3)}$. 
If the O-plane and the brane are parallel  along $T^2_{(1)}$, one has to choose $(b_1,b_2)=(1/2,0)$, otherwise $(b_1,b_2)=(0,1/2)$. 
\end{itemize}
After highlighting these new subtleties appearing due to the non-factorised nature of the background lattice $A_3 \times A_1 \times B_2$, we are ready to present the annulus and M\"obius strip amplitudes in the following sections, which are
needed to determine the one-loop beta function coefficients and  gauge threshold corrections. A detailed derivation of the prefactors $c_{\mathcal{A}/\mathcal{M}}$ and the Jacobi theta functions and lattice sums appearing below can be found in \cite{Gmeiner:2009fb}.

%%%%%%%%%%%%%%%%%%%%%%%%%%%%%%%%%%%%%%%%%%%%%%%%%%%%%%
\subsection{Contributions to the annulus amplitudes with \textbf{1} insertion}\label{Annulus-1-insertion}

In this section we present the annulus amplitudes between two stacks of D6-branes with a \textbf{1} projector insertion in the tree channel. The following cases need to be distinguished:
\begin{itemize}
\item \textbf{Two branes intersect at three non-trivial angles }$\pi(\phi^1,\phi^2,\phi^3=-\phi^1-\phi^2)$. Apart from the factor $\kappa$, defined in equation \eqref{defkappa}, the amplitude is formally the same as in the usual factorisable toric backgrounds and is given by:
\begin{equation}\label{Eq:TA-1-3angles}
\begin{split}
\mathcal{T}_{\mathcal{A}}&:=\kappa c_\mathcal{A}^\textbf{1}\int  \text{d}ll^\varepsilon\Theta_\text{osc}^{\textbf{1},\;(\phi^1,\phi^2,\phi^3)}(l)=\kappa c_\mathcal{A}^\textbf{1}\int  \text{d}ll^\varepsilon\sum_{\alpha\beta}(-1)^{2(\alpha+\beta)}\frac{\vartheta^{\prime\prime}{\alpha\brack \beta}(0,2il)}{\eta^3(2il)}\prod_{i=1}^3\frac{\vartheta{\alpha\brack \beta}}{\vartheta{1/2\brack 1/2}}(\phi^i,2il)\\
&=-2\pi \kappa c_\mathcal{A}^\textbf{1}\int  \text{d}ll^\varepsilon\sum_{i=1}^3\frac{\vartheta^\prime_1}{\vartheta_1}(\phi^i,2il)\\
&=-2\pi^2\kappa c_\mathcal{A}^\textbf{1}\left(\sum_{i}\cot{(\pi\phi^i)}\int\text{d}l
+\ln\left(\frac{M_s}{\mu}\right)^2\left(\sum_i\frac{\text{sgn}\,\phi^i}{2}\right) -\frac{1}{2}\sum_i\ln\left(\frac{\Gamma(\mid\phi^i\mid)}{\Gamma(1-\mid\phi^i\mid)}\right)^{\text{sgn}\;\phi^i}\right)
\end{split}
\end{equation}
with the prefactor $c^\textbf{1}_\mathcal{A}=\frac{N_b I_{ab}}{4\pi^2}$ containing the toroidal intersection number $I_{ab}$ 
 and the usual identification of the dimensional regularisation parameter with the string scale $M_s$
  via \mbox{$\frac{1}{\varepsilon}+\gamma_E-\ln2 \equiv \ln\left(\frac{M_s}{\mu}\right)^2$}, as detailed e.g. in~\cite{Gmeiner:2009fb}.

\item \textbf{Two branes intersect in one line}. One of the angles vanishes. We will distinguish two cases: The parallel one-cycles are on the $B_2$-type torus $T^2_{(3)}$ or inside its complement $T^4_{(3)}$. 

\textbf{On the two-torus $T^2_{(3)}$}: $\pi(\phi,-\phi,0)$. As the D6-branes are parallel on the torus $T^2_{(3)}$, we can simply use the usual formulae from the factorised case without any modification (except for the global factor $\kappa$). The annulus amplitude is then given by:
\begin{equation}\label{amplitude1}
\begin{split}
\mathcal{T}_{\mathcal{A}}&:=\kappa c_\mathcal{A}^\textbf{1}\int  \text{d}ll^\varepsilon\tilde{\mathcal{L}}_{ab}^{\mathcal{A},(3)}\Theta_\text{osc}^{\textbf{1},\;(\phi,-\phi,0)}(l)=4\pi^2 \kappa c_\mathcal{A}^\textbf{1}\int  \text{d}ll^\varepsilon\tilde{\mathcal{L}}_{ab}^{\mathcal{A},(3)}\\
&=4\pi^2\kappa c_\mathcal{A}^\textbf{1}\left(\int\text{d}l
+\frac{\delta_{\sigma^3_{ab},0}\delta_{\tau^3_{ab},0}}{V_{ab}^{(3)}}     
\ln\left(\frac{M_s}{\mu}\right)^2-\frac{\Lambda(\sigma^3_{ab},\tau^3_{ab},\mathsf{v}_3,V_{ab}^{(3)})}{V_{ab}^{(3)}} \right).
\end{split}
\end{equation}
The lattice contribution to the gauge thresholds is given by:
\begin{equation}\label{lambdacontribution}
\begin{split}
\Lambda(\sigma^3,\tau^3,\mathsf{v}_3,V_{ab}^{(3)})=&
\delta_{\sigma^3,0}\delta_{\tau^3,0}\ln\,(2\pi V_{ab}^{(3)}\eta^4(i\mathsf{v}_3))\\
&+(1-\delta_{\sigma^3,0}\delta_{\tau^3,0})\ln\,
\left\vert e^{-\pi (\sigma^3)^2 \mathsf{v}_3/4 }\frac{\vartheta_1(\frac{\tau^3}{2}-i\frac{\sigma^3}{2}\mathsf{v}_3,i\mathsf{v}_3)}{\eta(i\mathsf{v}_3)}\right\vert^2\,.
\end{split}
\end{equation} 
Here $\mathsf{v}_3$  is the volume of the $B_2$-type two-torus in units of $\alpha'$ 
and $V_{ab}^{(3)}=\frac{(L_a^{(3)})^2}{\mathsf{v}_3}$ is the (normalised) square of the length of the one-cycle on the two-torus $T^2_{(3)}$
as specified in  terms of the factorised wrapping numbers in equation~\eqref{Eq:product-Vs}.
 As before, the prefactor $c_\mathcal{A}^\textbf{1}$ can be transferred from the factorisable case:
\begin{equation}
c_\mathcal{A}^\textbf{1}=-\frac{N_b V_{ab}^{(3)}I_{ab}^{(1\cdot2)}}{8\pi^2}\,.
\end{equation}

\textbf{Inside the four-torus $T^4_{(3)}$}. When the D-branes are parallel inside the $T^4_{(3)}$-torus the amplitude will depend on the two-torus on which the branes are parallel and the type of cycles that the branes wrap. For concreteness we will present the amplitudes for the case where the branes are parallel along the second two-torus, i.e. $\pi(\phi,0,-\phi)$, and comment briefly on the case of branes parallel on the first two-torus at the end.

When the D-branes wrap three-cycles of type 2 or 4 according to the classification in section \ref{sec:cycleclassification}, we obtain an analogous expression to the one above:
\begin{equation}\label{Eq:TA-1-Type24}
\begin{split}
\mathcal{T}_{\mathcal{A}}&:=\kappa c_\mathcal{A}^\textbf{1}\int  \text{d}ll^\varepsilon\tilde{\mathcal{L}}_{ab}^{\mathcal{A},(2)}\Theta_\text{osc}^{\textbf{1},\;(\phi,0,-\phi)}(l)=4\pi^2 \kappa c_\mathcal{A}^\textbf{1}\int  \text{d}ll^\varepsilon\tilde{\mathcal{L}}_{ab}^{\mathcal{A},(2)}\\
&=4\pi^2\kappa c_\mathcal{A}^\textbf{1}\left(\int\text{d}l
+\frac{\delta_{\sigma^2_{ab},0}\delta_{\tau^2_{ab},0}}{V_{ab}^{(2)}}     
\ln\left(\frac{M_s}{\mu}\right)^2-\frac{\Lambda(\sigma^2_{ab},\tau^2_{ab},\mathsf{v}_2,V_{ab}^{(2)})}{V_{ab}^{(2)}} \right)
\end{split}
\end{equation}
with
\begin{equation}
c_\mathcal{A}^\textbf{1}=-\frac{N_b V_{ab}^{(2)} I_{ab}^{(1\cdot3)}}{8\pi^2}\, ,
\end{equation}
and $V_{ab}^{(2)}$ the square of the (normalised) one-cycle length as specified in equation~\eqref{Eq:product-Vs} for the factorised description of the $A_3 \times A_1 \times B_2$ lattice; and the function $\Lambda$ is defined as in equation \eqref{lambdacontribution}.

When the three-cycles wrapped by the branes are of type 1 or 3, we encounter the situation mentioned above in which the amplitude is given by the combination of two kinds of winding sum contributions. The amplitude can then be written as:
\begin{equation}\label{Eq:TA-1-13}
\begin{split}
\mathcal{T}_{\mathcal{A}}=4\pi^2c_\mathcal{A}^\textbf{1}\bigg(&2\int\text{d}l
+\frac{(\delta_{\sigma^2_{ab},0}+\delta_{\sigma^2_{ab},1})\delta_{\tau^2_{ab},0}}{V_{ab}^{(2)}}     
\ln\left(\frac{M_s}{\mu}\right)^2\\&-\frac{\Lambda(\sigma^2_{ab},\tau^2_{ab},\mathsf{v}_2,V_{ab}^{(2)})+\Lambda(1-\sigma^2_{ab},\tau^2_{ab},\mathsf{v}_2,V_{ab}^{(2)})}{V_{ab}^{(2)}} \bigg) \, ,
\end{split}
\end{equation}
with the coefficient $c_\mathcal{A}^\textbf{1}$ and the function $\Lambda$ again defined as above.

When the branes are parallel along the first two-torus, the formulae are completely analogous (with the obvious exchange of superscripts $1\leftrightarrow2$ in the relevant quantities), being valid for wrapped cycles of types 3 or 4, and types 1 or 2, respectively.

\item \textbf{Two branes are completely parallel} $\pi(0,0,0)$.
If both D6-branes are completely parallel, there is no contribution to the tadpole, the beta function or the gauge threshold corrections: 
\begin{equation}\label{Eq:TA-1-parallel}
\mathcal{T}_{\mathcal{A}}=\kappa c_\mathcal{A}^\textbf{1}\int  \text{d}ll^\varepsilon\prod_{i=1}^3\tilde{\mathcal{L}}_{ab}^{(i)}\Theta_\text{osc}^{\textbf{1},\;(0,0,0)}(l)=0.
\end{equation}
\end{itemize}

%%%%%%%%%%%%%%%%%%%%%%%%%%%%%%%%%%%%%%%%%%%%
\subsection{Contributions to the annulus amplitudes with $\mathbb{Z}_2$ insertion}\label{Annulus-Z2-insertion}

In this section we present the annulus amplitudes between two stacks of D6-branes with a non-trivially acting $\mathbb{Z}_2$ projector insertion in the tree channel.
The following cases have to be distinguished:
\begin{itemize}
\item \textbf{Two branes intersect at three non-trivial angles }$\pi(\phi^1,\phi^2,\phi^3=-\phi^1-\phi^2)$. As in the case of the \textbf{1} insertion, we only need to add the factor $\kappa$ to the expression for the amplitude in the usual factorisable toric backgrounds. In other words, the amplitude is given by:
\begin{equation}\label{Eq:TA-Z2-3angles}
\begin{split}
\mathcal{T}_{\mathcal{A}}&:=\kappa c_\mathcal{A}^{\mathbb{Z}_2}\int  \text{d}ll^\varepsilon\Theta_\text{osc}^{{\mathbb{Z}_2},\;(\phi^1,\phi^2,\phi^3)}(l)=-2\pi \kappa c_\mathcal{A}^{\mathbb{Z}_2}\int  \text{d}ll^\varepsilon \left(\frac{\vartheta'_1}{\vartheta_1}(\phi^{(2)},2il)+\sum_{i=1,3}\frac{\vartheta'_4}{\vartheta_4}(\phi^{(i)},2il)\right)\\
&=-2\pi^2\kappa c_\mathcal{A}^{\mathbb{Z}_2}\bigg[\cot(\pi\phi^{(2)})\int_0^\infty dl+\sum_{i=1,3}(\text{sgn}(\phi^{(i)})-2\phi^{(i)})\ln2\\
&+\sum_{i=1}^3\left(\ln\left(\frac{M_s}{\mu}\right)^2\frac{\text{sgn}(\phi^{(i)})}{2}-\frac{1}{2}\ln\left(\frac{\Gamma(|\phi^{(i)}|)}{\Gamma(1-|\phi^{(i)}|)}\right)^{\text{sgn}(\phi^{(i)})}\right)\bigg] \, ,
\end{split}
\end{equation}
where $c_\mathcal{A}^{\mathbb{Z}_2}=\frac{N_b I_{ab}^{\mathbb{Z}_2}}{4\pi^2}$ with the $\Z_2$-invariant intersection number $I_{ab}^{\mathbb{Z}_2}$ computed as detailed in section~\ref{ShiftSymm}.
\item \textbf{Two branes intersect in one line}. One needs to distinguish two situations. Either the branes are parallel along the $\mathbb{Z}_2$-invariant two-torus (in our case, the second two-torus $T^2_{(2)}$), or they are parallel on a non-invariant one (in our case, the first and third two-tori). We will discuss each of these situations separately.

\textbf{Two branes are parallel on the first or third two-torus.} The amplitude does not depend on whether the branes are parallel on the first two-torus, i.e. $\pi(0,\phi,-\phi)$, or on the third two-torus, i.e. $\pi(-\phi,\phi,0)$. As usual, we only need to add the factor $\kappa$ to the formulae in \cite{Gmeiner:2009fb} to obtain the correct amplitude:
\begin{equation}\label{Eq:TA-Z2-t3}
\begin{split}
\mathcal{T}_{\mathcal{A}}&:=\kappa c_\mathcal{A}^{\mathbb{Z}_2}\int  \text{d}ll^\varepsilon\Theta_\text{osc}^{{\mathbb{Z}_2},\;(0,\phi,-\phi)}(l)=-2\pi \kappa c_\mathcal{A}^{\mathbb{Z}_2}\int  \text{d}ll^\varepsilon \left(\frac{\vartheta'_1}{\vartheta_1}(\phi,2il)+\frac{\vartheta'_4}{\vartheta_4}(-\phi,2il)\right)\\
&=-2\pi^2\kappa c_\mathcal{A}^{\mathbb{Z}_2}\left(\cot(\pi\phi)\int_0^\infty dl+(\text{sgn}(-\phi)+2\phi)\ln2\right) \, ,
\end{split}
\end{equation}
where again $c_\mathcal{A}^{\mathbb{Z}_2}=\frac{N_b I_{ab}^{\mathbb{Z}_2}}{4\pi^2}$. 

\textbf{Two branes are parallel in the second two-torus,} i.e. $\pi(\phi,0,-\phi)$. The amplitude will again be different depending on which type of three-cycles the D-branes are wrapped on. For wrapped cycles of type 2 or 4 in the classification of section \ref{sec:cycleclassification}, the amplitude is given by:
\begin{equation}\label{Eq:TA-Z2-24}
\begin{split}
\mathcal{T}_{\mathcal{A}}&:=\kappa c_\mathcal{A}^{\mathbb{Z}_2}\int \text{d}ll^\varepsilon\tilde{\mathcal{L}}_{ab}^{\mathcal{A},(2)}\Theta_\text{osc}^{\textbf{1},\;(\phi,0,-\phi)}(l)=4\pi^2\kappa c_\mathcal{A}^{\mathbb{Z}_2}\int  \text{d}ll^\varepsilon\tilde{\mathcal{L}}_{ab}^{\mathcal{A},(2)}\\
&=4\pi^2\kappa c_\mathcal{A}^{\mathbb{Z}_2}\left(\int\text{d}l
+\frac{\delta_{\sigma^2_{ab},0}\delta_{\tau^2_{ab},0}}{V_{ab}^{(2)}}     
\ln\left(\frac{M_s}{\mu}\right)^2-\frac{\Lambda(\sigma^2_{ab},\tau^2_{ab},\mathsf{v}_2,V_{ab}^{(2)})}{V_{ab}^{(2)}} \right) \, ,
\end{split}
\end{equation}
with
\begin{equation}
c_\mathcal{A}^{\mathbb{Z}_2}=-\frac{N_bV_{ab}^{(2)}I_{ab}^{\mathbb{Z}_2,(1\cdot3)}}{8\pi^2}\, ,
\end{equation}
and the function $\Lambda$ defined in equation \eqref{lambdacontribution}.

When the branes wrap cycles of type 1 or 3, the amplitude consists again of two kinds of contributions, and it can be written as:
\begin{equation}\label{Eq:TA-Z2-13}
\begin{split}
\mathcal{T}_{\mathcal{A}}=4\pi^2c_\mathcal{A}^{\mathbb{Z}_2}\bigg(&2\int\text{d}l
+\frac{(\delta_{\sigma^2_{ab},0}+\delta_{1-\sigma^2_{ab},0})\delta_{\tau^2_{ab},0}}{V_{ab}^{(2)}}     
\ln\left(\frac{M_s}{\mu}\right)^2\\&-\frac{\Lambda(\sigma^2_{ab},\tau^2_{ab},\mathsf{v}_2,V_{ab}^{(2)})+\Lambda(1-\sigma^2_{ab},\tau^2_{ab},\mathsf{v}_2,V_{ab}^{(2)})}{V_{ab}^{(2)}} \bigg) \, ,
\end{split}
\end{equation}
with the coefficient $c_\mathcal{A}^{\mathbb{Z}_2}$ and the function $\Lambda$ again as above.
\item \textbf{Two branes are completely parallel }$\pi(0,0,0)$.
Unlike for the \textbf{1} insertion, the annulus amplitude between two completely parallel branes is non-zero when we have a non-trivial $\mathbb{Z}_2$ insertion. As in the case immediately prior to this one, it will depend on the types of cycles wrapped by the D-branes. For cycles of type 2 or 4 we only need to add the factor $\kappa$, and the amplitude is  given by:
\begin{equation}\label{Eq:TA-Z2-parallel24}
\begin{split}
\mathcal{T}_{\mathcal{A}}&:=\kappa c_\mathcal{A}^{\mathbb{Z}_2}\int  \text{d}ll^\varepsilon\tilde{\mathcal{L}}_{ab}^{\mathcal{A},(2)}\Theta_\text{osc}^{\mathbb{Z}_2,\;(0,0,0)}(l)=4\pi^2\kappa c_\mathcal{A}^{\mathbb{Z}_2}\int  \text{d}ll^\varepsilon\tilde{\mathcal{L}}_{ab}^{\mathcal{A},(2)}\\
&=4\pi^2\kappa c_\mathcal{A}^{\mathbb{Z}_2}\left(\int\text{d}l
+\frac{\delta_{\sigma^2_{ab},0}\delta_{\tau^2_{ab},0}}{V_{ab}^{(2)}}     
\ln\left(\frac{M_s}{\mu}\right)^2-\frac{\Lambda(\sigma^2_{ab},\tau^2_{ab},\mathsf{v}_2,V_{ab}^{(2)})}{V_{ab}^{(2)}} \right) \, ,
\end{split}
\end{equation}
with
\begin{equation}
c_\mathcal{A}^{\mathbb{Z}_2}=-\frac{N_b V_{ab}^{(2)}I_{ab}^{\mathbb{Z}_2,(1\cdot3)}}{8\pi^2}\, ,
\end{equation}
and the function $\Lambda$ formally defined in equation \eqref{lambdacontribution}.

For cycles of type 1 and 3 we have once more two different kinds of winding sum contributions to the annulus amplitude, which can be written as:
\begin{equation}\label{Eq:TA-Z2-parallel13}
\begin{split}
\mathcal{T}_{\mathcal{A}}=4\pi^2c_\mathcal{A}^{\mathbb{Z}_2}\bigg(&2\int\text{d}l
+\frac{(\delta_{\sigma^2_{ab},0}+\delta_{\sigma^2_{ab},1})\delta_{\tau^2_{ab},0}}{V_{ab}^{(2)}}     
\ln\left(\frac{M_s}{\mu}\right)^2\\&-\frac{\Lambda(\sigma^2_{ab},\tau^2_{ab},\mathsf{v}_2,V_{ab}^{(2)})+\Lambda(1-\sigma^2_{ab},\tau^2_{ab},\mathsf{v}_2,V_{ab}^{(2)})}{V_{ab}^{(2)}} \bigg) \, ,
\end{split}
\end{equation}
with $c_\mathcal{A}^{\mathbb{Z}_2}$ and the function $\Lambda$ again as above.
\end{itemize}

%%%%%%%%%%%%%%%%%%%%%%%%%%%%%%%%%%%%%%%%%%%%%%%%%%%%%%%%%%%%%%%%%%
\subsection{Contributions to the M\"obius strip amplitudes}\label{MS-insertions}

In this section we present the tree-channel M\"obius strip amplitudes between a stack of D6-branes and the $\Omega\mathcal{R}\mathcal{Q}^{-k}$ orientifold planes.

\begin{itemize}
\item \textbf{Brane and O-plane intersect at three non-trivial angles }$\pi(\phi^1,\phi^2,\phi^3)$. As in the previous subsections, the only difference with respect to the amplitude in the case of the usual factorisable toroidal backgrounds is the multiplication by the factor $\kappa$:
\begin{equation}\label{Eq:TM-3angles}
\begin{split}
\mathcal{T}_{\mathcal{M}}&:=\kappa c_\mathcal{M}\int \text{d}ll^\varepsilon\sum_{(\alpha,\beta)}(-1)^{2(\alpha+\beta)}\frac{\vartheta^{\prime\prime}{\alpha\brack \beta}(0,2il-\frac{1}{2})}{\eta^3(2il-\frac{1}{2})}\prod_{i=1}^3\frac{\vartheta{\alpha\brack \beta}}{\vartheta{1/2\brack 1/2}}(\phi^i,2il)\\
&=-2\pi\kappa  c_\mathcal{M}\int \text{d}ll^\varepsilon\sum_{i=1}^3\frac{\vartheta^\prime_1}{\vartheta_1}(\phi^i,2il-\frac{1}{2})\\
&=-2\pi^2\kappa c_\mathcal{M}\sum_{i=1}^3\bigg(\cot(\pi\phi^{(i)})\int_0^\infty dl+\ln\left(\frac{M_s}{\mu}\right)^2\frac{\text{sgn}(\phi^{(i)})}{4}\left[1+2H\left(|\phi^{(i)}|-\frac{1}{2}\right)\right]\\
&-\frac{\ln2}{4}\text{sgn}(\phi^{(i)})\left[1-2H\left(|\phi^{(i)}|-\frac{1}{2}\right)\right]\\
&-\frac{1}{4}\ln\left(\frac{\Gamma(|\phi^{(i)}|}{\Gamma(1-|\phi^{(i)}|)}\right)^{\text{sgn}(\phi^{(i)})}-\frac{1}{4}\ln\frac{\Gamma(\phi^{(i)}+\frac{1}{2}-\text{sgn}(\phi^{(i)})H(|\phi^{(i)}|-\frac{1}{2}))}{\Gamma(-\phi^{(i)}+\frac{1}{2}+\text{sgn}(\phi^{(i)})H(|\phi^{(i)}|-\frac{1}{2}))}\bigg) \, ,
\end{split}
\end{equation}
with $c_\mathcal{M}=-\frac{\tilde{I}_a^{\Omega\mathcal{R}\mathcal{Q}^{-k}}}{\pi^2}$ and $\tilde{I}_a^{\Omega\mathcal{R}\mathcal{Q}^{-k}}$ the intersection number among the toroidal three-cycle wrapped by the D6-brane $a$ and the $\Omega\mathcal{R}\mathcal{Q}^{-k}$ invariant O6-plane (see section~\ref{ShiftSymm} for the definition of the tilde); $H(x)$ denotes the Heavyside step function:
\begin{equation}
 H(x)=\begin{cases}
       1\quad 0<x\\
       \frac{1}{2}\quad x=0\\
       0\quad x<0.
      \end{cases}
\end{equation}

\item \textbf{The brane and the O-plane intersect along one line}. As in the case of the annulus amplitude with \textbf{1} insertion, we need to distinguish between the brane and the O-plane intersecting in the $B_2$-type torus $T^2_{(3)}$ and the $A_3\times A_1$-type  torus $T^4_{(3)}$.

\textbf{Brane and O-plane are parallel along the third torus}, i.e. $\pi(\phi,-\phi,0)$. Apart from the factor $\kappa$ one does not need to formally modify the amplitude for the usual toroidal background:
\begin{equation}\label{Eq:TM-t3}
\begin{split}
{}\hspace{-5mm}
\mathcal{T}_{\mathcal{M}}&:=\kappa c_\mathcal{M}\int \text{d}ll^\varepsilon\tilde{\mathcal{L}}_{a,\Omega\mathcal{R}\mathcal{Q}^{-k}}^{\mathcal{M},(3)}\Theta_\text{osc}^{\Omega\mathcal{R}\mathcal{Q}^{-k},\;(\phi,-\phi,0)}(2\pi l-\frac{1}{2})=4\pi^2\kappa c_\mathcal{M}\int  \text{d}ll^\varepsilon\tilde{\mathcal{L}}_{a,\Omega\mathcal{R}\mathcal{Q}^{-k}}^{\mathcal{M},(3)}\\
&=4\pi^2 c_{\mathcal{M}}\left(\int\text{d}l+(-1)^{2b_3\sigma^3_a\tau^3_a}\frac{\delta_{\sigma^3_{aa'},0}\delta_{\tau^3_{aa'},0}}{2\tilde{V}_{aa'}^{(3)}} \ln\left(\frac{M_s}{\mu}\right)^2-\frac{\widehat{\Lambda}(\sigma^3_{a},\tau^3_{a},\tilde{\mathsf{v}}_3,2\tilde{V}_{aa'}^{(3)})}{2\tilde{V}_{aa'}^{(3)}} \right) \, 
\end{split}
\end{equation} 
where $c_\mathcal{M}=\frac{\tilde{I}_a^{\Omega\mathcal{R}\mathcal{Q}^{-k}(1\cdot2)}\tilde{V}_{aa'}^{(3)}}{2\pi^2}$. It was noted in \cite{Forste:2010gw,Honecker:2012qr} that, when the brane and the O-plane are parallel on a tilted torus and the brane has both non-trivial Wilson line and discrete displacement, one needs to introduce an extra sign factor $(-1)^{2b_3\sigma^3_a\tau^3_a}$ in the contribution to the beta function coefficient in order to match the counting of massless states from the explicit construction via Chan-Paton labels, but the tadpole contribution should be independent of the value of $\sigma^3_a\tau^3_a$.
 As a consequence, the correct form of the lattice sum $\widehat{\Lambda}$ is to date not known,\footnote{See the recent discussion in section~4.1 of~\cite{Honecker:2017air} for further details. The subtlety can be traced back to the fact that while T-duality along one basic direction of an untilted torus easily matches discrete Wilson lines of D6-branes to positions of D5-branes and preserves a fully factorised form of the Kaluza-Klein and winding sums, T-duality along a tilted torus is not expected to have this simple shape, and dual considerations with a non-trivial background closed string $B$-field and non-trivial $\Z_2$-twisted sectors have to our best knowledge not been performed.
 \label{foot:MS-subtleties}} 
 but we suppose that whenever the sign is trivial, i.e. $(-1)^{2b_3\sigma^3_a\tau^3_a}=1$, the function $\widehat{\Lambda}(\sigma^3_a,\tau^3_a, \ldots)$ should be the same as the function $\Lambda (\sigma^3_{aa'},\tau^3_{aa'}, \ldots)$ defined in \eqref{lambdacontribution}.

\item \textbf{Brane and O-plane are parallel along part of the four-torus}. The amplitude will again depend on the two-torus where the brane and the O-plane are parallel and the types of three-cycles they wrap. As before, we will consider here the case of the brane and the O-plane being parallel on the second two-torus, i.e. $\pi(\phi,0,-\phi)$, and briefly comment on the case of them being parallel on the first two-torus, i.e. $\pi(0,\phi,-\phi)$, at the end.

When the brane and the O-plane wrap cycles of type 2 or 4 the M\"obius strip amplitude is given by:
\begin{equation}\label{Eq:TM-24}
\begin{split}
\mathcal{T}_{\mathcal{M}}&:=\kappa c_\mathcal{M}\int \text{d}ll^\varepsilon\tilde{\mathcal{L}}_{a,\Omega\mathcal{R}\mathcal{Q}^{-k}}^{\mathcal{M},(2)}\Theta_\text{osc}^{\Omega\mathcal{R}\mathcal{Q}^{-k},\;(\phi,0,-\phi)}(2\pi l-\frac{1}{2})=4\pi^2\kappa c_\mathcal{M}\int  \text{d}ll^\varepsilon\tilde{\mathcal{L}}_{a,\Omega\mathcal{R}\mathcal{Q}^{-k}}^{\mathcal{M},(2)}\\
&=4\pi^2 \kappa c_{\mathcal{M}}\left(\int\text{d}l+(-1)^{2b_2\sigma^2_a\tau^2_a}\frac{\delta_{\sigma^2_{aa'},0}\delta_{\tau^2_{aa'},0}}{2\tilde{V}_{aa'}^{(2)}} \ln\left(\frac{M_s}{\mu}\right)^2-\frac{\widehat{\Lambda}(\sigma^2_{a},\tau^2_{a},\tilde{\mathsf{v}}_2,2\tilde{V}_{aa'}^{(2)})}{2\tilde{V}_{aa'}^{(2)}} \right) \, ,
\end{split}
\end{equation}
where $c_\mathcal{M}=\frac{\tilde{I}_a^{\Omega\mathcal{R}\mathcal{Q}^{-k}(1\cdot3)}\tilde{V}_{aa'}^{(2)}}{2\pi^2}$, and  for the function $\widehat{\Lambda}$ see the comment above. 

When the brane and the O-plane wrap cycles of type 1 or 3, we have once again an amplitude composed of two different kinds of contributions. We can write this M\"obius strip amplitude as:
\begin{equation}\label{Eq:TM-13}
\begin{split}
\mathcal{T}_{\mathcal{M}}=&4\pi^2 c_{\mathcal{M}}\left(2\int\text{d}l+\frac{(-1)^{2b_2\sigma^2_a\tau^2_a}\delta_{\sigma^2_{aa'},0}\delta_{\tau^2_{aa'},0}}{2\tilde{V}_{aa'}^{(2)}} \ln\bigg(\frac{M_s}{\mu}\right)^2\\
&-\frac{\widehat{\Lambda}(\sigma^2_{a},\tau^2_{a},\tilde{\mathsf{v}}_2,2\tilde{V}_{aa'}^{(2)})+\widehat{\Lambda}(1-\sigma^2_{a},\tau^2_{a},\tilde{\mathsf{v}}_2,2\tilde{V}_{aa'}^{(2)})}{2\tilde{V}_{aa'}^{(2)}} \bigg).
\end{split}
\end{equation}
When the branes are parallel along the first two-torus, the formulae are completely analogous (with the obvious exchange of superscripts $1\leftrightarrow2$ in the relevant quantities), being valid for wrapped cycles of types 3 or 4, and types 1 or 2, respectively.

\item \textbf{Brane and O-plane are completely parallel}

If the D6-brane and the O-plane are completely parallel, there is no contribution from the amplitude to the tadpole, the beta function or the gauge threshold corrections:
\begin{equation}\label{Eq:TM-parallel}
\mathcal{T}_{\mathcal{M}}=0.
\end{equation}
\end{itemize}

%%%%%%%%%%%%%%%%%%%%%%%%%%%%%%%%%%%%%%%%%
\subsection{Summary of contributions to the beta function $b_{SU(N_a)}$}\label{Summary-b-SUN}

In order to distinguish  the gauge group enhancement $SU(N_a)\rightarrow USp(2N_a)$ or 
\linebreak
\mbox{$SU(N_a)\rightarrow SO(2N_a)$} and to compute the non-chiral spectrum of massless open strings, as anticipated at the beginning of section~\ref{amplitudes} we extract the corresponding beta function coefficients from the amplitudes 
computed in sections~\ref{Annulus-1-insertion}, \ref{Annulus-Z2-insertion} 
%
%{\color{red}\bf $\leftarrow$ introduce label at the start of that subsection}
%
and~\ref{MS-insertions}. The results are for convenience summarised in table \ref{tab:contributions-to-beta-function}.
\renewcommand{\arraystretch}{1.8}
\begin{table}[h!]
\hspace{-.7cm}
\resizebox{17.5cm}{!}{
\begin{tabular}{|c|c|c|c|}
\hline
\multicolumn{4}{|c|}{\textbf{Contributions to  $b_{SU(N_a)}$ from annulus and M\"obius strip amplitudes}}\\
\hline
\hline
$\pi(\phi^1_{ab},\phi^2_{ab},\phi^3_{ab})$&types& Annulus & M\"obius strip\\
\hline
$\begin{array}{c}
\vspace{-0.4cm}(\phi^1,\phi^2,\phi^3=-\sum_{i=1}^2\phi^i)\\
0\,<\,\mid\phi^i\mid,\mid\phi^j\mid\,\leq\,\mid\phi^k\mid\,<\,1\vspace{-.4cm}\\
\sgn(\phi^i)=\sgn(\phi^j)\neq \sgn(\phi^m)
\end{array}$&all& \hspace{1cm}$-\frac{\kappa N_b (I_{ab}+I^{\Z_2}_{ab})}{2}\frac{\sum_{i=1}^3\sgn(\phi_{ab}^i)}{2}$ \hspace{1.1cm} \eqref{Eq:TA-1-3angles},\eqref{Eq:TA-Z2-3angles} & 
$-\frac{\kappa \tilde{I}^{\Omega\mathcal{R}\mathcal{\mathcal{Q}}^{-k}}_a\sgn(\phi^m)}{2}\left[1+2H(\mid\phi^m\mid-\frac{1}{2})\right]$ \eqref{Eq:TM-3angles}\\
\hline
$\pi(\phi,-\phi,0)$&all & \hspace{1cm}$-\frac{\kappa N_b I_{ab}^{(1\cdot 2)}}{2}\delta_{\sigma^3_{ab,0}}\delta_{\tau^3_{ab,0}}$ \hspace{1.9cm} \eqref{amplitude1},\eqref{Eq:TA-Z2-t3}& 
\hspace{.2cm}$\kappa \delta_{\sigma^3_{aa^\prime,0}}\delta_{\tau^3_{aa^\prime,0}} (-1)^{2b_3\sigma_a^3\tau_a^3} \tilde{I}_{a}^{\Omega\mathcal{R}\mathcal{Q}^{-k,(1\cdot 2)}}$ \eqref{Eq:TM-t3}\\
\hline
$\pi(\phi,0,-\phi)$&$\begin{array}{c}
2,\,4\\1,\,3
\end{array}$&
$\begin{array}{cr}
-\frac{\kappa N_b (I_{ab}^{(1\cdot 3)}+I_{ab}^{\Z_a,(1\cdot 3)})}{2}\delta_{\sigma^2_{ab,0}}\delta_{\tau^2_{ab,0}}&\eqref{Eq:TA-1-Type24},\eqref{Eq:TA-Z2-24}\\
-\frac{N_b (I_{ab}^{(1\cdot 3 }+I_{ab}^{\Z_a,(1\cdot 3)})}{2}(\delta_{\sigma^2_{ab,0}}+\delta_{\sigma^2_{ab,1}})\delta_{\tau^2_{ab,0}}&
\eqref{Eq:TA-1-13},\eqref{Eq:TA-Z2-13}
\end{array}$ &
$\begin{array}{cr}
\kappa \delta_{\sigma^2_{aa^\prime,0}}\delta_{\tau^2_{aa^\prime,0}} (-1)^{2b_2\sigma_a^2\tau_a^2} \tilde{I}_{a}^{\Omega\mathcal{R}\mathcal{Q}^{-k,(1\cdot 3)}}&\eqref{Eq:TM-24}\\
\delta_{\sigma^2_{aa^\prime,0}}\delta_{\tau^2_{aa^\prime,0}} (-1)^{2b_2\sigma_a^2\tau_a^2}\tilde{I}_{a}^{\Omega\mathcal{R}\mathcal{Q}^{-k,(1\cdot 3)}}&\eqref{Eq:TM-13}
\end{array}$ 
 \\
\hline
$\pi(0,\phi,-\phi)$ & $\begin{array}{c}3,\,4\\1,\,2\end{array}$&
$\begin{array}{cr}
-\frac{\kappa N_b I_{ab}^{(2\cdot 3)}}{2}\delta_{\sigma^1_{ab,0}}\delta_{\tau^1_{ab,0}}&\eqref{Eq:TA-1-Type24},\eqref{Eq:TA-Z2-t3}\\
\hspace{0.7cm} -\frac{N_b I_{ab}^{(2\cdot 3)}}{2}(\delta_{\sigma^1_{ab,0}}+\delta_{\sigma^1_{ab,1}})\delta_{\tau^1_{ab,0}}& \hspace{0.7cm}\eqref{Eq:TA-1-13},\eqref{Eq:TA-Z2-t3}
\end{array}$& 
$\begin{array}{cr}
\kappa \delta_{\sigma^1_{aa^\prime,0}}\delta_{\tau^1_{aa^\prime,0}} (-1)^{2b_1\sigma_a^1\tau_a^1} \tilde{I}_{a}^{\Omega\mathcal{R}\mathcal{Q}^{-k,(2\cdot 3)}}& \eqref{Eq:TM-24}\\
\delta_{\sigma^1_{aa^\prime,0}} \delta_{\tau^1_{aa^\prime,0}} (-1)^{2b_1\sigma_a^1\tau_a^1} \tilde{I}_{a}^{\Omega\mathcal{R}\mathcal{Q}^{-k,(2\cdot 3)}}& \eqref{Eq:TM-13}
\end{array}$
\\
\hline
$\pi(0,0,0)$ & $\begin{array}{c}1,\,3\\4,\,4\\\text{else}\end{array}$&
$\begin{array}{cr}
\hspace{0.4cm}-\frac{N_b I_{ab}^{\Z_a,(1\cdot 3)}}{2}(\delta_{\sigma^2_{ab,0}}+\delta_{\sigma^2_{ab,1}})\delta_{\tau^2_{ab,0}}& \hspace{0.5cm}\eqref{Eq:TA-1-parallel},\eqref{Eq:TA-Z2-parallel13}\\
-\frac{N_b I_{ab}^{\Z_a,(1\cdot 3)}}{2}\delta_{\sigma^2_{ab,0}}\delta_{\tau^2_{ab,0}}&\hspace{-0.4cm} \text{see section \ref{USP_or_SO}}  \\
-\frac{\kappa N_b I_{ab}^{\Z_a,(1\cdot 3)}}{2}\delta_{\sigma^2_{ab,0}}\delta_{\tau^2_{ab,0}}& \eqref{Eq:TA-1-parallel},\eqref{Eq:TA-Z2-parallel24}
\end{array}$& \hspace{2.9cm} $-$\hspace{2.8cm}  \eqref{Eq:TM-parallel} \\
\hline
\end{tabular}
}
\caption{The annulus and M\"obius strip contributions to the beta function coefficients. In the first column, we list the relative angles among pairs of three-cycles and in the second the types of three-cycles according to the classification in section~\protect\ref{ShiftSymm}. 
The third  and last column comprise the contributions to  the beta function coefficients from the annulus and M\"obius strip amplitudes, respectively,
with the number in parenthesis referencing to the equation(s) containing the amplitude(s) $\mathcal{T}$ from which the corresponding expression is extracted. 
The intersection numbers $\tilde{I}^{\Omega\mathcal{RQ}^{-k},(i)}_a$ among the brane $a$ and the $\Omega\mathcal{RQ}^{-k}$ invariant O-plane are weighted with the number $2(1-b_i)$ per two torus (see section \ref{ShiftSymm}).
}\label{tab:contributions-to-beta-function}
\end{table}

%%%%%%%%%%%%%%%%%%%%%%%%%%%%%%%%%%%%%%%%%
\subsection{Identification of enhanced gauge groups}\label{USP_or_SO}

Whenever a stack of $N$ branes and its orientifold image coincide, the $U(N)$ gauge group is enhanced to either $SO(2N)$ or $USp(2N)$. The way to discern the correct enhanced gauge group is to compute the beta function coefficients via one-loop open string amplitudes, in particular the contributions from the M\"obius strip amplitudes in the section~\ref{MS-insertions}, 
and compare the results with the field theoretically derived beta function coefficients for $USp(2N)$ and $SO(2N)$ in equations \eqref{usp_theory} and \eqref{so_theory}, respectively.

For each of the eight lattice orientations in section \ref{orientifolding}, there are three toroidal three-cycles giving rise to orientifold invariant fractional three-cycles, which are presented in tables \ref{tab:r1}-\ref{tab:r4}, for the involutions $\mathcal{R}_{1}$, $\mathcal{R}_{2}$, $\mathcal{R}_{3}$ and $\mathcal{R}_{4}$, respectively, for $u_1=0$, and tables \ref{tab:r105}-\ref{tab:r405} for $u_1=1/2$. Notice that brane $b$ is always parallel to one of the O-plane orbits, whereas no  orientifold invariant fractional three-cycle exists along the other O-plane orbit. 
The branes $a$ and $c$ are for each choice of involution $\mathcal{R}_i$ at relative angles $(\pm\frac{\pi}{2},\mp\frac{\pi}{2},0)$ or $(0,\pm\frac{\pi}{2},\mp\frac{\pi}{2})$ with respect to one of the O-planes, i.e. they are in particular perpendicular to the O-plane along the two-torus $T_2^{(2)}$ where $\Z_2$ acts trivially. These configurations are not pointwise, but in homology, invariant under the orientifold involution and give rise to gauge group enhancement to $USp(2M_{a,c})$ with one chiral multiplet in the antisymmetric representation stemming from the $aa$ or $cc$ sector,  in complete analogy to the factorisable $T^6/(\Z_{2N} \times \OR)$ orientifolds~\cite{Gmeiner:2009fb,Honecker:2011sm}.
The names of the stacks ($a,b,c$) have been chosen such that, for two topologically conjectured to be dual backgrounds, cycles which map to each other under the transformations specified in section~\ref{duality} have the same name.

For simplicity, the intersection numbers in the tables already include the correction factor $\kappa$ and the number of O6-planes. The minus sign that appears when the brane and the O-plane are antiparallel should also be taken into account when computing the beta function coefficient. The global factor $\frac{1}{2}$ in the contribution to the beta function coefficient arises from the fact that in the formulae of sections~\ref{Annulus-1-insertion}-\ref{Summary-b-SUN}, 
we  implicitly sum over D6-branes and their orientifold images (i.e. the expression for the amplitude in the e.g. $xx$ sector also includes the contribution from the $x'x'$ sector), but for orientifold invariant branes no such sum appears.

For the conjectured duality among lattice orientations to hold, we expect to obtain the same massless open string spectrum for related branes in dual backgrounds. 
Looking at the tables one can see that there are many cases where the duality holds naively, but for it to be fully preserved minor modifications are required, in the form of factors of two (shown in red in the tables) in the $\mathbb{Z}_2$ invariant intersection numbers in the $xx$ sector when the cycle wrapped by the D-brane is of type 4 in the classification of section \ref{sec:cycleclassification}, i.e. only if the cycle does not possess an image under the shift symmetry in the factorised description. 
This extra factor of two is not required only for the validity of the conjectured duality. In fact, one can check that if the factor of two were not present, instead of getting either a multiplet in the symmetric (=adjoint for $USp$) or a multiplet in the antisymmetric representation of the enhanced gauge group (depending on whether the brane and the O-plane are parallel or orthogonal along the $\mathbb{Z}_2$-invariant two-torus, respectively) one obtains a non-integer number of multiplets in each representation. To illustrate this better, let us consider brane $b$ in table \ref{tab:r1}. With the extra factor of two, the contribution to the beta function coefficient from the $bb$ sector is consistent with the vector and one chiral multiplet in the symmetric representation, i.e., an $USp(2M_b)$ gauge group and one chiral multiplet in the adjoint representation encoding the (continuous) Wilson line and displacement moduli along the $\Z_2$ invariant two-torus. If the factor of two were not present, the contribution to the beta function coefficient from the $bb$ sector would correspond to an $USp(2M_b)$ gauge group   with $\frac{3}{2}$ multiplets in the symmetric representation and $\frac{1}{2}$ multiplets in the antisymmetric representation, which makes no sense.

After taking into account this minor modification for branes of type 4, which can be traced back to the fact that dividing out images under the shift symmetry is subtle,
comparing the results for the contributions to the beta function coefficients for each orientifold invariant fractional three-cycle with the field theoretical expressions in equations \eqref{usp_theory} and \eqref{so_theory}, we conclude that in all cases the $U(N)$ group is enhanced to an $USp(2N)$ group.
It is noteworthy that Euclidean D2-branes wrapping the same three-cycles provide $O(1)$ instantons which might lead to non-vanishing non-perturbtive corrections to the superpotential.

To finish this section, we will briefly comment on the choices of $(b_1,b_2)$ we made at the end of section \ref{ShiftSymm}. Similarly to the factor of two above, taking the choice $(1/2,0)$ versus $(0,1/2)$
leads to non-integer numbers of multiplets in symmetric and antisymmetric representations. As an example let us consider the $a(\mathcal{Q}a)$ sector of brane $a$ in table \ref{tab:r1}. Using $(b_1,b_2)=(0,1/2)$ as mentioned in section \ref{ShiftSymm}, we obtain a contribution to the beta function coefficient of $2M_a-2$, which is consistent with two multiplets in the antisymmetric representation. However, if one uses $(b_1,b_2)=(1/2,0)$, applying the formulae in sections \ref{Annulus-1-insertion}-\ref{MS-insertions} to the contribution to the beta function coefficient gives $2M_a-1$, which would correspond to the unphysical solution of $n_A=\frac{3}{2}$ multiplets in the antisymmetric and $n_S=\frac{1}{2}$ multiplets in the symmetric representations of the enhanced gauge group.

\renewcommand{\arraystretch}{1.5}
\begin{table}[h!]
\hspace{-1cm}
\resizebox{18cm}{!}{
\normalsize{
\begin{tabular}{|c|c|c|c|c|c|c|c|c|c|c|}
\hline
\multicolumn{11}{|c|}{\textbf{Configurations of toroidal three-cycles supporting $USP(2N)$ gauge group on orientifold with the lattice type \textbf{A$_\textbf{a}$AB}}}\\
\hline \hline
brane &wrap. num & $(\tau^1,\tau^3)$ & $(\sigma^1,\sigma^3)$ & $I^{\Z_2}_{xx}$ &$I_{x(\mathcal{Q}x)}$ & $I^{\Z_2}_{x(\mathcal{Q}x)}$& $\tilde{I}^{\mathcal{R}}_{x}$ & $\tilde{I}^{\mathcal{R}\mathcal{Q}^{-3}}_{x}$& $\tilde{I}^{\mathcal{R}\mathcal{Q}^{-2}}_{x}$& $\tilde{I}^{\mathcal{R}\mathcal{Q}^{-1}}_{x}$\\ 
\hline 
$a$ & (1,0)(-1,0)(0,1) & $(\tau^1,\tau^3)$ & $(-;\sigma^3)$ & (2,0,2)&(2,-0,1)&(2,-0,1)&(-2,-0,-1)&(1,-4,0)&(2,0,-1)
&(0,-2,2)\\ 
\hline 
\multicolumn{11}{|c|}{$b_a\supset\frac{1}{2}\left[\left(-M_aI^{\Z_2,(1\cdot 3)}_{aa}+\tilde{I}_{a}^{\mathcal{R}\mathcal{Q}^{-3}(1\cdot 2)}+\tilde{I}_{a}^{\mathcal{R}\mathcal{Q}^{-1},(2\cdot 3)}\right)+\left(-M_a(I^{(1\cdot 3)}_{a(\mathcal{Q}a)}+I^{\Z_2,(1\cdot 3)}_{a(\mathcal{Q}a)})+\tilde{I}_{a}^{\mathcal{R}(1\cdot 3)}+\tilde{I}_{a}^{\mathcal{R}\mathcal{Q}^{-2},(1\cdot 3)}
\right)\right]=-2M_a-4+2M_a-2$}\\
\hline\hline
$b$ & (1,1)(-1,0)(1,2)& $(-,\tau^3)$ & $(\sigma^1,\sigma^3)$ with $\tau^3\sigma^3=0$ &  (\textcolor{red}{2}$\cdot$ 1,0,2)&(1,-0,2)&(1,-0,2)&(-2,-0,-2)&(-1,2,2)&(0,0,0)&(-1,-2,2)\\ 
\hline 
& & $(-,\tau^3)$ & $(\sigma^1,\sigma^3)$ with $\tau^3\sigma^3=1$  & (\textcolor{red}{2}$\cdot$ 1,0,2)&(1,-0,2)&(1,-0,-2)&(-2,-0,-2)&(-1,2,2)&(0,0,0)&(-1,-2,2)\\  
\hline
\multicolumn{7}{|r}{$b_b\supset\frac{1}{2}\left[\left(-M_bI^{\Z_2,(1\cdot 3)}_{bb}+\tilde{I}_{b}^{\mathcal{R}(1\cdot 3)}\right)+\left(-M_b(I^{(1\cdot 3)}_{b(\mathcal{Q}b)}+I^{\Z_2,(1\cdot 3)}_{b(\mathcal{Q}b)})
\right)\right]=$}&\multicolumn{4}{l|}{$-\textcolor{red}{2}M_b-2+2M_b$ \,for $\tau^3\sigma^3=0$}\\
\multicolumn{7}{|r}{}&\multicolumn{4}{l|}{$-\textcolor{red}{2}M_b-2$ \qquad\quad for $\tau^3\sigma^3=1$}\\
\hline\hline
$c$ & (1,-1)(0,-1)(1,2) & $(-,\tau^3)$ & $(\sigma^1,\sigma^3)$ with $\tau^3 \sigma^3=0$ &(2,0,2)&(2,-0,2)&(2,-0,2)&(0,2,-2)&(-2,0,2)&(1,-4,0)&(2,-0,2)\\ 
\hline 
\multicolumn{11}{|c|}{$b_c\supset\frac{1}{2}\left[\left(-M_cI^{\Z_2,(1\cdot 3)}_{cc}+\tilde{I}_{c}^{\mathcal{R}(2\cdot 3)}+\tilde{I}_{c}^{\mathcal{R}\mathcal{Q}^{-2},(1\cdot 2)}\right)+\left(-M_c(I^{(1\cdot 3)}_{c(\mathcal{Q}c)}+I^{\Z_2,(1\cdot 3)}_{c(\mathcal{Q}c)})+\tilde{I}_{c}^{\mathcal{R}\mathcal{Q}^{-3}(1\cdot 3)}+\tilde{I}_{c}^{\mathcal{R}\mathcal{Q}^{-1},(1\cdot 3)}
\right)\right]=-2M_c-4+4M_c-4$}\\
\hline\hline
\multicolumn{11}{|c|}{O-planes}\\
\hline
\multicolumn{11}{|c|}{\quad$\mathcal{R}$:\;\;$(1,-1)(1,0)(1,0)$\qquad\vline\quad $\mathcal{R}\mathcal{Q}^{-3}$:\;\; $(0,-1)(0,-1)(0,1)$\qquad\vline\quad $\mathcal{R}\mathcal{Q}^{-2}$:\;\;$(1,1)(-1,0)(1,2)$\qquad\vline\quad $\mathcal{R}\mathcal{Q}^{-1}$:\;\;$(1,0)(0,1)(-1,-1)$ }\\
\hline
\end{tabular}
}
}
\caption{Configurations of  toroidal three-cycles giving rise to the $\mathcal{R}_1$-invariant fractional three-cycles, their  intersection numbers per two-torus $T^2_{(i)}$ and the contribution to the  beta function coefficient for $u_1=0$. $-0_i$ denotes branes being anti-parallel on the two-torus $T^2_{(i)}$. The necessary extra factors of two for branes of type 4 (i.e. without image under the shift symmetry)  mentioned in the main text are marked in red. For the sake of comparison, in the last line we list  the wrapping numbers of the O-planes. 
}\label{tab:r1}
\end{table}
\nopagebreak
\vspace{-3cm}
\renewcommand{\arraystretch}{1.5}
\begin{table}[h!]
\hspace{-1cm}
\resizebox{18cm}{!}{
\normalsize{
\begin{tabular}{|c|c|c|c|c|c|c|c|c|c|c|}
\hline
\multicolumn{11}{|c|}{\textbf{Configurations of toroidal three-cycles supporting $USP(2N)$ gauge group on orientifold with the lattice type \textbf{A$_\textbf{a}$AA}}}\\
\hline \hline
brane &wrap. num & $(\tau^1,\tau^3)$ & $(\sigma^1;\sigma^3)$ &  $I^{\Z_2}_{xx}$ &$I_{x(\mathcal{Q}x)}$ & $I^{\Z_2}_{x(\mathcal{Q}x)}$& $\tilde{I}^{\mathcal{R}}_{x}$ & $\tilde{I}^{\mathcal{R}\mathcal{Q}^{-3}}_{x}$& $\tilde{I}^{\mathcal{R}\mathcal{Q}^{-2}}_{x}$& $\tilde{I}^{\mathcal{R}\mathcal{Q}^{-1}}_{x}$\\ 
\hline 
\hline 
$a$ & (1,1)(0,-1)(0,1) & $(-,\tau^3)$ & $(\sigma^1,\sigma^3)$ &(2,0,2)&(2,-0,1)&(2,-0,1)&(1,-4,0)&(2,0,-1)&(0,-2,2)&(2,-0,1) \\ 
\hline 
\multicolumn{11}{|c|}{$b_a\supset \frac{1}{2}\left[\left(-M_aI^{\Z_2,(1\cdot 3)}_{aa}+\tilde{I}_{a}^{\mathcal{R}(1\cdot 2)}+\tilde{I}_{a}^{\mathcal{R}\mathcal{Q}^{-2},(2\cdot 3)}\right)+\left(-M_a(I^{(1\cdot 3)}_{a(\mathcal{Q}a)}+I^{\Z_2,(1\cdot 3)}_{a(\mathcal{Q}a)})+\tilde{I}_{a}^{\mathcal{R}\mathcal{Q}^{-3}(1\cdot 3)}+\tilde{I}_{a}^{\mathcal{R}\mathcal{Q}^{-1},(1\cdot 3)}
\right)\right]=-2M_a-4+2M_a-2$}\\
\hline\hline
$b$ & (0,1)(0,-1)(1,2)& $(\tau^1,\tau^3)$ & $(-,\sigma^3)$ with $\tau^3\sigma^3=0$ & (2,0,2)&(1,-0,2)&(1,-0,2)&(1,-2,2)&(0,0,0)&(1,2,2)&(2,-0,2) \\ 
\hline 
& & $(\tau^1,\tau^3)$ & $(-,\sigma^3)$ with $\tau^3\sigma^3=1$ &(2,0,2)&(1,-0,2)&(1,-0,-2)&(1,-2,2)&(0,0,0)&(1,2,2)&(2,-0,2) \\
\hline
\multicolumn{7}{|r}{$b_b\supset\frac{1}{2}\left[\left(-M_bI^{\Z_2,(1\cdot 3)}_{bb}+\tilde{I}_{b}^{\mathcal{R}\mathcal{Q}^{-1}(1\cdot 3)}\right)+\left(-M_b(I^{(1\cdot 3)}_{b(\mathcal{Q}b)}+I^{\Z_2,(1\cdot 3)}_{b(\mathcal{Q}b)}
\right)\right]=$}&\multicolumn{4}{l|}{$-2M_b-2+2M_b$ for $\tau^3\sigma^3=0$}\\
\multicolumn{7}{|r}{}&\multicolumn{4}{l|}{$-2M_b-2$ \hspace{1.05cm} for $\tau^3\sigma^3=1$}\\
\hline\hline
$c$ & (1,0)(1,0)(1,2) & $(\tau^1,\tau^3)$ & $(-,\sigma^3)$ with $\tau^3 \sigma^3=0$
&(2,0,2)&(2,-0,2)&(2,-0,2)&(-2,0,2)&(1,-4,0)&(2,-0,2)&(-0,2,2) \\ 
\hline 
\multicolumn{11}{|c|}{$b^c\supset\frac{1}{2}\left[\left(-M_cI^{\Z_2,(1\cdot 3)}_{cc}+\tilde{I}_{c}^{\mathcal{R}\mathcal{Q}^{-3}(1\cdot 2)}+\tilde{I}_{c}^{\mathcal{R}\mathcal{Q}^{-1},(2\cdot 3)}\right)+\left(-M_c(I^{(1\cdot 3)}_{c(\mathcal{Q}c)}+I^{\Z_2,(1\cdot 3)}_{c(\mathcal{Q}c)})+\tilde{I}_{c}^{\mathcal{R}(1\cdot 3)}+\tilde{I}_{c}^{\mathcal{R}\mathcal{Q}^{-2},(1\cdot 3)}
\right)\right]=-2M_c-4+4M_c-4 $}\\
\hline\hline
\multicolumn{11}{|c|}{O-planes}\\
\hline
\multicolumn{11}{|c|}{\quad$\mathcal{R}$:\;\;$(1,-1)(1,0)(0,1)$\qquad\vline\quad $\mathcal{R}\mathcal{Q}^{-3}$:\;\; $(0,1)(0,-1)(1,2)$\qquad\vline\quad $\mathcal{R}\mathcal{Q}^{-2}$:\;\;$(1,1)(-1,0)(-1,-1)$\qquad\vline\quad $\mathcal{R}\mathcal{Q}^{-1}$:\;\;$(-1,0)(0,1)(-1,0)$ }\\
\hline
\end{tabular} }
}
\caption{Configurations of toroidal three-cycles giving rise to the $\mathcal{R}_2$-invariant fractional three-cycles, their  intersection numbers per two-torus $T^2_{(i)}$ and  the contribution to the beta function coefficient for $u_1=0$. For the remaining notation see the caption to table \ref{tab:r1}.}\label{tab:r2}
\end{table}

\newpage

%%%%%%%%%%%%%%%%%%%%%%%%%%%%%%%%%%%%%%%%%%%%%%%%%%%%%%%%%%%%%%%%%%%%%
\renewcommand{\arraystretch}{1.5}
\begin{table}[h!]
\hspace{-1cm}
\resizebox{18cm}{!}{
\normalsize{
\begin{tabular}{|c|c|c|c|c|c|c|c|c|c|c|}
\hline
\multicolumn{11}{|c|}{\textbf{Configurations of toroidal three-cycles supporting $USP(2N)$ gauge group on orientifold with the lattice type \textbf{A$_\textbf{a}$BA}}}\\
\hline \hline
brane &wrap. num & $(\tau^1,\tau^3)$ & $(\sigma^1,\sigma^3)$ & $I^{\Z_2}_{xx}$ &$I_{x(\mathcal{Q}x)}$ & $I^{\Z_2}_{x(\mathcal{Q}x)}$& $\tilde{I}^{\mathcal{R}}_{x}$ & $\tilde{I}^{\mathcal{R}\mathcal{Q}^{-3}}_{x}$& $\tilde{I}^{\mathcal{R}\mathcal{Q}^{-2}}_{x}$& $\tilde{I}^{\mathcal{R}\mathcal{Q}^{-1}}_{x}$\\ 
\hline 
\hline 
$a$ & (1,0)(0,-1)(0,1) & $(\tau^1,\tau^3)$ & $(-,\sigma^3)$ 
&(2,0,2)&(1,-0,1)&(1,-0,1)&(1,-4,0)&(1,0,-1)&(0,-2,2)&(1,-0,1) \\ 
\hline 
\multicolumn{11}{|c|}{$b_a\supset\frac{1}{2}\left[\left(-M_aI^{\Z_2,(1\cdot 3)}_{aa}+\tilde{I}_{a}^{\mathcal{R}(1\cdot 2)}+\tilde{I}_{a}^{\mathcal{R}\mathcal{Q}^{-2},(2\cdot 3)}\right)+\left(-M_a(I^{(1\cdot 3)}_{a(\mathcal{Q}a)}+I^{\Z_2,(1\cdot 3)}_{a(\mathcal{Q}a)})+\tilde{I}_{a}^{\mathcal{R}\mathcal{Q}^{-3}(1\cdot 3)}+\tilde{I}_{a}^{\mathcal{R}\mathcal{Q}^{-1},(1\cdot 3)}
\right)\right]=-2M_a-4+M_a-1$}\\
\hline\hline
$b$ & (1,1)(0,-1)(1,2)& $(-,\tau_3)$ & $(\sigma^1,\sigma^3)$ with $\tau^3\sigma^3=0$ &(2,0,2)&(2,-0,2)&(2,-0,2)&(2,-2,2)&(0,0,0)&(2,2,2)&(2,-0,2) \\ 
\hline 
& & $(-,\tau_3)$ & $(\sigma^1,\sigma^3)$ with $\tau^3\sigma^3=1$ &(2,0,2)&(2,-0,2)&(2,-0,-2)&(2,-2,2)&(0,0,0)&(2,2,2)&(2,-0,2) \\
\hline
\multicolumn{7}{|r}{$b_b\supset \frac{1}{2}\left[\left(-M_bI^{\Z_2,(1\cdot 3)}_{bb}+\tilde{I}_{b}^{\mathcal{R}\mathcal{Q}^{-1}(1\cdot 3)}\right)+\left(-M_b(I^{(1\cdot 3)}_{b(\mathcal{Q}b)}+I^{\Z_2,(1\cdot 3)}_{b(\mathcal{Q}b)})
\right)\right]=$}&\multicolumn{4}{l|}{$-2M_b-2+4M_b$ for $\tau^3\sigma^3=0$}\\
\multicolumn{7}{|r}{}&\multicolumn{4}{l|}{$-2M_b-2$ \hspace{1.05cm} for $\tau^3\sigma^3=1$}\\
\hline\hline
$c$ & (1,-1)(1,0)(1,2) & $(\tau^1,\tau^3)$ & $(\sigma^1,\sigma^3)$ with $\tau^3 \sigma^3=0$  &(\textcolor{red}{2}$\cdot$ 1,0,2)&(1,-0,2)&(1,-0,2)&(-1,0,2)&(1,-4,0)&(1,-0,2)&(-0,2,2)\\ 
\hline 
\multicolumn{11}{|c|}{$b_c\supset\frac{1}{2}\left[\left(-M_cI^{\Z_2,(1\cdot 3)}_{cc}+\tilde{I}_{c}^{\mathcal{R}\mathcal{Q}^{-3}(1\cdot 2)}+\tilde{I}_{c}^{\mathcal{R}\mathcal{Q}^{-1},(2\cdot 3)}\right)+\left(-M_c(I^{(1\cdot 3)}_{c(\mathcal{Q}c)}+I^{\Z_2,(1\cdot 3)}_{c(\mathcal{Q}c)})+\tilde{I}_{c}^{\mathcal{R}(1\cdot 3)}+\tilde{I}_{c}^{\mathcal{R}\mathcal{Q}^{-2},(1\cdot 3)}
\right)\right]=-\textcolor{red}{2}M_c-4+2M_c-2$}\\
\hline\hline
\multicolumn{11}{|c|}{O-planes}\\
\hline
\multicolumn{11}{|c|}{\quad$\mathcal{R}$:\;\;$(0,-1)(1,0)(0,1)$\qquad\vline\quad $\mathcal{R}\mathcal{Q}^{-3}$:\;\; $(1,1)(0,-1)(1,2)$\qquad\vline\quad $\mathcal{R}\mathcal{Q}^{-2}$:\;\;$(1,0)(-1,0)(-1,-1)$\qquad\vline\quad $\mathcal{R}\mathcal{Q}^{-1}$:\;\;$(-1,1)(0,1)(-1,0)$}\\
\hline
\end{tabular} }
}
\caption{Configurations of toroidal three-cycles giving rise to the $\mathcal{R}_3$-invariant fractional three-cycles, their  intersection numbers per two-torus $T^2_{(i)}$ and  the contribution to the beta function coefficient for $u_1=0$. For the remaining notation see the caption to table \ref{tab:r1}.}\label{tab:r3}
\end{table}

%%%%%%%%%%%%%%%%%%%%%%%%%%%%%%%%%%%%%%%%%%%%%%%%%%%%%%%%%%%%%%%%%%%%%%%%%%%%%%%%%%%%%%%%%%%%%%%%%%%%%%%%%%%%%%%%%%%%%%%%%%%%%%%%%%%%%%%%%%%%%%%%%%%%%%%%%%%%%%%%%%%%%%%%%%%%%%%%%%%%%%%%%%%%%%%%%%%%%%%%%%%%%%%%%%%%%%%%%%%%%%%%%%%%%%%%%%%%%%%%%%%%%%%%%%%%%%%%%%%%%%%%%%%%%%%%%%%%

\renewcommand{\arraystretch}{1.5}
\begin{table}[h!]
\hspace{-1cm}
\resizebox{18cm}{!}{
\normalsize{
\begin{tabular}{|c|c|c|c|c|c|c|c|c|c|c|}
\hline
\multicolumn{11}{|c|}{\textbf{Configurations of toroidal three-cycles supporting $USP(2N)$ gauge group on orientifold with the lattice type \textbf{A$_\textbf{a}$BB}}}\\
\hline \hline
brane &wrap. num & $(\tau^1,\tau_3)$ & $(\sigma^1;\sigma^3)$ &  $I^{\Z_2}_{xx}$ &$I_{x(\mathcal{Q}x)}$ & $I^{\Z_2}_{x(\mathcal{Q}x)}$& $\tilde{I}^{\mathcal{R}}_{x}$ & $\tilde{I}^{\mathcal{R}\mathcal{Q}^{-3}}_{x}$& $\tilde{I}^{\mathcal{R}\mathcal{Q}^{-2}}_{x}$& $\tilde{I}^{\mathcal{R}\mathcal{Q}^{-1}}_{x}$\\ 
\hline 
\hline 
$a$ & (1,-1)($-1$,0)(0,1) & $(-,\tau^3)$ & $(\sigma^1,\sigma^3)$ &($\textcolor{red}{2}\cdot$1,0,2)&(1,-0,1)&(1,-0,1)&(1,-0,1)&(1,-4,0)&(-1,0,1)&(0,-2,2)\\ 
\hline 
\multicolumn{11}{|c|}{$b_a\supset\frac{1}{2}\left[\left(-M_aI^{\Z_2,(1\cdot 3)}_{aa}+\tilde{I}_{a}^{\mathcal{R}\mathcal{Q}^{-3}(1\cdot 2)}+\tilde{I}_{a}^{\mathcal{R}\mathcal{Q}^{-1},(2\cdot 3)}\right)+\left(-M_a(I^{(1\cdot 3)}_{a(\mathcal{Q}a)}+I^{\Z_2,(1\cdot 3)}_{a(\mathcal{Q}a)})+\tilde{I}_{a}^{\mathcal{R}(1\cdot 3)}+\tilde{I}_{a}^{\mathcal{R}\mathcal{Q}^{-2},(1\cdot 3)}
\right)\right]=-\textcolor{red}{2}M_a-4+M_a-1$}\\
\hline\hline
$b$ & (1,0)(-1,0)(1,2)& $(\tau^1,\tau^3)$ & $(-,\sigma^3)$ with $\tau^3\sigma^3=0$ &(2,0,2)&(2,-0,2)&(2,-0,2)&(2,-0,2)&(2,-2,2)&(0,0,0)&(2,2,2) \\ 
\hline 
& & $(\tau^1,\tau^3)$ & $(-,\sigma^3)$ with $\tau^3\sigma^3=1$ &(2,0,2)&(2,-0,2)&(2,-0,-2)&(2,-0,2)&(2,-2,2)&(0,0,0)&(2,2,2)   \\ 
\hline
\multicolumn{7}{|r}{$b_b\supset \frac{1}{2}\left[\left(-M_bI^{\Z_2,(1\cdot 3)}_{bb}+\tilde{I}_{b}^{\mathcal{R}(1\cdot 3)}\right)+\left(-M_b(I^{(1\cdot 3)}_{b(\mathcal{Q}b)}+I^{\Z_2,(1\cdot 3)}_{b(\mathcal{Q}b)})
\right)\right]$}&\multicolumn{4}{l|}{$=-2M_b-2+4M_b$  for $\tau^3\sigma^3=0$}\\
\multicolumn{7}{|r}{}&\multicolumn{4}{l|}{$=-2M_b-2$ \hspace{1.05cm}  for $\tau^3\sigma^3=1$}\\
\hline\hline
$c$ & (0,-1)(0,-1)(1,2) & $(\tau^1,\tau^3)$ & $(-,\sigma^3)$ with $\tau^3 \sigma^3=0$ &(2,0,2)&(1,-0,2)&(1,-0,2)&(-0,2,2)&(-1,0,2)&(1,4,-0)&(1,-0,2) \\ 
\hline 
\multicolumn{11}{|c|}{$b^c\supset \frac{1}{2}\left[\left(-M_cI^{\Z_2,(1\cdot 3)}_{cc}+\tilde{I}_{c}^{\mathcal{R}(2\cdot 3)}+\tilde{I}_{c}^{\mathcal{R}\mathcal{Q}^{-2},(1\cdot 2)}\right)+\left(-M_c(I^{(1\cdot 3)}_{c(\mathcal{Q}c)}+I^{\Z_2,(1\cdot 3)}_{c(\mathcal{Q}c)})+\tilde{I}_{c}^{\mathcal{R}\mathcal{Q}^{-3}(1\cdot 3)}+\tilde{I}_{c}^{\mathcal{R}\mathcal{Q}^{-1},(1\cdot 3)}
\right)\right]=-2M_c-4+2M_c-2$}\\
\hline\hline
\multicolumn{11}{|c|}{O-planes}\\
\hline
\multicolumn{11}{|c|}{\quad$\mathcal{R}$:\;\;$(0,1)(1,0)(-1,0)$\qquad\vline\quad $\mathcal{R}\mathcal{Q}^{-3}$:\;\; $(-1,-1)(0,-1)(0,1)$\qquad\vline\quad $\mathcal{R}\mathcal{Q}^{-2}$:\;\;$(-1,0)(-1,0)(-1,-2)$\qquad\vline\quad $\mathcal{R}\mathcal{Q}^{-1}$:\;\;$(1,-1)(0,1)(-1,-1)$ }\\
\hline
\end{tabular} }
}
\caption{Configurations of toroidal three-cycles giving rise to the $\mathcal{R}_4$-invariant fractional three-cycles, their  intersection numbers per two-torus $T^2_{(i)}$ and  the contribution to the beta function coefficient for $u_1=0$. For the remaining notation see the caption to table \ref{tab:r1}.}\label{tab:r4}
\end{table}

%%%%%%%%%%%%%%%%%%%%%%%%%%%%%%%%%%%%%%%%%%%%%%%%%%%%%%%%%%%%%%%%%%%%%%
\newpage
\renewcommand{\arraystretch}{1.5}
\begin{table}[h!]
\hspace{-1cm}
\resizebox{18cm}{!}{
\normalsize{
\begin{tabular}{|c|c|c|c|c|c|c|c|c|c|c|}
\hline
\multicolumn{11}{|c|}{\textbf{Configurations of toroidal three-cycles supporting $USP(2N)$ gauge group on orientifold with the lattice type \textbf{A$_\textbf{b}$AB}}}\\
\hline \hline
brane &wrap. num & $(\tau^1,\tau^3)$ & $(\sigma^1,\sigma^3)$ & $I^{\Z_2}_{xx}$ &$I_{x(\mathcal{Q}x)}$ & $I^{\Z_2}_{x(\mathcal{Q}x)}$& $\tilde{I}^{\mathcal{R}}_{x}$ & $\tilde{I}^{\mathcal{R}\mathcal{Q}^{-3}}_{x}$& $\tilde{I}^{\mathcal{R}\mathcal{Q}^{-2}}_{x}$& $\tilde{I}^{\mathcal{R}\mathcal{Q}^{-1}}_{x}$\\ 
\hline 
$a$ & (1,0)(-1,0)(0,1) & $(\tau^1,\tau^3)$ & $(-,\sigma^3)$ & (2,0,2)&(2,-0,1)&(2,-0,1)&(-2,-0,-1)&(1,-4,0)&(2,0,-1)
&(0,-2,2)\\ 
\hline 
\multicolumn{11}{|c|}{$b_a\supset\frac{1}{2}\left[\left(-M_aI^{\Z_2,(1\cdot 3)}_{aa}+\tilde{I}_{a}^{\mathcal{R}\mathcal{Q}^{-3}(1\cdot 2)}+\tilde{I}_{a}^{\mathcal{R}\mathcal{Q}^{-1},(2\cdot 3)}\right)+\left(-M_a(I^{(1\cdot 3)}_{a(\mathcal{Q}a)}+I^{\Z_2,(1\cdot 3)}_{a(\mathcal{Q}a)})+\tilde{I}_{a}^{\mathcal{R}(1\cdot 3)}+\tilde{I}_{a}^{\mathcal{R}\mathcal{Q}^{-2},(1\cdot 3)}
\right)\right]=-2M_a-4+2M_a-2$}\\
\hline\hline
$b$ & (1,1)(-1,0)(1,2)& $(-,\tau^3)$ & $(\sigma^1,\sigma^3)$ with $\tau^3\sigma^3=0$ & (\textcolor{red}{2}$\cdot$ 1,0,2)&(1,-0,2)&(1,-0,2)&(-2,-0,-2)
&(-1,2,2)&(0,0,0)&(-1,-2,2)\\ 
\hline 
&  & $(-,\tau^3)$ &$(\sigma^1,\sigma^3)$ with $\tau^3\sigma^3=1$ & (\textcolor{red}{2}$\cdot$ 1,0,2)&(1,-0,2)&
(1,-0,-2)&(-2,-0,-2)&(-1,2,2)&(0,0,0)&(-1,-2,2)\\  
\hline
\multicolumn{7}{|r}{$b_b\supset\frac{1}{2}\left[\left(-M_bI^{\Z_2,(1\cdot 3)}_{bb}+\tilde{I}_{b}^{\mathcal{R}(1\cdot 3)}\right)+\left(-M_b(I^{(1\cdot 3)}_{b(\mathcal{Q}b)}+I^{\Z_2,(1\cdot 3)}_{b(\mathcal{Q}b)})
\right)\right]=$}&\multicolumn{4}{l|}{$-\textcolor{red}{2}M_b-2+2M_b$  for $\tau^3\sigma^3=0$}\\
\multicolumn{7}{|r}{}&\multicolumn{4}{l|}{$-\textcolor{red}{2}M_b-2$ \hspace{1.05cm}  for $\tau^3\sigma^3=1$}\\
\hline\hline
$c$ & (1,-1)(1,-2)(1,2) & $(-,\tau^3)$ & $(\sigma^1,\sigma^3)$ with $\tau^3 \sigma^3=0$  &(\textcolor{red}{2}$\cdot$ 1,0,2)&(1,-0,2)&(1,-0,2)&(0,2,-2)&(-1,0,2)&(-1,4,0)&(1,-0,2)\\ 
\hline 
\multicolumn{11}{|c|}{$b_c\supset\frac{1}{2}\left[\left(-M_cI^{\Z_2,(1\cdot 3)}_{cc}+\tilde{I}_{c}^{\mathcal{R}(2\cdot 3)}+\tilde{I}_{c}^{\mathcal{R}\mathcal{Q}^{-2},(1\cdot 2)}\right)+\left(-M_c(I^{(1\cdot 3)}_{c(\mathcal{Q}c)}+I^{\Z_2,(1\cdot 3)}_{c(\mathcal{Q}c)})+\tilde{I}_{c}^{\mathcal{R}\mathcal{Q}^{-3}(1\cdot 3)}+\tilde{I}_{c}^{\mathcal{R}\mathcal{Q}^{-1},(1\cdot 3)}
\right)\right]=-\textcolor{red}{2}M_c-4+2M_c-2$}\\
\hline\hline
\multicolumn{11}{|c|}{O-planes}\\
\hline
\multicolumn{11}{|c|}{\quad$\mathcal{R}$:\;\;$(1,-1)(1$,0$)$(1,0)\qquad\vline\quad $\mathcal{R}\mathcal{Q}^{-3}$:\;\; $(0,-1)(1,-2)(0,1)$\qquad\vline\quad $\mathcal{R}\mathcal{Q}^{-2}$:\;\;$(1,1)(-1,0)(1,2)$\qquad\vline\quad $\mathcal{R}\mathcal{Q}^{-1}$:\;\;$(1,0)(-1,2)(-1,-1)$ }\\
\hline 
\end{tabular} }
}
\caption{Configurations of toroidal three-cycles giving rise to the $\mathcal{R}_1$-invariant fractional three-cycles, their  intersection numbers per two-torus $T^2_{(i)}$ and  the contribution to the beta function coefficient for $u_1=\frac{1}{2}$. For the remaining notation see the caption to table \ref{tab:r1}.}\label{tab:r105}
\end{table}

%%%%%%%%%%%%%%%%%%%%%%%%%%%%%%%%%%%%%%%%%%%%%%%%%%%%%%%%%%%%%%%%%%%%%%%%
\renewcommand{\arraystretch}{1.5}
\begin{table}[h!]
\hspace{-1cm}
\resizebox{18cm}{!}{
\normalsize{
\begin{tabular}{|c|c|c|c|c|c|c|c|c|c|c|}
\hline
\multicolumn{11}{|c|}{\textbf{Configurations of toroidal three-cycles supporting $USP(2N)$ gauge group on orientifold with the lattice type \textbf{A$_\textbf{b}$AA}}}\\
\hline \hline
brane &wrap. num & $(\tau^1,\tau^3)$ & $(\sigma^1,\sigma^3)$ & $I^{\Z_2}_{xx}$ &$I_{x(\mathcal{Q}x)}$ & $I^{\Z_2}_{x(\mathcal{Q}x)}$& $\tilde{I}^{\mathcal{R}}_{x}$ & $\tilde{I}^{\mathcal{R}\mathcal{Q}^{-3}}_{x}$& $\tilde{I}^{\mathcal{R}\mathcal{Q}^{-2}}_{x}$& $\tilde{I}^{\mathcal{R}\mathcal{Q}^{-1}}_{x}$\\ 
\hline 
\hline 
$a$ & (1,1)(1,-2)(0,1) & $(-,\tau^3)$ & $(\sigma^1,\sigma^3)$ &(\textcolor{red}{2}$\cdot$ 1,0,2)&
(1,-0,1)&(1,-0,1)&(-1,4,0)&(1,0,-1)&(0,-2,2)&(1,-0,1) \\ 
\hline 
\multicolumn{11}{|c|}{$b_a\supset \frac{1}{2}\left[\left(-M_aI^{\Z_2,(1\cdot 3)}_{aa}+\tilde{I}_{a}^{\mathcal{R}(1\cdot 2)}+\tilde{I}_{a}^{\mathcal{R}\mathcal{Q}^{-2},(2\cdot 3)}\right)+\left(-M_a(I^{(1\cdot 3)}_{a(\mathcal{Q}a)}+I^{\Z_2,(1\cdot 3)}_{a(\mathcal{Q}a)})+\tilde{I}_{a}^{\mathcal{R}\mathcal{Q}^{-3}(1\cdot 3)}+\tilde{I}_{a}^{\mathcal{R}\mathcal{Q}^{-1},(1\cdot 3)}
\right)\right]=-\textcolor{red}{2}M_a-4+M_a-1$}\\
\hline\hline
$b$ & (0,1)(1,-2)(1,2)& $(\tau_1,\tau_3)$ & $(-,\sigma^3)$ with $\tau^3\sigma^3=0$ &(2,0,2)&(2,-0,2)&(2,-0,2)&(1,-4,2)&(0,0,0)&(1,4,2)&(2,-0,2) \\ 
\hline 
& &$(\tau_1,\tau_3)$ & $(-,\sigma^3)$ with $\tau^3\sigma^3=1$ &(2,0,2)&(2,-0,2)&(2,-0,-2)&(1,-4,2)&(0,0,0)&(1,4,2)&(2,-0,2) \\ \hline
\multicolumn{7}{|r}{$b_b\supset\frac{1}{2}\left[\left(-M_bI^{\Z_2,(1\cdot 3)}_{bb}+\tilde{I}_{b}^{\mathcal{R}\mathcal{Q}^{-1}(1\cdot 3)}\right)+\left(-M_b(I^{(1\cdot 3)}_{b(\mathcal{Q}b)}+I^{\Z_2,(1\cdot 3)}_{b(\mathcal{Q}b)})
\right)\right]=$}&\multicolumn{4}{l|}{$-2M_b-2+4M_b$ for $\tau^3\sigma^3=0$}\\
\multicolumn{7}{|r}{}&\multicolumn{4}{l|}{$-2M_b-2$ \hspace{1.05cm} for $\tau^3\sigma^3=1$}\\
\hline\hline
$c$ & (1,0)(1,0)(1,2) & $(\tau^1,\tau^3)$ & $(-,\sigma^3)$ with $\tau^3 \sigma^3=0$ 
&(2,0,2)&(2,-0,2)&(2,-0,2)&(-2,0,2)&(1,-4,0)&(2,-0,2)&(-0,2,2) \\ 
\hline 
\multicolumn{11}{|c|}{$b_c\supset\frac{1}{2}\left[\left(-M_cI^{\Z_2,(1\cdot 3)}_{cc}+\tilde{I}_{c}^{\mathcal{R}\mathcal{Q}^{-3}(1\cdot 2)}+\tilde{I}_{c}^{\mathcal{R}\mathcal{Q}^{-1},(2\cdot 3)}\right)+\left(-M_c(I^{(1\cdot 3)}_{c(\mathcal{Q}c)}+I^{\Z_2,(1\cdot 3)}_{c(\mathcal{Q}c)})+\tilde{I}_{c}^{\mathcal{R}(1\cdot 3)}+\tilde{I}_{c}^{\mathcal{R}\mathcal{Q}^{-2},(1\cdot 3)}
\right)\right]=-2M_c-4+4M_c-4 $}\\
\hline\hline
\multicolumn{11}{|c|}{O-planes}\\
\hline
\multicolumn{11}{|c|}{\quad$\mathcal{R}$:\;\;$(1,-1)(1$,0$)$(0,1$)$\qquad\vline\quad $\mathcal{R}\mathcal{Q}^{-3}$:\;\; $(0,1)(1,-2)(1,2)$\qquad\vline\quad $\mathcal{R}\mathcal{Q}^{-2}$:\;\;$(1,1)(-1,0)(-1,-1)$\qquad\vline\quad $\mathcal{R}\mathcal{Q}^{-1}$:\;\;$(-1,0)(-1,2)(-1,0)$ }\\
\hline 
\end{tabular} }
}
\caption{Configurations of toroidal three-cycles giving rise to the $\mathcal{R}_2$-invariant fractional three-cycles, their  intersection numbers per two-torus $T^2_{(i)}$ and  the contribution to the beta function coefficient for $u_1=\frac{1}{2}$. For the remaining notation see the caption to table \ref{tab:r1}.}\label{tab:r205}
\end{table}

%%%%%%%%%%%%%%%%%%%%%%%%%%%%%%%%%%%%%%%%%%%%%%%%%%%%%%%%%%%%%%%%%%%%%%%%
\renewcommand{\arraystretch}{1.5}
\begin{table}[h!]
\hspace{-1cm}
\resizebox{18cm}{!}{
\normalsize{
\begin{tabular}{|c|c|c|c|c|c|c|c|c|c|c|}
\hline
\multicolumn{11}{|c|}{\textbf{Configurations of toroidal three-cycles supporting $USP(2N)$ gauge group on orientifold with the lattice type \textbf{A$_\textbf{b}$BA}}}\\
\hline \hline
brane &wrap. num & $(\tau^1,\tau^3)$ & $(\sigma^1,\sigma^3)$ & $I^{\Z_2}_{xx}$ &$I_{x(\mathcal{Q}x)}$ & $I^{\Z_2}_{x(\mathcal{Q}x)}$& $\tilde{I}^{\mathcal{R}}_{x}$ & $\tilde{I}^{\mathcal{R}\mathcal{Q}^{-3}}_{x}$& $\tilde{I}^{\mathcal{R}\mathcal{Q}^{-2}}_{x}$& $\tilde{I}^{\mathcal{R}\mathcal{Q}^{-1}}_{x}$\\ 
\hline 
\hline 
$a$ & (1,0)(1,-2)(0,1) & $(\tau^1,\tau^3)$ & $(-,\sigma^3)$ 
&(2,0,2)&(2,-0,1)&(2,-0,1)&(1,-4,0)&(2,0,-1)
&(0,-2,2)&(2,-0,1) \\ 
\hline 
\multicolumn{11}{|c|}{$b_a\supset\frac{1}{2}\left[\left(-M_aI^{\Z_2,(1\cdot 3)}_{aa}+\tilde{I}_{a}^{\mathcal{R}(1\cdot 2)}+\tilde{I}_{a}^{\mathcal{R}\mathcal{Q}^{-2},(2\cdot 3)}\right)+\left(-M_a(I^{(1\cdot 3)}_{a(\mathcal{Q}a)}+I^{\Z_2,(1\cdot 3)}_{a(\mathcal{Q}a)})+\tilde{I}_{a}^{\mathcal{R}\mathcal{Q}^{-3}(1\cdot 3)}+\tilde{I}_{a}^{\mathcal{R}\mathcal{Q}^{-1},(1\cdot 3)}
\right)\right]=-2M_a-4+2M_a-2$}\\
\hline\hline
$b$ & (1,1)(1,-2)(1,2)& $(-,\tau^3)$ & $(\sigma^1,\sigma^3)$ with $\tau^3\sigma^3=0$ 
&(\textcolor{red}{2}$\cdot$ 1,0,2)&(1,-0,2)&(1,-0,2)&(1,-2,2)
&(0,0,0)&(1,2,2)&(2,-0,2) \\ 
\hline 
& &$(-,\tau^3)$ & $(\sigma^1,\sigma^3)$ with $\tau^3\sigma^3=1$  &(\textcolor{red}{2}$\cdot$ 1,0,2)&(1,-0,2)&(1,-0,-2)&(1,-2,2)
&(0,0,0)&(1,2,2)&(2,-0,2) \\ 
\hline 
\multicolumn{7}{|r}{$b_b\supset \frac{1}{2}\left[\left(-M_bI^{\Z_2,(1\cdot 3)}_{bb}+\tilde{I}_{b}^{\mathcal{R}\mathcal{Q}^{-1}(1\cdot 3)}\right)+\left(-M_b(I^{(1\cdot 3)}_{b(\mathcal{Q}b)}+I^{\Z_2,(1\cdot 3)}_{b(\mathcal{Q}b)})
\right)\right]=$}&\multicolumn{4}{l|}{$-\textcolor{red}{2}M_b-2+2M_b$ for $\tau^3\sigma^3=0$}\\
\multicolumn{7}{|r}{}&\multicolumn{4}{l|}{$-\textcolor{red}{2}M_b-2$ \hspace{1.05cm} for $\tau^3\sigma^3=1$}\\
\hline\hline
$c$ & (1,-1)(1,0)(1,2) & $(-,\tau_3)$ & $(\sigma^1,\sigma^3)$ with $\tau^3 \sigma^3=0$  &(\textcolor{red}{2}$\cdot$ 1,0,2)&(1,-0,2)&(1,-0,2)&(-1,0,2)&(1,-4,0)&(1,-0,2)&(-0,2,2)\\ 
\hline 
\multicolumn{11}{|c|}{$b_c\supset\frac{1}{2}\left[\left(-M_cI^{\Z_2,(1\cdot 3)}_{cc}+\tilde{I}_{c}^{\mathcal{R}\mathcal{Q}^{-3}(1\cdot 2)}+\tilde{I}_{c}^{\mathcal{R}\mathcal{Q}^{-1},(2\cdot 3)}\right)+\left(-M_c(I^{(1\cdot 3)}_{c(\mathcal{Q}c)}+I^{\Z_2,(1\cdot 3)}_{c(\mathcal{Q}c)})+\tilde{I}_{c}^{\mathcal{R}(1\cdot 3)}+\tilde{I}_{c}^{\mathcal{R}\mathcal{Q}^{-2},(1\cdot 3)}\right)\right]=-\textcolor{red}{2}M_c-4+2M_c-2$}\\
\hline\hline
\multicolumn{11}{|c|}{O-planes}\\
\hline
\multicolumn{11}{|c|}{\quad$\mathcal{R}$:\;\;(0,-1)(1,0$)$(0,1$)$\qquad\vline\quad $\mathcal{R}\mathcal{Q}^{-3}$:\;\;$(1,1)(1,-2)(1,2)$\qquad\vline\quad $\mathcal{R}\mathcal{Q}^{-2}$:\;\;$(1,0)(-1,0)(-1,-1)$\qquad\vline\quad $\mathcal{R}\mathcal{Q}^{-1}$:\;\;$(-1,1)(-1,2)(-1,0)$ }\\
\hline 
\end{tabular} }
}
\caption{Configurations of toroidal three-cycles giving rise to the $\mathcal{R}_3$-invariant fractional three-cycles, their  intersection numbers per two-torus $T^2_{(i)}$ and  the contribution to the beta function coefficient for $u_1=\frac{1}{2}$. For the remaining notation see the caption to table \ref{tab:r1}.}\label{tab:r305}
\end{table}

%%%%%%%%%%%%%%%%%%%%%%%%%%%%%%%%%%%%%%%%%%%%%%%%%%%%%%%%%%%%%%%%%%%%%%%%%%%%%%%%%%%%%%%%%%%%%%%%%%%%%%%%%%%%%%%%%%%%%%%%%%%%%%%%%%%%%%%%%%%%%%%%%%%%%%%%%%%%%%%%%%%%%%%%%%%%%%%%%%%%%%%%%%%%%%%%%%%%%%%%%%%%%%%%%%%%%%%%%%%%%%%%%%%%%%%%%%%%%%%%%%%%%%%%%%%%%%%%%%%%%%%%%%%%%%%%%%%%

\renewcommand{\arraystretch}{1.5}
\begin{table}[h!]
\hspace{-1cm}
\resizebox{18cm}{!}{
\normalsize{
\begin{tabular}{|c|c|c|c|c|c|c|c|c|c|c|}
\hline
\multicolumn{11}{|c|}{\textbf{Configurations of toroidal three-cycles supporting $USP(2N)$ gauge group on orientifold with the lattice type \textbf{A$_\textbf{b}$BB}}}\\
\hline \hline
brane &wrap. num & $(\tau^1,\tau^3)$ & $(\sigma^1,\sigma^3)$ & $I^{\Z_2}_{xx}$ &$I_{x(\mathcal{Q}x)}$ & $I^{\Z_2}_{x(\mathcal{Q}x)}$& $\tilde{I}^{\mathcal{R}}_{x}$ & $\tilde{I}^{\mathcal{R}\mathcal{Q}^{-3}}_{x}$& $\tilde{I}^{\mathcal{R}\mathcal{Q}^{-2}}_{x}$& $\tilde{I}^{\mathcal{R}\mathcal{Q}^{-1}}_{x}$\\ 
\hline 
\hline 
$a$ & (1,-1)($-1$,0)(0,1) & $(-,\tau^3)$ & $(\sigma^1,\sigma^3)$ &(\textcolor{red}{2}$\cdot$ 1,0,2)&(1,-0,1)&
(1,-0,1)&(1,-0,1)&(1,-4,0)&(1,0,-1)&(0,-2,2)\\ 
\hline 
\multicolumn{11}{|c|}{$b_a\supset\frac{1}{2}\left[\left(-M_aI^{\Z_2,(1\cdot 3)}_{aa}+\tilde{I}_{a}^{\mathcal{R}\mathcal{Q}^{-3}(1\cdot 2)}+\tilde{I}_{a}^{\mathcal{R}\mathcal{Q}^{-1},(2\cdot 3)}\right)+\left(-M_a(I^{(1\cdot 3)}_{a(\mathcal{Q}a)}+I^{\Z_2,(1\cdot 3)}_{a(\mathcal{Q}a)})+\tilde{I}_{a}^{\mathcal{R}(1\cdot 3)}+\tilde{I}_{a}^{\mathcal{R}\mathcal{Q}^{-2},(1\cdot 3)}
\right)\right]=-\textcolor{red}{2}M_a-4+M_a-1$}\\
\hline\hline
$b$ & (1,0)(-1,0)(1,2)& $(\tau^1,\tau^3)$ & $(-,\sigma^3)$ with $\tau^3\sigma^3=0$ &(2,0,2)&(2,-0,2)&(2,-0,2)&(2,-0,2)&(2,-2,2)&(0,0,0)&(2,2,2) \\ 
\hline 
& & $(\tau^1,\tau^3)$ & $(-,\sigma^3)$ with $\tau^3\sigma^3=1$ &(2,0,2)&(2,-0,2)&(2,-0,-2)&(2,-0,2)&(2,-2,2)&(0,0,0)&(2,2,2)   \\ 
\hline
\multicolumn{7}{|r}{$b_b\supset \frac{1}{2}\left[\left(-M_bI^{\Z_2,(1\cdot 3)}_{bb}+\tilde{I}_{b}^{\mathcal{R}(1\cdot 3)}\right)+\left(-M_b(I^{(1\cdot 3)}_{b(\mathcal{Q}b)}+I^{\Z_2,(1\cdot 3)}_{b(\mathcal{Q}b)})
\right)\right]=$}&\multicolumn{4}{l|}{$-2M_b-2+4M_b$ for $\tau^3\sigma^3=0$}\\
\multicolumn{7}{|r}{}&\multicolumn{4}{l|}{$-2M_b-2$ \hspace{1.05cm} for $\tau^3\sigma^3=1$}\\
\hline\hline
$c$ & (0,-1)(1,-2)(1,2) &$(\tau^1,\tau^3)$ & $(-,\sigma^3)$ with $\tau^3\sigma^3=0$
&(2,0,2)&(2,-0,2)&(2,-0,2)&(-0,2,2)&(-2,0,2)&(-1,4,0)&(2,-0,2) \\ 
\hline
\multicolumn{11}{|c|}{$b_c\supset \frac{1}{2}\left[\left(-M_cI^{\Z_2,(1\cdot 3)}_{cc}+\tilde{I}_{c}^{\mathcal{R}(2\cdot 3)}+\tilde{I}_{c}^{\mathcal{R}\mathcal{Q}^{-2},(1\cdot 2)}\right)+\left(-M_c(I^{(1\cdot 3)}_{c(\mathcal{Q}c)}+I^{\Z_2,(1\cdot 3)}_{c(\mathcal{Q}c)})+\tilde{I}_{c}^{\mathcal{R}\mathcal{Q}^{-3}(1\cdot 3)}+\tilde{I}_{c}^{\mathcal{R}\mathcal{Q}^{-1},(1\cdot 3)}
\right)\right]=-2M_c-4+4M_c-4$}\\
\hline\hline
\multicolumn{11}{|c|}{O-planes}\\
\hline
\multicolumn{11}{|c|}{\quad$\mathcal{R}$:\;\;(0,1)$(1,0)$(-1,0)\qquad\vline\quad $\mathcal{R}\mathcal{Q}^{-3}$:\;\;$(-1,-1)(1,-2)(0,1)$\qquad\vline\quad $\mathcal{R}\mathcal{Q}^{-2}$:\;\;$(1,0)(-1,0)(1,2)$\qquad\vline\quad $\mathcal{R}\mathcal{Q}^{-1}$:\;\;$(1,-1)(-1,2)(-1,-1)$}\\
\hline 
\end{tabular} }
}
\caption{Configurations of toroidal three-cycles giving rise to the $\mathcal{R}_4$-invariant fractional three-cycles, their  intersection numbers per two-torus $T^2_{(i)}$ and  the contribution to the beta function coefficient for $u_1=\frac{1}{2}$. For the remaining notation see the caption to table \ref{tab:r1}.}\label{tab:r405}
\end{table}

%%%%%%%%%%%%%%%%%%%%%%%%%%%%%%%%%%%%
\section{Example}\label{example}
In this section, we present an example of a supersymmetric globally consistent intersecting D6-model on the non-factorisable orientifold $T^6/(\mathbb{Z}_4\times\Omega\mathcal{R})$. We will use the formulae from section~\ref{amplitudes} 
to compute the full massless  open string spectrum, the beta function coefficients and the one-loop gauge threshold corrections.

\subsection{Brane configuration and intersection numbers}
The case we consider is a four-generation Pati-Salam model with gauge group \linebreak
\mbox{$U(4)_a\times USp(2)_b\times USp(2)_c\times USp(4)_h$} that we found to satisfy all untwisted and twisted RR tadpole cancellation conditions in \cite{Berasaluce-Gonzalez:2016kqb}. The K-theory constraints are trivially satisfied for this choice of gauge group.
The different D6-brane stacks and their toroidal wrapping numbers, discrete displacements and Wilson lines on $T_{(1)}^2\times T_{(3)}^2$, and $\mathbb{Z}_2$ eigenvalues $\tau_0$ are listed in table \ref{tab:geom-setup1}.
\begin{table}[h!]
\begin{center}
\begin{tabular}{|c||c|c|c|c|}
\hline
\multicolumn{5}{|c|}{\textbf{D6-branes configuration for a four generation PS-model}}\\
\multicolumn{5}{|c|}{\textbf{on the $A_3\times A_1\times B_2$-orientifold}}\\
\hline
\hline 
Stack& Angle w.r.t. $\Omega\mathcal{R}$ & $(n^1,m^1;n^2,m^2;n^3,m^3)$ & $(\sigma^1,\sigma^3)$ & $(\tau^0,\tau^1,\tau^3)$ \\ 
\hline 
a& $\pi(0,\,\frac{1}{2},\,-\frac{1}{2})$  & $(1,-1;0,-1;1,2)$&  (1,1)& $(1,0,1)$ \\ 
\hline 
b& $\pi(-\frac{1}{2},\,1\,,-\frac{1}{2})$ & $(1,1;-1,0;1,2)$ & (1,1) & $(0,0,1)$ \\ 
c & &$(1,1;-1,0;1,2)$& (1,1) & $(1,0,1)$ \\ 
h & &$(1,1;-1,0;1,2)$&  (0,0) & $(0,0,0)$ \\ 
\hline 
\end{tabular} 
\caption{Geometrical setup of the supersymmetric Pati-Salam model example with four generation on the \textbf{A$_\textbf{a}$AB} lattice orientation of
 the $T^6/(\Z_4 \times \OR)$ orientifold.}\label{tab:geom-setup1}
\end{center}
\end{table}

As long as no continuous displacements or Wilson lines are switched on along the second two-torus $T^2_{(2)}$ where the $\mathbb{Z}_2$ action is trivial, branes $b$, $c$ and $h$ are orientifold invariant, and the gauge groups are enhanced to $USp(2)$ for the left- and right-symmetric branes $b$ and $c$, and $USp(4)$ for the `hidden' stack $h$.

The Abelian $U(1)_a$ symmetry acquires a mass proportional to the string scale $M_\text{string}$, since not all its St\"uckelberg couplings vanish. At energies below this scale, $U(1)_a$ is broken to a discrete $\mathbb{Z}_2$ symmetry by non-perturbative effects. In our particular case, this discrete symmetry corresponds to the $\mathbb{Z}_2$ arising from the K-theory constraints.

\begin{table}[h!]
\begin{center}
\begin{tabular}{|c||c|c|c|c|}
\hline
\multicolumn{5}{|c|}{\textbf{ Non $\kappa$-weighted toroidal intersection numbers for PS example}}\\
\hline
\hline 
$x$ & $\quad I_{\cdot a}=I_{\cdot a^\prime}\quad $ & $\quad I_{\cdot (\mathcal{Q}a)}=I_{\cdot (\mathcal{Q}a)^\prime}\quad $ & $\quad I_{\cdot y}= I_{\cdot y^\prime}\quad $ & $I_{\cdot (\mathcal{Q}y)}=I_{\cdot (\mathcal{Q}y^\prime)}$ 
\\ 
\hline 
a & $0_{123}$ & $(-4)\cdot 0_2$ & $(-2)\cdot 0_3$ & $(-2)\cdot 0_1$ \\ 
\hline 
$y=b,c,h$ &  \multicolumn{2}{|c|}{}  & $0_{123}$ & $(-4)\cdot 0_2$\\ 
\hline 
\end{tabular} 
\caption{Pure (non weighted with $\kappa$) toroidal intersection numbers of the supersymmetric PS model on $T^6/(\Z_4 \times \OR)$ with D6-brane configuration specified in table \ref{tab:geom-setup1}. $0_i$ denotes that the branes are parallel on $T^2_{(i)}$. }\label{tab:int-numbers ohne k}
\end{center}
\end{table}
%%%%%%%%%%%%%%%%%%%%%%%%%%%%%%%%%%%%%%%%%%%%%%%%%%%%
\begin{table}[h!]
\begin{center}
\resizebox{16cm}{!}{
\begin{tabular}{|c||c|c|c|c|c|c|c|c|c|c|}
\hline
\multicolumn{11}{|c|}{\textbf{Non $\kappa$-weighted  $\Z_2$ invariant intersection numbers for PS  example}}\\
\hline
\hline 
$x$ & $I^{\Z_2}_{\cdot a}$ & $I^{\Z_2}_{\cdot (\mathcal{Q}a)}$ & $I^{\Z_2}_{\cdot a^\prime}$ &  $I^{\Z_2}_{\cdot (\mathcal{Q}a^\prime)}$ & $I^{\Z_2}_{\cdot b}$ & $I^{\Z_2}_{\cdot (\mathcal{Q}b)}$ & $I^{\Z_2}_{\cdot c}$ & $I^{\Z_2}_{\cdot (\mathcal{Q}c)}$ & $I^{\Z_2}_{\cdot h}$ & $I^{\Z_2}_{\cdot (\mathcal{Q}h)}$ 
\\ 
\hline 
a & $4\cdot 0_2$ & $4\cdot 0_2$ & $(-4)\cdot 0_2$ & $(-4)\cdot 0_2$& $4$ & 4&$-4$&$-4$& $\emptyset$ & $\emptyset$ \\ 
\hline 
$b$ &  \multicolumn{4}{|c|}{}  & $4\cdot0_{2}$ & $4\cdot0_{2}$ & $(-4)\cdot 0_2$ & $(-4)\cdot 0_2$ & $\emptyset$ & $\emptyset$ \\ 
\hline 
$c$ & \multicolumn{6}{|c|}{}  &$4\cdot 0_2$ & $4\cdot 0_2$ & $\emptyset$ & $\emptyset$ \\
\hline
$h$ & \multicolumn{8}{|c|}{}  &$4\cdot 0_2$ & $(-4)\cdot 0_2$ \\
\hline
\end{tabular}}
\caption{Pure (non weighted with $\kappa$) $\Z_2$ invariant intersection numbers of the supersymmetric PS model on $T^6/(\Z_4 \times \OR)$ defined in table \ref{tab:geom-setup1}. 
$\emptyset$ occurs when branes are parallel but carry relative Wilson lines or are displaced on a two-torus. Since the branes $b,\,c$ and $h$ are orientifold invariant, $I^{\Z_2}_{\cdot (\mathcal{Q}^ky^\prime)}= I^{\Z_2}_{\cdot (\mathcal{Q}^k y)}$.}\label{tab:Z2-invariant-int-numbers ohne k}
\end{center}
\end{table}

\begin{table}[h!]
\begin{center}
\begin{tabular}{|c||c|c|c|c|}
\hline
\multicolumn{5}{|c|}{\textbf{Non $\kappa$-weighted toroidal intersection numbers for PS example}}\\
\hline
\hline 
$x$ & $I^{\Omega\mathcal{R}}_{\cdot x}$ & $I^{\Omega\mathcal{R}\mathcal{Q}^{-3}}_{\cdot x}$ & $I^{\Omega\mathcal{R}\mathcal{Q}^{-2}}_{\cdot x}$ & $I^{\Omega\mathcal{R}\mathcal{Q}^{-1}}_{\cdot x}$
\\ 
\hline 
a & $\quad (-2)\cdot 0_1\quad $ & $\quad (-1)\cdot 0_2\quad $ & $\quad (-2)\cdot 0_3\quad $ & $(-1)\cdot 0_2$ \\ 
\hline 
$\quad b,c,h\quad $ &  $(-4)\cdot 0_2$& $-1$ & $0_{123}$ & $1$\\ 
\hline 
\end{tabular} 
\caption{Pure (non weighted with $\kappa$) toroidal intersection numbers with the O6-planes of the supersymmetric PS model on $T^6/(\Z_4 \times \OR)$ with D6-brane configuration displayed in table \ref{tab:geom-setup1}. 
}\label{tab:oplanes-int-numbers ohne k}
\end{center}
\end{table}

\subsection{Spectrum}
Using the intersection numbers in the tables \ref{tab:int-numbers ohne k} and \ref{tab:Z2-invariant-int-numbers ohne k} among D6-branes and in table \ref{tab:oplanes-int-numbers ohne k} with the O6-planes and the formulae summarised in sections \ref{Summary-b-SUN} for $SU(N)$ and \ref{USP_or_SO} for $USp(2N)$ gauge groups, we can calculate the beta function coefficient for each gauge group of the Pati-Salam model with D6-brane configuration displayed in table \ref{tab:geom-setup1}:
\begin{equation}\label{beta-functions}
\begin{split}
b_{SU(N_a)}&=[aa+aa^\prime]+[ab+ab^\prime]
+[ac+ac^\prime]+[ah+ah^\prime]\\
&=(-2N_a+6N_a-8)+(M_b+M_b)+(M_c+M_c)+0\,,\\
b_{USp(2M_b)}&=[bb]+[ba]+[bc]+[bh]\\
&=(-2M_b)+(2N_a)+(3M_c)+(M_h)\,,\\
b_{USp(2M_c)}&=[cc]+[ca]+[cb]+[ch]\\
&=(-2M_c)+(2N_a)+(3M_b)+(M_h)\,,\\
b_{USp(2M_h)}&=[hh]+[ha]+[hb]+[hc]\\
&=(-2M_h-2+2M_h)+(0)+(M_b)+(M_c)\,.
\end{split}
\end{equation}  
Inserting the ranks of  the gauge group $SU(4)_a\times USp(2)_b\times USp(2)_c\times USp(4)_h$, the beta function coefficients take the following values:
\begin{equation}\label{beta-values}
b_{SU(4)_a}=12\,,\quad b_{USp(2)_b}=b_{USp(2)_c}=9\,,\quad b_{USp(4)_4}=0\,.
\end{equation}
Comparing the results \eqref{beta-functions} with the field theoretical formulae \eqref{beta-SUN} and \eqref{usp_theory}, we can determine the complete massless matter spectrum. The chosen notation has the representations of $\bf{SU(4)_a}$, $\bf{USp(2)_b}$, $\bf{USp(2)_c}$ and $\bf{USp(4)_h}$ in bold, and the charge under the massive $U(1)_a$ as a subscript.

The massless open string spectrum consists of the gauge group $SU(4)_a\times USp(2)_b\times USp(2)_c\times USp(4)_h$ and two kinds of matter spectra $[C]+[V]$:
\begin{itemize}
\item the chiral spectrum coming from non-vanishing intersection numbers is given by:
\begin{equation}\label{chiral-spectrum}
[C]=4\times\left[ {\bf(4,\bar{2},1,1)}_{+1}
+{\bf(\bar{4},1,2,1)}_{-1}\right],
\end{equation}
\item the vector like spectrum is given by:
\begin{equation}\label{non-chiral-spectrum}
\begin{split}
[V]=& {\bf(15,1,1,1)}_0+{\bf(1,3_S,1,1)}_0+{\bf(1,1,3_S,1)}_0+
2\times{\bf(1,1,1,10_S)}_0+
{\bf(1,1,1,6_A)}_0\\
+&\big(5\times{\bf(6_A,1,1,1)}_{+2}+{\bf(10_S,1,1,1)}_{+2}+\textit{c.c.}\big)
\\+&
 3\times{\bf(1,2,2,1)}_0+{\bf(1,2,1,4)}_0+{\bf(1,1,2,4)}_0
\end{split}
\end{equation}
\end{itemize}
Since this four-generation model preserves ${\cal N}=1$ supersymmetry, the vector-like and chiral spectrum contain a rich enough variety of charged scalars 
to account for the breaking of the Pati-Salam gauge group $SU(4) \times SU(2)_L \times SU(2)_R  \to [SU(3) \times SU(2)_L \times SU(2)_R \times U(1)_{B-L} \to] SU(3) \times SU(2)_L \times U(1)_Y \to SU(3) \times U(1)_{\text{e.-m.}}$. But extracting the associated superpotential couplings,
e.g. of the type ${\bf(6_A,1,1,1)}_{+2} \cdot  {\bf(\bar{4},1,2,1)}_{-1} \cdot {\bf(\bar{4},1,2,1)}_{-1}$, 
 from geometric considerations of worldsheet instantons goes well beyond the scope of this article.

The massless closed string spectrum can be read off from table \ref{tab:closed-string-spectrum}.

\subsection{Threshold corrections}

Before computing the one-loop gauge threshold corrections due to massive string excitations for the different gauge groups in our Pati-Salam model using the formulae from section \ref{amplitudes}, we briefly summarise the tree-level gauge couplings,
\begin{equation}\label{eq:g-tree-example}
\frac{16\pi^2}{g^2_{x,\textup{string}}}=\frac{M_\text{Planck}}{M_\text{string}}\times\left\lbrace\begin{array}{ll}
\sqrt{\frac{2R_1}{\sqrt{-a}R_3}}=2\sqrt{u_2}\overset{u_2=1}{=}2&\quad\text{for }x=a\,,\\
\sqrt{\frac{\sqrt{-a}R_3}{2R_1}}=\frac{1}{2\sqrt{u_2}}\overset{u_2=1}{=}\frac{1}{2}&\quad\text{for }x=b,c,h
\end{array}\right.
\end{equation}
and remind the reader that the field theoretical one-loop running with the energy scale due to massless strings is specified by the beta function coefficients in equation \eqref{beta-values}.

The full gauge threshold correction for the $SU(4)_a$ factor is given by:
\begin{equation}
\Delta_{SU(4)_a}=\sum_b N_b\left(\tilde{\Delta}_{ab}^{\textup{total}}+\tilde{\Delta}_{ab'}^{\textup{total}}\right) + \tilde{\Delta}_{a,\Omega\mathcal{R}}^{\textup{total}},
\end{equation}
with
\begin{equation}
\begin{split}
\tilde{\Delta}_{ab}^{\textup{total}}&=\tilde{\Delta}_{ba}^{\textup{total}}=\sum_{k=0}^{1}\left(\tilde{\Delta}_{a(\mathcal{Q}^kb)}^{{\bf 1}}+\tilde{\Delta}_{a(\mathcal{Q}^kb)}^{\mathbb{Z}_2}\right),\\
\tilde{\Delta}_{a,\Omega\mathcal{R}}^{\textup{total}}&=\sum_{k=0}^3\Delta_{a,\Omega\mathcal{R}\mathcal{Q}^{-k}}.
\end{split}
\end{equation}

For convenience we have defined $\tilde{\Delta}_{ab}^{\textup{total}}:=\Delta_{ab}^{\textup{total}}/N_b$, since this quantity has several symmetries in its two subscripts,
\begin{equation}
\tilde{\Delta}^{\textup{total}}_{ab}=\tilde{\Delta}^{\textup{total}}_{ba}=\tilde{\Delta}^{\textup{total}}_{a'b'}=\tilde{\Delta}^{\textup{total}}_{b'a'}.
\end{equation}

The gauge threshold for a symplectic gauge factor $USp(2M_x)$ is given by:
\begin{equation}
\Delta_{USp(2M_x)}=\sum_{z \in \{a,b,c,d\}} N_z \tilde{\Delta}^{\textup{total}}_{zx}+\frac{1}{2}\Delta^{\textup{total}}_{x,\Omega\mathcal{R}},
\end{equation}
where for all other symplectic gauge factors $USp(2M_y)$ one has to identify $N_z=M_y$.

The tree-channel annulus contributions with trivial projector insertion {\bf 1} to the threshold corrections, with $z_i\in\{b,c\}$, are given by:
\begin{equation}
\begin{split}
\tilde{\Delta}^{{\bf 1}}_{aa}&=\tilde{\Delta}^{{\bf 1}}_{aa'}=\tilde{\Delta}^{{\bf 1}}_{z_iz_j}=\tilde{\Delta}^{{\bf 1}}_{z_ih}=\tilde{\Delta}^{{\bf 1}}_{hh}=0,\\
\tilde{\Delta}^{{\bf 1}}_{a(\mathcal{Q}a)}&=\tilde{\Delta}^{{\bf 1}}_{a(\mathcal{Q}a')}=-2\Lambda(0,0,\mathsf{v}_2,u_2)-2\Lambda(1,0,\mathsf{v}_2,u_2),\\
\tilde{\Delta}^{{\bf 1}}_{az_i}&=-\Lambda(0,0,\mathsf{v}_3,2),\\
\tilde{\Delta}^{{\bf 1}}_{a(\mathcal{Q}z_i)}&=-\Lambda(0,0,\mathsf{v}_1,2),\\
\tilde{\Delta}^{{\bf 1}}_{ah}&=-\Lambda(1,1,\mathsf{v}_3,2),\\ 
\tilde{\Delta}^{{\bf 1}}_{a(\mathcal{Q}h)}&=-\Lambda(1,0,\mathsf{v}_1,2),\\
\tilde{\Delta}^{{\bf 1}}_{z_i(\mathcal{Q}z_j)}&=\tilde{\Delta}^{{\bf 1}}_{h(Qh)}=-\Lambda(0,0,\mathsf{v}_2,1/u_2),\\
\tilde{\Delta}^{{\bf 1}}_{z_i(\mathcal{Q}h)}&=-\Lambda(0,0,\mathsf{v}_2,1/u_2) \, ,
\end{split}
\end{equation}
with the lattice sum $\Lambda$ as defined in equation~\eqref{lambdacontribution}.

The M\"obius strip contributions to the threshold corrections with $y\in\{b,c,h\}$ read:
\begin{eqnarray}
\begin{rcases}
\Delta^{\Omega\mathcal{R}}_{a}=4\widehat{\Lambda}(1,0,\mathsf{v}_1,4),\\
\Delta^{\Omega\mathcal{R}\mathcal{Q}^{-2}}_{a}=4\widehat{\Lambda}(1,1,\mathsf{v}_3,4),\\
\Delta^{\Omega\mathcal{R}\mathcal{Q}^{-1}}_{a}=\Delta^{\Omega\mathcal{R}\mathcal{Q}^{-3}}_{a}=4\widehat{\Lambda}(0,0,\mathsf{v}_2,2u_2)\\\hspace{3.35cm}+4\widehat{\Lambda}(1,0,\mathsf{v}_2,2u_2),
\end{rcases}\Rightarrow &\Delta^{\textup{total}}_{a,\Omega\mathcal{R}}=
\begin{cases}
4\widehat{\Lambda}(1,0,\mathsf{v}_1,4)\\
+4\widehat{\Lambda}(1,1,\mathsf{v}_3,4)\\
+8\widehat{\Lambda}(0,0,\mathsf{v}_2,2u_2)\\
+8\widehat{\Lambda}(1,0,\mathsf{v}_2,2u_2).
\end{cases}\nonumber\\
\textup{ } &\\
\begin{rcases}
\Delta^{\Omega\mathcal{R}}_{y}=4\widehat{\Lambda}(0,0,2\mathsf{v}_2,2/u_2),\\
\Delta^{\Omega\mathcal{R}\mathcal{Q}^{-2}}_{y}=0,\\
\Delta^{\Omega\mathcal{R}\mathcal{Q}^{-1}}_{y}=-\Delta^{\Omega\mathcal{R}\mathcal{Q}^{-3}}_{y}=\frac{\ln 2}{2},
\end{rcases}\Rightarrow &\Delta^{\textup{total}}_{y,\Omega\mathcal{R}}=
\begin{cases}
\ln2\\
+4\widehat{\Lambda}(0,0,2\mathsf{v}_2,2/u_2) \, ,
\end{cases}\nonumber
\end{eqnarray}
with the subtleties of the Kaluza-Klein and winding sums $\widehat{\Lambda}$ on tilted tori as discussed in section~\ref{MS-insertions}.

The non-trivial contributions to the annulus amplitudes with $\mathbb{Z}_2$-projector insertions are:
\begin{equation}
\begin{split}
\tilde{\Delta}^{\mathbb{Z}_2}_{aa}&=-\tilde{\Delta}^{\mathbb{Z}_2}_{aa'}=2\Lambda(0,0,\mathsf{v}_2,u_2)+2\Lambda(1,0,\mathsf{v}_2,u_2),\\
\tilde{\Delta}^{\mathbb{Z}_2}_{a(\mathcal{Q}a)}&=-\tilde{\Delta}^{\mathbb{Z}_2}_{a(\mathcal{Q}a')}=2\Lambda(0,0,\mathsf{v}_2,u_2)+2\Lambda(1,0,\mathsf{v}_2,u_2),\\
\tilde{\Delta}^{\mathbb{Z}_2}_{bb}&=\tilde{\Delta}^{\mathbb{Z}_2}_{cc}=\tilde{\Delta}^{\mathbb{Z}_2}_{hh}={\color{red} 2}\Lambda(0,0,\mathsf{v}_2,1/u_2),\\
\tilde{\Delta}^{\mathbb{Z}_2}_{b(\mathcal{Q}b)}&=\tilde{\Delta}^{\mathbb{Z}_2}_{c(\mathcal{Q}c)}=-\tilde{\Delta}^{\mathbb{Z}_2}_{h(\mathcal{Q}h)}=\Lambda(0,0,\mathsf{v}_2,1/u_2),\\
\tilde{\Delta}^{\mathbb{Z}_2}_{bc}&=\tilde{\Delta}^{\mathbb{Z}_2}_{b(\mathcal{Q}c)}=-\Lambda(0,0,\mathsf{v}_2,1/u_2) \, ,\\
\end{split}
\end{equation}
where the red coloured factor of {\color{red} 2} on the third line corresponds to the one we argued about in section \ref{USP_or_SO}, 
which is needed to obtain meaningful results for the associated beta function coefficients.

The annulus contributions for fixed stacks of D6-branes can then be written as:
\begin{equation}
\begin{split}
\tilde{\Delta}^{\textup{total}}_{aa}&=2\Lambda(0,0,\mathsf{v}_2,u_2)+2\Lambda(1,0,\mathsf{v}_2,u_2),\\
\tilde{\Delta}^{\textup{total}}_{aa'}&=-6\Lambda(0,0,\mathsf{v}_2,u_2)-6\Lambda(1,0,\mathsf{v}_2,u_2),\\
\tilde{\Delta}^{\textup{total}}_{ab}&=\tilde{\Delta}^{\textup{total}}_{ac}=-\Lambda(0,0,\mathsf{v}_1,2)-\Lambda(0,0,\mathsf{v}_3,2),\\
\tilde{\Delta}^{\textup{total}}_{ah}&=-\Lambda(1,0,\mathsf{v}_1,2)-\Lambda(1,1,\mathsf{v}_3,2),\\
\tilde{\Delta}^{\textup{total}}_{bb}&=\tilde{\Delta}^{\textup{total}}_{cc}=2\Lambda(0,0,\mathsf{v}_2,1/u_2),\\
\tilde{\Delta}^{\textup{total}}_{hh}&=0,\\
\tilde{\Delta}^{\textup{total}}_{bc}&=-3\Lambda(0,0,\mathsf{v}_2,1/u_2),\\
\tilde{\Delta}^{\textup{total}}_{bh}&=\tilde{\Delta}^{\textup{total}}_{ch}=-\Lambda(0,0,\mathsf{v}_2,1/u_2).
\end{split}
\end{equation}

The complete threshold correction for the $SU(4)_a$ gauge group is thus given by:
\begin{equation}\label{fullthreSU4a}
\begin{split}
\Delta_{SU(4)_a}=&4(\tilde{\Delta}^{\textup{total}}_{aa}+\tilde{\Delta}^{\textup{total}}_{aa'})+2\tilde{\Delta}^{\textup{total}}_{ab}+2\tilde{\Delta}^{\textup{total}}_{ac}+4\tilde{\Delta}^{\textup{total}}_{ah}+\Delta^{\textup{total}}_{a,\Omega\mathcal{R}}\\
=& -4\Lambda(0,0,\mathsf{v}_1,2) -4\Lambda(1,0,\mathsf{v}_1,2)+4\widehat{\Lambda}(1,0,\mathsf{v}_1,4)
\\
& -16\Lambda(0,0,\mathsf{v}_2,u_2)-16\Lambda(1,0,\mathsf{v}_2,u_2) +8 \widehat{\Lambda}(0,0,\mathsf{v}_2,2u_2)+8 \widehat{\Lambda}(1,0,\mathsf{v}_2,2u_2)\\
&-4\Lambda(0,0,\mathsf{v}_3,2)-4\Lambda(1,1,\mathsf{v}_3,2)+4\widehat{\Lambda}(1,1,\mathsf{v}_3,4) \, ,
\end{split}
\end{equation}
and the complete threshold corrections to the $USp$ gauge groups read:
\begin{equation}\label{fullthreSP2b}
\begin{split}
\Delta_{USp(2)_b}=&4\tilde{\Delta}^{\textup{total}}_{ba}+\tilde{\Delta}^{\textup{total}}_{bb}+\tilde{\Delta}^{\textup{total}}_{bc}+2\tilde{\Delta}^{\textup{total}}_{bh}+\frac{1}{2}\Delta^{\textup{total}}_{b,\Omega\mathcal{R}}\\
=& -4\Lambda(0,0,\mathsf{v}_1,2) -3\Lambda(0,0,\mathsf{v}_2,1/u_2)+2 \widehat{\Lambda}(0,0,2\mathsf{v}_2,2/u_2)
-4\Lambda(0,0,\mathsf{v}_3,2)
+ \frac{\ln 2}{2}\\
=&4\tilde{\Delta}^{\textup{total}}_{ca}+\tilde{\Delta}^{\textup{total}}_{cb}+\tilde{\Delta}^{\textup{total}}_{cc}+2\tilde{\Delta}^{\textup{total}}_{ch}+\frac{1}{2}\Delta^{\textup{total}}_{c,\Omega\mathcal{R}} = \Delta_{USp(2)_c} \, ,
\end{split}
\end{equation}
\begin{equation}
\begin{split}\label{fullthreSP4h}
\Delta_{USp(4)_h}=&4\tilde{\Delta}^{\textup{total}}_{ha}+\tilde{\Delta}^{\textup{total}}_{hb}+\tilde{\Delta}^{\textup{total}}_{hc}+2\tilde{\Delta}^{\textup{total}}_{hh}+\frac{1}{2}\Delta^{\textup{total}}_{h,\Omega\mathcal{R}}\\
=& -4\Lambda(1,0,\mathsf{v}_1,2) -2\Lambda(0,0,\mathsf{v}_2,1/u_2)+2\widehat{\Lambda}(0,0,2\mathsf{v}_2,2/u_2)
-4\Lambda(1,1,\mathsf{v}_3,2) +\frac{\ln 2}{2}
\, .
\end{split}
\end{equation}

\begin{figure}[h!]
\begin{center}
\includegraphics[width=17cm]{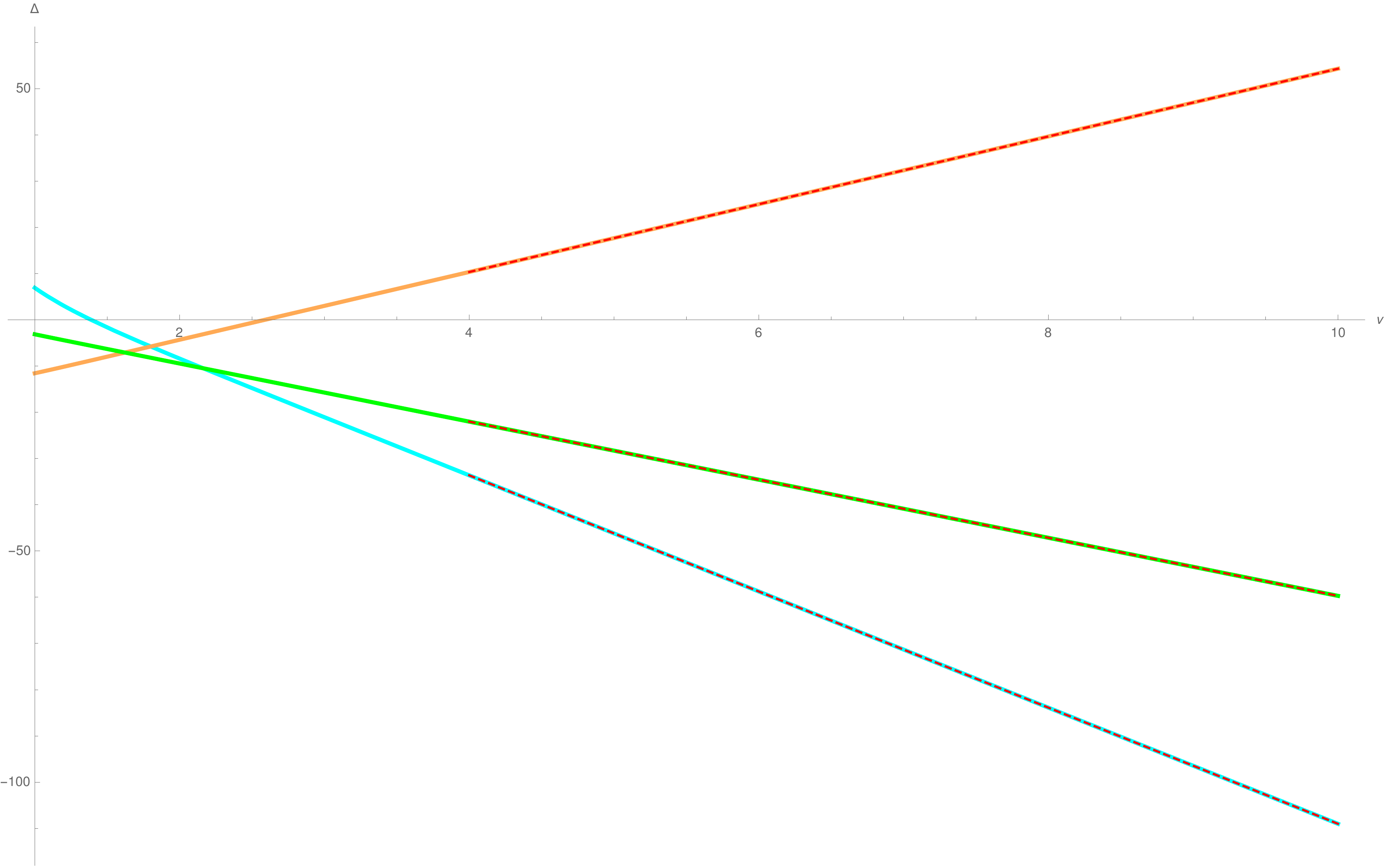}
\caption{Gauge threshold corrections and their asymptotic behaviour in dependence of the two-torus volume $\mathsf{v}$ for the $T^6/(\mathbb{Z}_4 \times \OR)$ example
under the assumption that the M\"obius strip contribution does not depend on the discrete displacement and Wilson line parameters.
 The contribution to $SU(4)_a$ is shown in solid blue, to  $USp(2)_b$ and $USp(2)_c$ in solid orange and to $USp(4)_h$ in solid green. 
For computational purposes, $u_2=1$ has been chosen.}
\label{fig:threshold2}
\end{center}
\end{figure}

A numeric evaluation of the gauge threshold corrections for the particular case of isotropic two-tori volumes $\mathsf{v}_1=\mathsf{v}_2=\mathsf{v}_3\equiv \mathsf{v}$ 
under the (unrealistic at least for $(\sigma,\tau)=(1,1)$) assumption that $\widehat{\Lambda} (\sigma,\tau,\mathsf{v}_i,V) = \Lambda (0,0,\mathsf{v}_i,V)$
is shown in figure \ref{fig:threshold2}.\footnote{As discussed e.g. in section 4.1 of~\cite{Honecker:2017air}, the correct shape of the Kaluza-Klein and winding sum 
$\widehat{\Lambda} (\sigma,\tau,\mathsf{v}_i,V)$  on tilted tori for $(\sigma,\tau) \neq (0,0)$ is to date not known, cf. also
our discussion in the context of the M\"obius strip amplitudes, in particular footnote~\ref{foot:MS-subtleties} in section~\ref{MS-insertions} of the present article.}
The results for $\Delta_{SU(4)_a}$ and $\Delta_{USp(4)_h}$ are independent of the complex structure parameter $u_2$ while those for $\Delta_{USp(2)_b}=\Delta_{USp(2)_c}$ vary slightly with it, but the overall behaviour is the same as for the depicted value $u_2=1$, which was chosen for computational purposes. Depending on the value of $\mathsf{v} \leqslant 2.5$ or $\mathsf{v} > 2.5$, the threshold corrections of the left- and right-symmetric gauge groups $USp(2)_b$ and $USp(2)_c$ can be either enhanced (negative $\Delta$) or reduced (positive $\Delta$).
The gauge couplings of $SU(4)_a$  and $USp(4)_h$ will for any value $\mathsf{v} > 1.5$ be enhanced.

To estimate the uncertainty due to the unknown correct shape of the lattice sums in the M\"obius strip,  
the asymptotic behaviour of the function $\Lambda(\sigma_{ab},\tau_{ab},\mathsf{v},V_{ab})$ for $\mathsf{v}>1$ can be used.
It depends on the values of $\sigma_{ab}$ and $\tau_{ab}$ (see e.g. \cite{Honecker:2012qr,Honecker:2017air}) and is for the annulus lattice sums given by:
\begin{equation}
\begin{split}
\Lambda(0,0,\mathsf{v},V)&\rightarrow \ln(2\pi V)-\frac{\pi \mathsf{v}}{3},\\
\Lambda(0,1,\mathsf{v},V)&\rightarrow 2\ln 2-\frac{\pi \mathsf{v}}{3},\\
\Lambda(1,\tau,\mathsf{v},V)&\rightarrow \frac{\pi \mathsf{v}}{6} \,
\end{split}
\end{equation}
while for the unknown M\"obius strip contributions we make the ansatz per two-torus $T^2_{(i)}$ as in~\cite{Honecker:2017air}:
\begin{equation}
\widehat{\Lambda}(\sigma, \tau, \mathsf{v}, V) \rightarrow - c^{(i)}_{\sigma,\tau} \frac{\pi \mathsf{v}}{3} + \text{const.} ,
\end{equation}
with $c^{(i)}_{\sigma,\tau} ={\cal O}(1)$ expected. For an untilted two-torus, such as the {\bf A}-type $B_2$ on the $\Z_4$ orientifold under consideration, $c^{(3), {\bf A}}_{\sigma,\tau} \equiv 1$ holds whenever only discrete displacement and Wilson line parameters are allowed, i.e. $\sigma,\tau \in \{0,1\}$.
Therefore, the asymptotic behaviours of the threshold corrections \eqref{fullthreSU4a}, \eqref{fullthreSP2b} and 
\eqref{fullthreSP4h} read:
\begin{equation}\label{eq:Delta-asymptotics-example}
\begin{split}
\Delta_{SU(4)_a}&\rightarrow
 \frac{2\pi}{3}\Bigl( (1- 2 c^{(1)}_{1,0}) \, \mathsf{v}_1 + 4 (1-  c^{(2)}_{0,0}-c^{(2)}_{1,0})  \, \mathsf{v}_2 + (1 - 2 c^{(3)}_{1,1}) \,  \mathsf{v}_3 \Bigr) 
 + \text{const.} \, , \\
\Delta_{USp(2)_b}=\Delta_{USp(2)_c}&\rightarrow 
\frac{4\pi}{3}(\mathsf{v}_1+\mathsf{v}_3) + (3-4 c^{(2)}_{0,0}) \frac{\pi}{3}\mathsf{v}_2
+ \text{const.} \, , \\
\Delta_{USp(4)_h}&\rightarrow -\frac{2\pi}{3}(\mathsf{v}_1+\mathsf{v}_3) + (1-2 c^{(2)}_{0,0}) \frac{2\pi}{3}\mathsf{v}_2
+ \text{const.} \, 
.
\end{split}
\end{equation}
It is noteworthy that the thresholds of the $USp$ factors only contain the unconfirmed asymptotics $c^{(2)}_{0,0} = {\cal O}(1)$ of the M\"obius strip sector 
along the second two-torus $T^2_{(2)}$, all with the same sign. The gauge couplings of the left- and right-symmetric groups $USp(2)_b \times USp(2)_c$ have identical tree-level values~\eqref{eq:g-tree-example},
identical one-loop  beta function coefficients~\eqref{beta-values} and also identical threshold corrections. The contributions of the first and third two-torus, $\mathsf{v}_1$ and 
$\mathsf{v}_3$, are twice as large and with opposite sign compared  to those of the `hidden' gauge group $USp(4)_h$. The left-/right-symmetric groups thus become more weakly coupled with increasing volumina along $T^2_{(1)} \times T^2_{(3)}$, whereas the gauge coupling of the `hidden' stack becomes stronger. This feature opens up the possibility of gaugino condensation in the `hidden' sector leading to supersymmetry breaking. 

The gauge threshold correction to the $SU(4)_a$ group in equation~\eqref{eq:Delta-asymptotics-example}
has an undetermined contribution from the M\"obius strip amplitudes along each two-torus. Clearly, the coefficients $c^{(i)}_{\sigma^i_a,\tau^i_a}$ are expected to dominate the shape of the asymptotics, highlighting the importance of determining all details in a phenomenological appealing model meticulously.

%%%%%%%%%%%%%%%%%%%%%%%%%%%%%%%%%
\section{Conclusion}\label{conclusion}

In this article, we initiated the development of CFT methods to study the low-energy effective action of Type IIA orientifolds with fractional D6-branes on non-factorisable toroidal orbifold
backgrounds. 
To our best knowledge, the only preceding work in this direction deals with holomorphic Yukawa couplings on the $D_6$ lattice~\cite{Forste:2014bfa}.
Based on the conclusion of~\cite{Blaszczyk:2011hs} that non-factorisable toroidal backgrounds can be formulated in a factorised form by imposing an extra shift symmetry, 
we found a set of factorisable coordinates for the $\Z_4 \times \OR$ orientifold with \mbox{$A_3 \times A_1 \times B_2$} lattice
and rewrote the classification of all possible anti-holomorphic involutions ${\cal R}_i$ accompanying the worldsheet parity operation, 
as well as the RR tadpole cancellation and the supersymmetry conditions from~\cite{Seifert:2015fwr,Berasaluce-Gonzalez:2016kqb}, 
adding the K-theory constraints here for the first time. 

After defining toroidal, bulk and exceptional wrapping numbers in the factorised picture and classifying three-cycles according to their (non)-invariance under the new shift symmetry, we were able to generalise the expressions for one-loop open string vacuum amplitudes (which upon worldsheet duality transform into closed string tree-level amplitudes)
 from factorisable toroidal orbifolds to non-factorisable ones. The shift symmetry enters on the one hand via the counting of independent D6-brane and O6-plane intersections and on the other hand via a duplication of some winding sums. 
By using the trick of magnetic gauging along the non-compact directions, we were able to derive for the first time the one-loop beta function coefficients of the gauge couplings and thereby determine the vector-like massless open string spectrum on the $A_3 \times A_1 \times B_2$ lattice
and pinpoint that all gauge group enhancements along orientifold-invariant fractional three-cycles lead to $USp(2N)$ gauge groups, which in turn allowed us to determine the K-theory constraints and selection rules for remnant discrete $\Z_n$ gauge symmetries from massive $U(1)$ factors for this class of non-factorisable $\Z_4 \times \OR$ orientifolds.

Furthermore, we computed here for the first time the massless closed string spectra of the $\Z_4 \times \OR$ orientifolds on the $A_3 \times A_1 \times B_2$ lattice.
The result in table~\ref{tab:closed-string-spectrum} on the one hand supports our conjecture on the existence of duality relations at the topological level, which was based on the counting 
of three-cycles of a given length in~\cite{Berasaluce-Gonzalez:2016kqb} and is now extended to discrete D6-brane data 
(wrapping numbers, $\Z_2$ eigenvalue, displacements and Wilson lines) in the factorised language. On the other hand, the spectrum in the ${\cal Q}+{\cal Q}^3$ sector shows the well-known behaviour that the K\"ahler modulus of the orbifold theory is for some $\Z_4$ fixed loci projected out by the orientifold action, and instead the vector from within the ${\cal N}=2$ supersymmetric multiplet is preserved. A full resolution of all $\Z_4$ singularities is thus only possible for the choice of the ${\bf A}_{\bf a}{\bf BA} \leftrightarrow {\bf A}_{\bf a}{\bf BB}$ background orientation.

As demonstrated for an explicit example, phenomenologically appealing string vacua with chiral matter on fractional D6-branes usually require non-trivial combinations of discrete Wilson line and displacement data on tilted tori, which in the T-dual formulation of magnetised fractional D-branes corresponds to a non-trivial $B$-field background.
For these, the M\"obius strip amplitudes have so far not been established, and further examples on other non-factorisable $\Z_N \times \OR$ orientifolds
will have to be investigated to verify if transferring naively the conjectured non-trivial sign factor in the M\"obius strip contribution to the beta function coefficient from the factorisable to the non-factorisable backgrounds is indeed correct. Two classes of examples, which are expected to be favourable for three-generation model building due to some $\Z_3$ subgroup are the $\Z_6 \times \OR$ orientifolds~\cite{Seifert:2017aa}, which again show the reduction of geometric moduli on non-factorisable tori: for $\Z_{6-I}$ these are
 $(h_{1,1}^{twisted} \, , \, h_{2,1}^{twisted}) = (24,5)$ and $(20,1)$ before orientifolding on the lattices  $A_2 \times (G_2)^2$ and $(G_2 \times (A_2)^2)^b$, respectively,
and for $\Z_{6-II}$  $(h_{1,1}^{twisted} \, , \, h_{2,1}^{twisted}) = (32,10)$, $(28,6)$, $(26,4)$ and $(22,0)$ for the $A_1^2 \times A_2 \times G_2$, $(A_1^2 \times A_2^2)^{\#}$, $A_2 \times D_4$, $A_1 \times A_5$ backgrounds, cf. e.g.~\cite{Erler:1992ki,Lust:2005dy}. 

In this article, we initiated the computation of vacuum amplitudes for fractional D6-branes on non-factorisable toroidal backgrounds. To derive Yukawa and higher order interaction terms, the CFT framework will have to be extended to vertex operator insertions, which to date are only known for pure bulk branes on factorisable toroidal backgrounds. However, following the 
method employed e.g. in~\cite{Akerblom:2007uc,Honecker:2011sm} the K\"ahler metrics at leading order can be read off by matching the threshold corrections in the string frame
computed in the present article
 to the one-loop expansions of gauge couplings in the supergravity frame, thereby providing the non-holomorphic part of the Yukawa couplings at leading order. Additionally, the holomorphic gauge kinetic function 
- which is also extracted by the string to supergravity frame matching - will reappear in the computation of D-brane instanton corrections, cf. e.g. the review~\cite{Blumenhagen:2009qh}. 

Finally, developing the low-energy effective field theory for more generic backgrounds than just the (orientifolded) factorised six-torus and orbifolds thereof using CFT methods, will provide important information also for corners of the string landscape where to date only supergravity methods are available. A step into this landscape can be provided by combining the resolution techniques studied 
in e.g.~\cite{Lust:2006zh,Blaszczyk:2011hs} with generalisations of the deformation methods initiated in~\cite{Blaszczyk:2014xla,Blaszczyk:2015oia,Koltermann:2015oyv,Honecker:2017air}.

\noindent
{\bf Acknowledgements:} 
The authors thank Igor Buchberger for useful discussions. \\
This work is partially supported by the {\it Cluster of Excellence `Precision Physics, Fundamental Interactions and Structure of Matter' (PRISMA)} DGF no. EXC 1098,
the DFG research grant HO 4166/2-2, and the DFG Research Training Group {\it `Symmetry Breaking in Fundamental Interactions'} GRK 1581.

\appendix

\section{Determination of the values of $\kappa$}\label{kappa}

In section~\ref{ShiftSymm} we introduced the parameter $\kappa$ to take into account the effects of the replicas that appear as a consequence of the shift symmetry without considering them directly. In this section we will explain how each of the possible values of $\kappa$ arises.

The main idea is that, instead of computing all possible contributions to a given amplitude separately and then dividing by two to account for the shift symmetry, we determine first which contributions are independent and only consider those. A possible way of selecting each contribution once (depending on the types of cycles wrapped by the branes) is as follows:
\begin{itemize}
\item None of the branes wraps a cycle of type 4. In this case we have two pairs of branes. We only need to compute the contributions of one copy of one of the pairs with the full second pair. Figure \ref{intersectingbranepairs} shows the case of branes wrapping $(2,1)\times(1,1)$ and $(1,-2)\times(0,1)$, where both stacks intersect at points in all two-tori. The intersection points of (solid blue, solid red) and (dashed blue, dashed red) are identified under the shift symmetry, and so are (solid blue, dashed red) and (dashed blue, solid red). Therefore, we take into account all contributions just by considering the intersection of the solid blue stack with both the solid and dashed red ones. And the total number is twice the intersection number between representatives of each pair, thus $\kappa=2$.
\begin{figure}[h]
\begin{center}
\includegraphics[width=17cm]{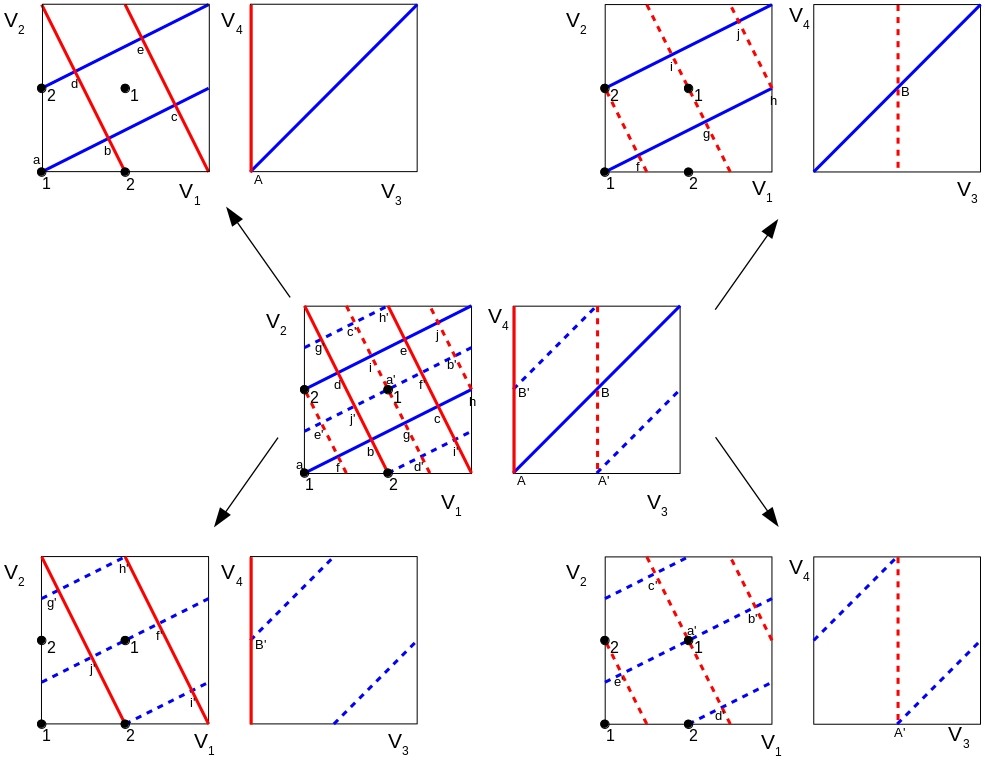}
\caption{Intersection pattern of branes wrapping $(2,1)\times(1,1)$ and $(1,-2)\times(0,1)$. 
Parallel solid and dashed lines in the central representation of $T^2_{(1)} \times T^2_{(2)}$ are identified under the shift symmetry. The intersection points of (solid blue, solid red) and (dashed blue, dashed red) are identified under the shift symmetry, and so are (solid blue, dashed red) and (dashed blue, solid red).
}
\label{intersectingbranepairs}
\end{center}
\end{figure}
\item One brane wraps a cycle of type 4. In this case we just take the contributions of the invariant brane and one of the remaining pair. This is a particular situation of the previous case, since the invariant cycle is equivalent to a pair of non-invariant cycles. In this case $\kappa=1$ (one can think of it as the 2 coming from $\kappa$ in the case above being absorbed by the wrapping numbers of the shift-invariant three-cycle).
\item Both branes wrap cycles of type 4. The trick here is to divide one of the invariant cycles in two halves, and treat it as a pair of replicas. There are two obvious ways of doing this, which are presented in figure \ref{type4cycles}. The first option corresponds to the case with half-integer wrapping numbers in the second two-torus that was mentioned in the main part of this work. The second way of splitting is useful when the two stacks are parallel on the second two-torus. As in the previous cases, the independent intersection points are computed by taking one-half of a cycle (in a given colour) and the remaining full cycle (without splitting into two colours). In this particular configuration, it is easy to see that only half the intersection points of the two cycles are independent, which gives $\kappa=1/2$. An example is shown in figure \ref{paralleltype4cycles}, where the cycles are parallel on the second two-torus, so we decompose one of the cycles using the second option in figure \ref{type4cycles} (the points where the colour changes have been shifted a bit so that they do not coincide with the intersection points). The pairs of intersection lines labelled by (1,5), (2,6), (3,7) and (4,8) along $T^2_{(1)}$ are identified under the shift symmetry. Therefore we only need to compute the contribution from one element of each pair, for instance, 2, 3, 4 and 5, which are the ones lying on the red half of the cycle.
\begin{figure}
\begin{center}
\includegraphics[width=15cm]{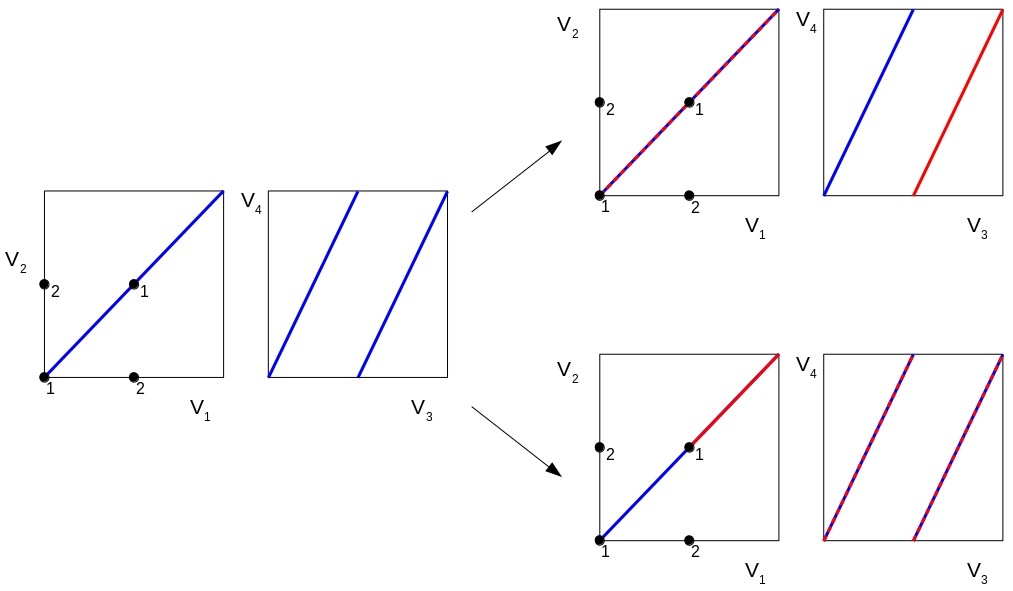}
\caption{Two different ways one can divide a shift symmetry invariant cycle into the two copies of the cycle in the non-factorisable picture. The first option corresponds to the case with half-integer wrapping number on the second two-torus that has been discussed in the main part.}
\label{type4cycles}
\end{center}
\end{figure}
\begin{figure}
\begin{center}
\includegraphics[width=12cm]{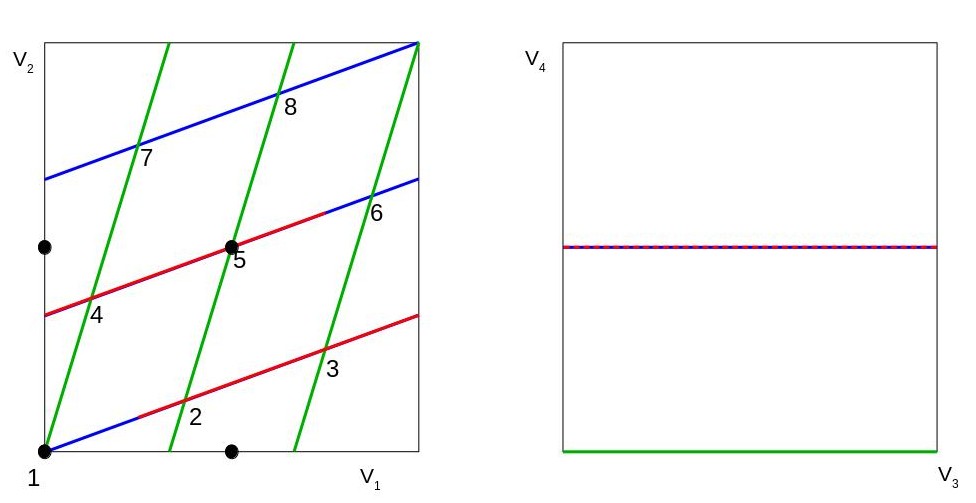}
\caption{Intersection pattern of branes wrapping $(1,3)\times(1,0)$ and $(3,1)\times(1,0)$. The pairs of intersection lines labelled by (1,5), (2,6), (3,7) and (4,8) along $T^2_{(1)}$ are identified under the shift symmetry, therefore we only need to compute the contributions coming from the points crossed by the red line.
}
\label{paralleltype4cycles}
\end{center}
\end{figure}
\end{itemize}
%%%%%%%%%%%%%%%%%%%%%%%%%%%%%%%%%%%%%%%%%%%%%%%%%%%%

\bibliographystyle{ieeetr}
\bibliography{refs_NonFactorisable}
\end{document}